\documentclass{aa}
\usepackage{graphicx,times,amssymb,longtable}
\begin{document}
   \title{Rotation and accretion of very low mass\\
   objects in the $\sigma$\,Ori cluster\thanks{Table 7 is only available in electronic form at the CDS via anonymous ftp 
   to cdsarc.u-strasbg.fr (130.79.128.5) or via http://cdsweb.u-strasbg.fr/cgi-bin/qcat?J/A+A/}}

   \author{Alexander Scholz\inst{1}\thanks{Visiting Astronomer at the German-Spanish Astronomical Centre, Calar Alto, 
      operated by the Max-Planck-Institute for Astronomy, Heidelberg, jointly with the Spanish
      National Commission for Astronomy}
          \and
          Jochen Eisl{\"o}ffel\inst{1}$^{**}$
          }

   \offprints{Alexander Scholz, scholz@tls-tautenburg.de}

   \institute{Th{\"u}ringer Landessternwarte Tautenburg,
              Sternwarte 5, D-07778 Tautenburg, Germany}

   \date{Received June 27, 2003; accepted December 27, 2003}

   \authorrunning{A. Scholz and J. Eisl{\"o}ffel}
   \titlerunning{Rotation and accretion of VLM objects in the $\sigma$\,Ori cluster}

   \abstract{We report on two photometric monitoring campaigns of Very Low Mass (VLM) objects in the
   young open cluster around $\sigma$\,Orionis. Our targets were pre-selected with multi-filter photometry 
   in a field of 0.36\,sqdeg. For 23 of these objects, spanning a mass range from 0.03 to 0.7\,$M_{\odot}$,
   we detect periodic variability. Of these, 16 exhibit low-level variability, with amplitudes of less
   than 0.2\,mag in the I-band, which is mostly well-approximated by a sine wave. These periodicities are probably 
   caused by photospheric spots co-rotating with the objects. In contrast, the remaining variable targets 
   show high-level variability with amplitudes ranging from 0.25 to 1.1\,mag, consisting of a periodic
   light variation onto which short-term fluctuations are superimposed. This variability pattern is very similar 
   to the photometric behaviour of solar-mass, classical T Tauri stars. Low-resolution spectra of a few of 
   these objects reveal strong H$\alpha$ and Ca-triplet emission, indicative of ongoing accretion processes.
   This suggests that 5-7\% of our targets still possess a circumstellar disk. In combination with previous results
   for younger objects, this translates into a disk lifetime of 3-4\,Myr, significantly shorter than for
   solar mass stars. The highly variable objects rotate on average slower than the low-amplitude variables,
   which is expected in terms of a disk-locking scenario. There is a trend towards faster rotation with 
   decreasing mass, which might be caused by shortening of the disk lifetimes or attenuation of magnetic 
   fields.
   
   }

   \maketitle
%

\section{Introduction}
\label{intro}

Open clusters are ideal environments to study stellar properties and evolution, 
because they contain a homogeneous population of objects with known distance, metallicity,
and age. Recent deep surveys have unveiled the population of several open clusters far
down into the substellar regime. Examples are the surveys by Moraux et al. (\cite{mbs03}),
Pinfield et al. (\cite{phj00}) and Zapatero Osorio et al. (\cite{zrm99}, \cite{zrm97}) in the 
\object{Pleiades}, and the work of Barrado y Navascu\'es et al. in \object{IC2391} (\cite{bsb01}) 
and \object{$\alpha$\,Per} (\cite{bbs02}). In two cases, namely in the clusters around \object{$\sigma$\,Ori} 
(Zapatero Osorio et al. \cite{zbm00}) and the Orion \object{Trapezium Cluster} (Lucas \& Roche \cite{lr00}), 
even isolated planetary mass objects were found. Thanks to these surveys, large 
samples of Very Low Mass (VLM) objects are known today, among which we conveniently will subsume all objects 
with masses below $0.4\,M_\odot$, including very low mass stars, brown dwarfs and free-floating 
planetary mass objects. All objects with masses below this $0.4\,M_\odot$ limit are thought to 
be fully convective (Chabrier \& Baraffe \cite{cb00}), giving a physical motivation for this 
definition of VLM objects.

Wide-field photometric monitoring is a powerful tool to investigate properties of large object samples: 
If an object exhibits asymmetrically distributed surface features, e.g. magnetically induced
star spots, its flux will be modulated with the rotation period. Hence, period search in 
the lightcurve allows the determination of the (projection-free) rotation period. The amplitude of 
the periodicity, in turn, depends on the properties of the star spots, allowing conclusions
about surface activity processes. Pointing to dense open cluster fields, such monitoring campaigns 
become very efficient, as one can register many objects contemporaneously within one field of view. 

Compared to solar-mass stars, there is a significant lack of known rotation periods for
VLM objects. Photometric monitoring studies delivered a small number of periods for evolved
VLM stars and Brown Dwarfs (Bailer-Jones \& Mundt \cite{bm99}, \cite{bm01}, 
Tinney \& Tolley \cite{tt99}, Mart\'{\i}n et al. \cite{mzl01}, Clarke et al. \cite{ctc02}, Gelino et 
al. \cite{gmh02}). Only three periods are known for VLM objects with ages between 50 and 100\,Myr 
(Mart\'{\i}n \& Zapatero Osorio \cite{mz97}, Terndrup et al. \cite{tkp99}).
Two periods for VLM members of $\sigma$\,Ori were published by Bailer-Jones \& Mundt (\cite{bm01}).
Recently, Joergens et al. (\cite{jfc03}) report on 5 measured periods for Brown Dwarfs and VLM stars
in the very young ChaI star forming region. Our own study in the
young open cluster IC4665 (age 36\,Myr) delivered several rotation periods for VLM objects 
(Eisl{\"o}ffel \& Scholz \cite{es02}). 

Observations of very young VLM objects are of special interest, because they can 
deliver clues about their formation process. Recent results suggest that
objects in the substellar regime form similar to stars. Several authors 
detected typical T Tauri star phenomena on VLM objects, e.g. outflow processes 
(Fern\'andez \& Comer\'on \cite{fc01}) and mid-infrared excesses attributed to the presence of a 
circumstellar disk (Natta \& Testi \cite{nt01}, Apai et al. \cite{aph02}, Jayawardhana et al. 
\cite{jas03}). Moreover, L\'opez Mart\'{\i} et al. (\cite{les03}) and Liu et al. (\cite{lnt03}) 
demonstrate a correlation between H$\alpha$ emission and mid-infrared excess for Brown Dwarfs down 
to $0.02\,M_\odot$, which they interpret as an indication for ongoing accretion. 

Solar-mass T Tauri stars show various types of photometric variability. Active accretion
processes often manifest themselves by large amplitude variations showing several possible
periods (e.g., Fern\'andez \& Eiroa \cite{fe96}, Bouvier et al. \cite{bck95}, Herbst et 
al. \cite{hmw00}). If VLM objects undergo a T Tauri phase as well, they should exhibit a 
similar photometric behaviour. Hence, photometric monitoring can deliver an independent 
contribution to the ongoing debate about VLM object formation.

The $\sigma$\,Ori cluster is an appropriate target for such a study, because it is relatively
nearby (350\,pc, B\'ejar et al. \cite{bzr99}), has negligible extinction ($E_\mathrm{B-V}=0.05$, 
B\'ejar et al. \cite{bzr99}) and an age of 3\,Myr (Zapatero Osorio et al. \cite{zbp02}). The extended 
work by B\'ejar et al. and Zapatero Osorio et al. revealed a rich VLM population. We monitored this 
cluster in two photometric time series. Complementary observations, presented in Sect.\,\ref{sel}, 
identified the cluster members in the time series field. We report on the monitoring
campaigns in Sect.\,\ref{mon} and the time series analysis in Sect.\,\ref{tsa}. The following 
sections describe the results of out lightcurve analysis: We first establish two origins for 
the observed variability (Sect.\,\ref{inter}). In Sect.\,\ref{rot}, we then concentrate on the 
investigation of the rotation periods. Section\,\ref{accr} contains the discussion 
of low-resolution spectra for highly variable objects. Finally, in Sect.\,\ref{conc}, we 
present our conclusions. 
 

\section{Selection of VLM objects in the $\sigma$\,Ori cluster}
\label{sel}

\begin{figure}
    \centering
    \resizebox{\hsize}{!}{\includegraphics[angle=-90]{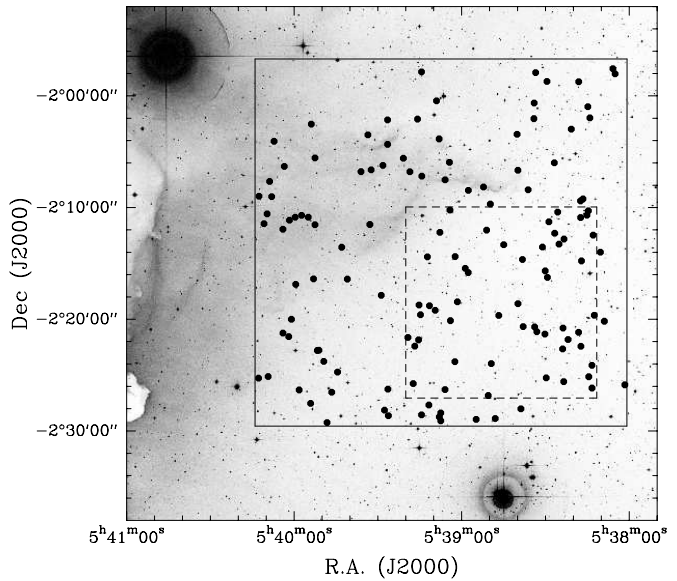}}
    \caption{Time series fields in the $\sigma$\,Ori cluster: The TLS field is indicated
    with solid, the CA field with dashed lines. The positions of the cluster member
    candidates (see Table\,7) are overplotted. $\sigma$\,Ori is the bright
    star near the bottom of the field.}
    \label{field}
\end{figure}

We searched for VLM members of the $\sigma$\,Ori cluster using the 2-m Schmidt 
camera at the Th{\"u}ringer Landessternwarte Tautenburg (TLS) and the 1.23-m telescope
on Calar Alto (CA) in combination with infrared photometry from 
2MASS\footnote{Catalogue available under {\it http://www.ipac.caltech.edu/2mass}}. 
Figure\,\ref{field} shows the $36'\times36'$ TLS field and the $17'\times17'$ CA field in the 
north of $\sigma$\,Ori. These survey images were obtained as part of the time series 
campaigns (see next section). The CA field was observed in the R- and I-band with exposure 
times of 30 and 600\,sec, respectively. For the TLS field, we used a 600-sec-exposure in the I-band. 
These images were reduced with the standard routines within the {\it ccdred} package
of IRAF\footnote{IRAF is distributed by National Optical Astronomy Observatories, which is 
operated by the Association of Universities for Research in Astronomy, Inc., under contract 
with the National Science Foundation.},
including overscan and bias subtraction, as well as flatfield correction.
The I-band fringe pattern caused by night sky emission was corrected with
a fringe mask constructed from blank field images. The sigma-clipped median 
of these frames ('superflat') was smoothed and subtracted from the original superflat,
delivering the required fringe mask. Because the fringe mask contains only an average 
fringe pattern, this mask must be scaled to properly subtract the fringes in the individual 
images. The scaling factor was determined as the ratio of the fringe amplitude in the fringe 
mask and in the respective image.

The pixel positions of the objects in the I-band images were determined with
the SExtractor software (Bertin \& Arnouts \cite{ba96}). The respective positions in the
R-band image were found by a linear transformation determined with several bright stars 
whose pixel positions were manually determined in R and I. The pixel positions
were transformed to sky coordinates with the known sky coordinates of 
unsaturated HST guide stars (Morrison et al. \cite{mrm01}) in our field using the 
{\it ccmap} routine of IRAF. The coordinate precision is $\pm1\farcs0$.

\begin{figure}[ht]
    \centering
    \resizebox{\hsize}{!}{\includegraphics[angle=-90]{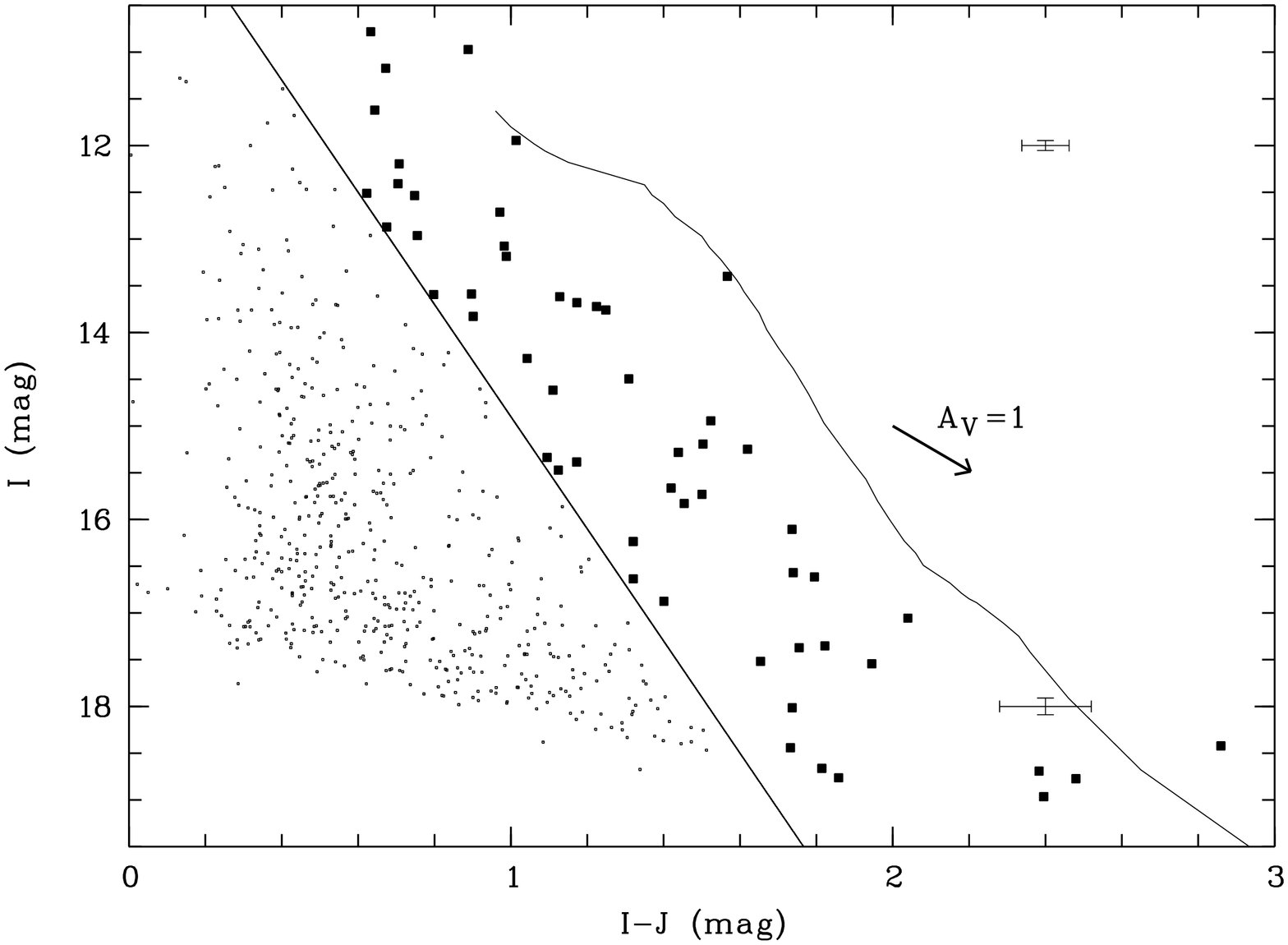}}
    \resizebox{\hsize}{!}{\includegraphics[angle=-90]{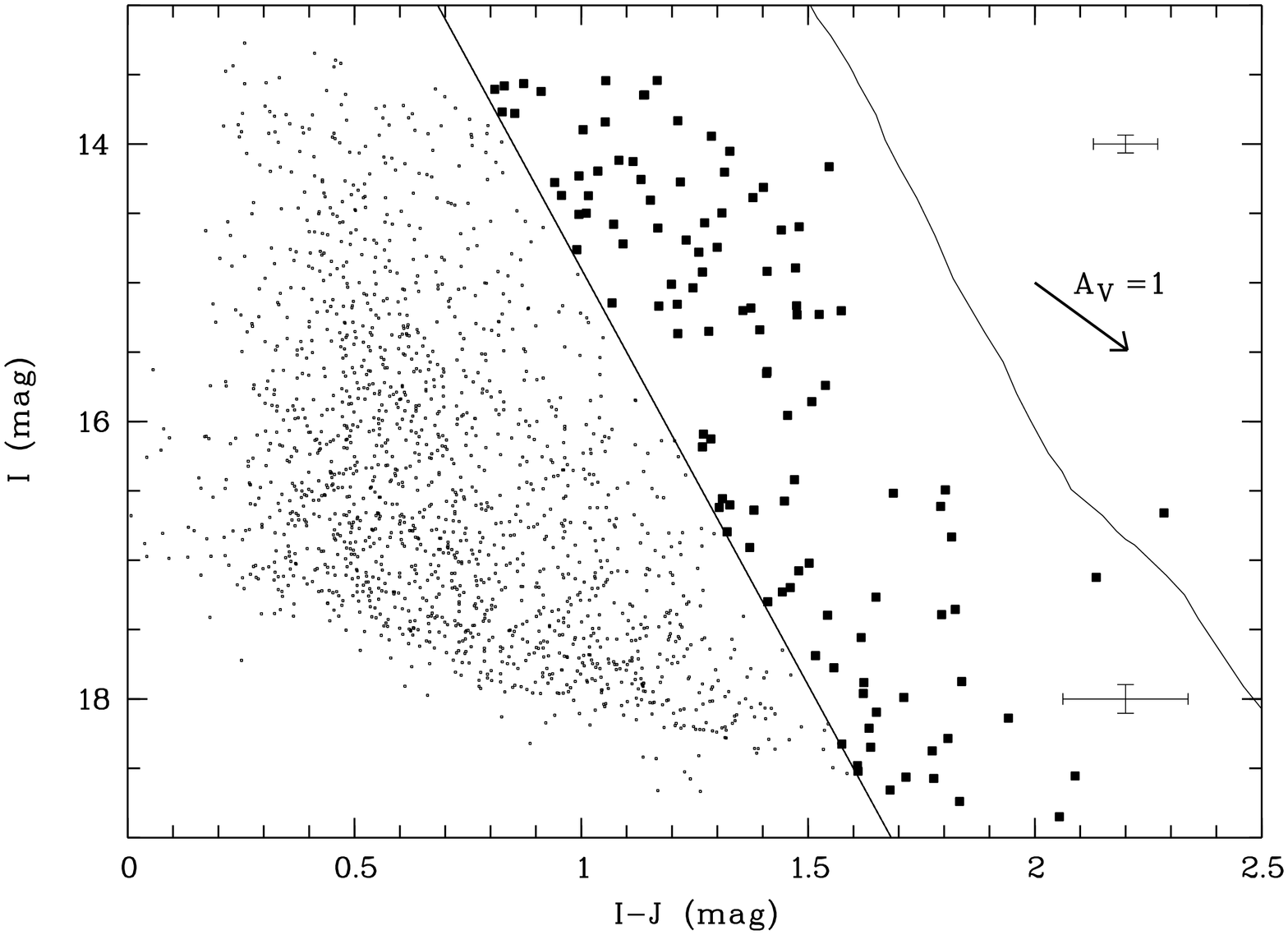}}
    \caption{ (I,I-J) colour magnitude diagram for the CA (upper panel) and the TLS field
    (lower panel). All probable cluster members are marked as larger dots. Error bars indicate 
    typical photometry errors for the candidates. The separation line between cluster member 
    candidates and field objects is shown as a straight line. The position of the 3\,Myr isochrone 
    from Baraffe et al. (\cite{bca98}) is indicated by the right-hand line. The arrow shows the 
    reddening vector for $A_V=1$\,mag.
    \label{cmdij}}
\end{figure}

Instrumental magnitudes for all objects in the TLS and CA catalogues were determined 
by PSF fitting (TLS) or aperture photometry (CA) using the {\it daophot} package within 
IRAF (Stetson \cite{s87}). The PSF fitting photometry was considered to be more appropriate 
for the TLS data, because the poor seeing turned the cluster into a 'crowded field'. 
Near-infrared photometry in J, H and K for both fields was obtained from the 2MASS database. 

In order to calibrate the CA images, we observed two standard fields (Landolt 
\cite{l92}) at an airmass comparable with that of the $\sigma$\,Ori field. I-band zero-point and colour 
coefficient were derived fitting the instrumental magnitudes and the catalogue magnitudes
with the following relation (I -- Landolt magnitude; i,r -- instrumental magnitudes):
\begin{equation}
I = i_\mathrm{CA} + ZP + C(r_\mathrm{CA}-i_\mathrm{CA})
\end{equation}
With the resulting parameters ($ZP=3.45$, $C=0.171$), the $\sigma$\,Ori magnitudes were 
converted to the Landolt system. The high colour coefficient points at significant 
differences between the CA and the Landolt I-band. Since the Landolt
standards are mainly stars with R-I$<$1.0, one has to be cautious to apply this colour 
correction to very red targets. For such objects, we may overestimate the
I-band flux.

The CA field completely overlaps with the TLS field.
Therefore, the TLS magnitudes were calibrated using objects contained in both 
catalogues. We noted a significant colour dependency of the zero-point offset between
CA and TLS I-band, therefore we applied a colour correction using the 2MASS J-band magnitudes.
The following transformation was used to convert TLS magnitudes into the Landolt system.
\begin{equation}
I = i_\mathrm{TLS} + ZP + C(i_\mathrm{TLS}-J)
\end{equation}
After this procedure, I-band magnitudes from both telescopes are available in the same
photometric system.

Cluster member candidates were selected from (I,I-J) colour magnitude diagrams (Fig.\,\ref{cmdij}). 
In both diagrams, a cumulation of objects to the right of the field stars is 
clearly visible. To determine the position of this cumulation precisely, we 
divided the colour magnitude diagrams (CMD) in horizontal bins of $\Delta I=1$\,mag. The
histograms of the I-J colour of these bins typically show a broad maximum on the left
and a second smaller maximum on the right side. This second peak indicates
the position of the cluster isochrone. We registered the I-J value of this peak for each bin
in both diagrams and fitted these values linearly. This fit, the empirical isochrone, 
shifted 0.3\,mag to the left to account for photometric errors and the uncertainty of the 
derived isochrone, was used as dividing line between cluster member candidates and field 
stars (see Fig.\,\ref{cmdij}, straight line). All objects to the right side of this 
separation line are potential low-mass cluster members. Some of these objects clearly appear
reddened in the CMD. Since extinction towards $\sigma$\,Ori is negligible (B\'ejar et al. 
\cite{bzr99}), this is a sign for intrinsic reddening, as found for objects with similar masses
in the same cluster (Oliveira et al. \cite{ojk02}). We selected 55 candidates from the 
CA and 112 from the TLS diagram. The TLS candidates contain 28 objects also identified with 
the CA photometry. The remaining CA candidates are either too bright for the TLS photometry 
or not resolved in the TLS image.

In Fig.\,\ref{cmdij}, we also indicate the position of the Baraffe et al. (\cite{bca98}) isochrone
for the $\sigma$\,Ori cluster. The I-J colours of our candidates are significantly smaller than 
those from the evolutionary track for the appropriate age. This difference is possibly caused 
by insufficient colour correction in the I-band (see above) as well as shortcomings in the modeled
BD atmospheres and colours. However, it is valid to compare the 2MASS photometry with the 
Baraffe et al. isochrone. Systematic mismatches between the 2MASS photometric system and the 
CIT system used for the model isochrones are safely below 3\% (Carpenter \cite{c01}) and thus 
negligible for our purposes. 

\begin{figure}[htbd]
    \centering
    \resizebox{\hsize}{!}{\includegraphics[angle=-90]{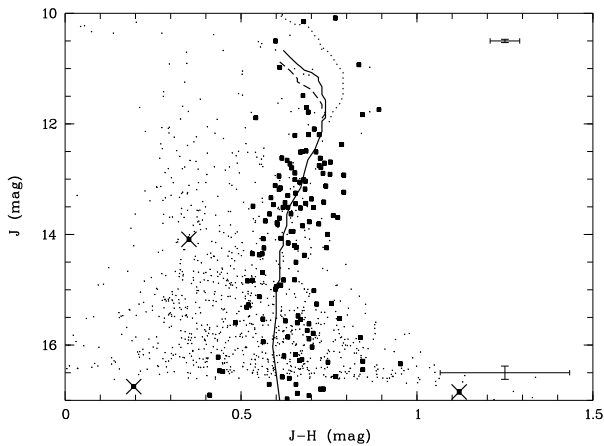}}
    \caption{(J,J-H) colour magnitude diagram for all candidates (plotted as filled squares). Error 
    bars indicate typical photometry errors for the candidates. The 3\,Myr isochrone for 
    $\sigma$\,Ori (Baraffe et al. \cite{bca98}) is shown as a solid line. For bright objects, the isochrones 
    for 1\,Myr (dotted line) and 5\,Myr (dashed line) differ significantly from the 3\,Myr track and
    are therefore also shown. Objects whose J-H colour deviates clearly from the cluster isochrone are 
    marked with crosses and excluded. The positions of 1000 arbitrarily selected field stars are shown
    as small dots.
    \label{cmdjh}}
\end{figure}

We used the R-band photometry from CA to verify the status
of the CA candidates. With one exception (which was rejected), all CA candidates with detection in
the R-band image are situated on the red side of the (I,R-I) CMD. As a further test, we compared
the (J-H) colours of our candidates with the models of Baraffe et al. (\cite{bca98}). The (J-H)
values were preferred over the (J-K) colour because the latter might be influenced
by excess radiation from a circumstellar disk, as already found for VLM objects of similar age
(Muench et al. \cite{mal01}, Oliveira et al. \cite{ojk02}). Figure\,\ref{cmdjh} shows the (J,J-H)
CMD for all our candidates and the Baraffe et al. isochrones for the $\sigma$\,Ori
cluster. Our VLM candidates fall around the 3\,Myr isochrone (solid line). 
The 1\,Myr (dotted line) and 5\,Myr (dashed line) isochrones are indistinguishable from the 3\,Myr 
track for $J>12$\,mag, but show clear deviations for brighter objects. We excluded one CA and two TLS 
candidates which are clear outliers and thus probably non-members of the cluster. It should be noted 
that the (J,J-H) CMD is not particularly well suited to discriminate between cluster members 
and field stars, as seen in Fig.\,\ref{cmdjh} from the 2MASS data for arbitrarily 
selected objects in the region around $\sigma$\,Ori. In contrast to the (I,I-J) CMD, there is no clear 
separation of field stars and cluster members. Therefore, the (J,J-H) CMD will only reveal obvious 
false detections and does not signficantly reduce the contamination rate of our sample (see below).

The remaining candidate list comprises 52 CA and 110 TLS objects,
including 27 double detections. Coordinates and photometry for these 135 objects are listed
in Table\,7. Masses for these candidates were estimated by fitting the 3 Myr isochrone 
of Baraffe et al. for the J- and H-band with a low degree polynomial and applying the fit to the 
J- and H-magnitudes. This gives us two mass estimates, one based on J-, the other on H-band photometry. 
We found both values to be very similar, and use their average as final mass (see Table\,7).
The theoretical isochrone covers the mass range from 0.02 to 1.4\,$M_{\odot}$. The mass
estimate for fainter or brighter objects is thus not possible with these models. For these
(very few) targets, we only give an upper or lower limit for the mass.

\begin{table}[htbd]
\caption[]{Cluster member candidates which were previously identified by B\'ejar et al. (\cite{bmz01})
as possible $\sigma$\,Ori members. We also list the I-band magnitudes given by B\'ejar et al. 
(\cite{bmz01}) as well as the difference to our I-band photometry.}
  \label{bejar} 
  \begin{tabular}{llcc}
  \hline
  \noalign{\smallskip}
  No. & Name & $I_C$ (mag) &  $\Delta$I\\
  \noalign{\smallskip}
  \hline
  \noalign{\smallskip}
   4 & SOri J053813.1-022410 (SOri\,13) & 16.41& 0.68\\
   9 & SOri J053817.1-022228 (SOri\,9)  & 15.81& 0.56\\
  22 & SOri J053829.5-022517 (SOri\,29) & 17.23& 0.66\\
  33 & SOri J053849.2-022358	        & 16.81& 0.98\\
  49 & SOri J053915.1-022152 (SOri\,38) & 17.64& 0.10\\
  73 & SOri J053848.0-022854 (SOri\,15) & 16.79& 0.83\\
  81 & SOri J053907.4-022908 (SOri\,20) & 17.32& 0.90\\
  83 & SOri J053907.9-022848 (SOri\,8)  & 15.74& 0.37\\
  86 & SOri J053911.7-022741 (SOri\,1)  & 15.08& 0.36\\
 106 & SOri J053944.4-022445 (SOri\,10) & 16.08& 0.34\\
 108 & SOri J053948.1-022914	        & 18.92& 0.82\\
  \noalign{\smallskip}
  \hline
  \end{tabular}
\end{table}

Our candidate list may be contaminated by fore- and background objects. However, the spectroscopic
follow-up observations of B\'ejar et al. (\cite{bmz01}) for an object sample selected with similar photometric
criteria suggest that contamination rates are low. A 7\,arcmin high stripe in the southern part of 
our field is also covered by the survey of B\'ejar et al., although they are only sensitive to objects
with $I>15$. In the overlapping region, we found 18 cluster member candidates in this magnitude range.
Of these candidates, 11 (61\%) are also identified by B\'ejar et al. (\cite{bmz01}). Assuming the 
completeness of their survey, this gives a first rough estimate of up to 40\% for the contamination 
rate. Table\,\ref{bejar} contains our identification numbers as well as object names and I-band photometry of 
B\'ejar et al. (\cite{bmz01}) for common candidates. Our I-band magnitudes for these objects are 
systematically smaller than those given in B\'ejar et al. (\cite{bmz01}). The offset between both 
photometric systems (given in the last column of Table\,\ref{bejar}) increases with the I-J colour, 
which may result from insufficient colour correction in the absolute photometry.
Since the objects in Table\,\ref{bejar} cover only a small range in I-J (8 of 11 objects have 
$0.4<I-J<0.7$), it is not possible to derive a colour correction that would bring all objects onto
the same system.

An alternative estimate for the number of contaminating field stars can be obtained with the Besancon Galaxy 
model\footnote{These simulations are available online under 
{\it http://www.obs-besancon.fr/www/modele/modele\_ang.html}} 
(Robin \& Cr\'ez\'e \cite{rc86}). The simulation delivers star counts depending on Galactic coordinates, colour 
and brightness. A simulated (I,R-I) CMD for a 1\,sq field centred on $\sigma$\,Ori contains 90 objects 
in a 0.4\,mag wide stripe around the cluster isochrone of Baraffe et al. (\cite{bca98}). Scaling to the 
appropriate field sizes, we calculate that 7 of the 52 CA candidates (13\%) and 32 of the 110 TLS (29\%) 
candidates are field stars, i.e. a significantly lower rate than in our first estimate. Since interstellar 
extinction variations might hamper the CMD simulation, we assume conservatively that about 30\% of our
candidates may not be members of the $\sigma$\,Ori cluster. However, the definite decision about the cluster 
membership of our candidates must be postponed until we have obtained follow-up spectroscopy for all of 
them. Our preliminary cluster member list in Table\,7 will be used in the following as target list 
for the variability study.

\section{Time Series Observations}
\label{mon}

\subsection{Observing campaigns and data reduction}

The region with our cluster member candidates was monitored in two I-band time
series campaigns, with the TLS Schmidt telescope and with the 1.23-m telescope
on Calar Alto (CA). Details of these runs are given in Table\,\ref{runs}. During the 
CA campaign, we took alternating short (30\,sec) and long (600\,sec) exposures to extend the
dynamical range of the time series photometry. In both campaigns, we pointed to the same 
position in the sky to within $\approx 10''$ in all monitoring nights. 

\begin{table}[htbd]
\caption[]{Monitoring campaigns}
  \label{runs} 
  \begin{tabular}{lll}
  \hline
  \noalign{\smallskip}
   & TLS Schmidt & CA1.23 \\
  \noalign{\smallskip}
  \hline
  \noalign{\smallskip}
  aperture & 1.34\,m & 1.23\,m \\
  CCD size & 2k$\times$2k& 2k$\times$2k\\
  pixel scale & 1$\farcs$26/pix & 0$\farcs$5/pix\\
  observing nights & 16--26 Jan 2001 & 16--27 Dec 2001\\
  mean seeing & 2$\farcs$5 & 1$\farcs$8\\
  CCD images & 77 & 78/78$^*$\\
  \noalign{\smallskip}
  \hline
  \end{tabular}
  
   $^*$\,The CA time series consists of 78 short and 78 long exposures,
   which were taken alternately.
\end{table}

\begin{figure}[htbd]
    \centering
    \resizebox{\hsize}{!}{\includegraphics[angle=-90]{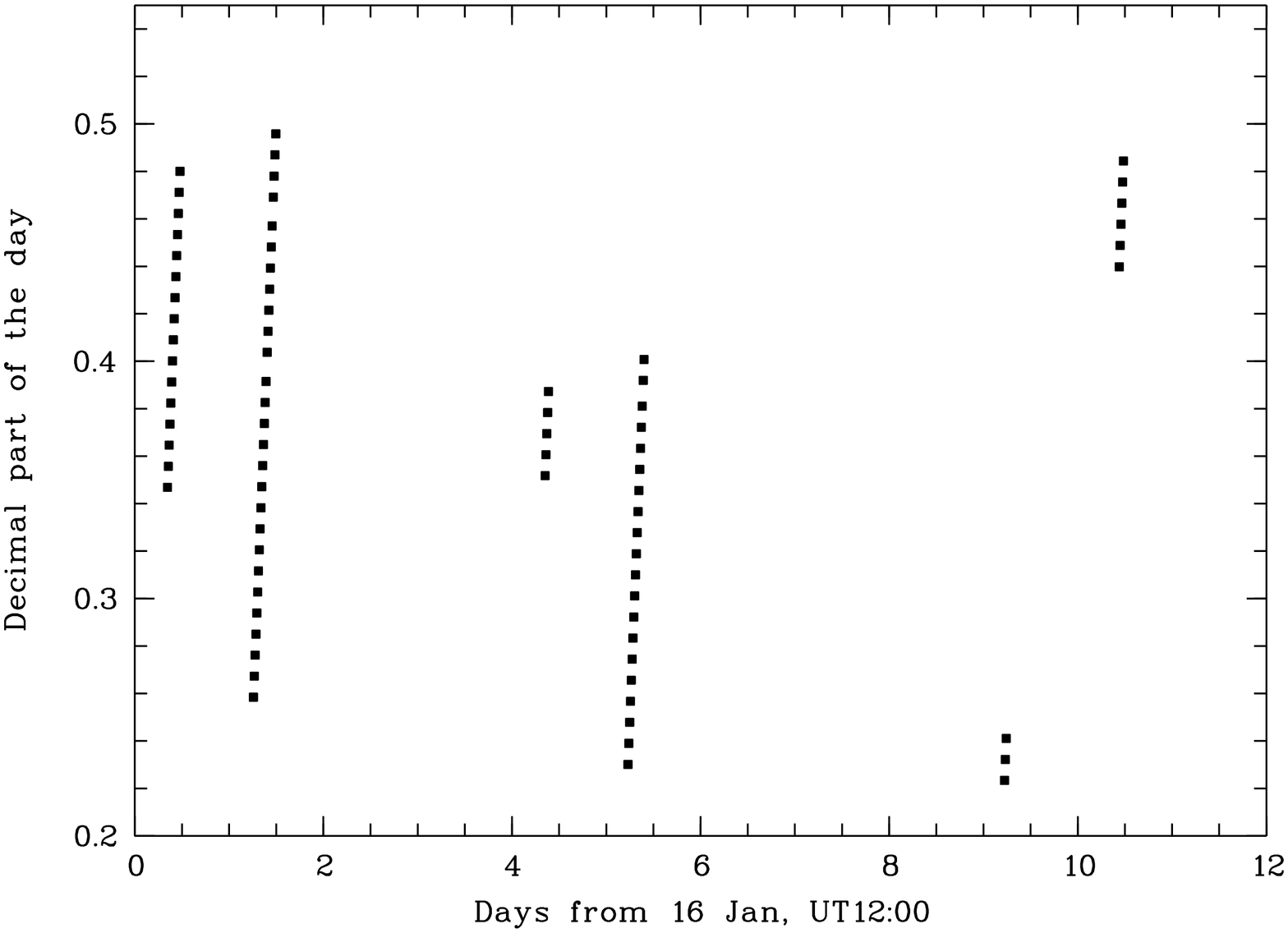}}
    \resizebox{\hsize}{!}{\includegraphics[angle=-90]{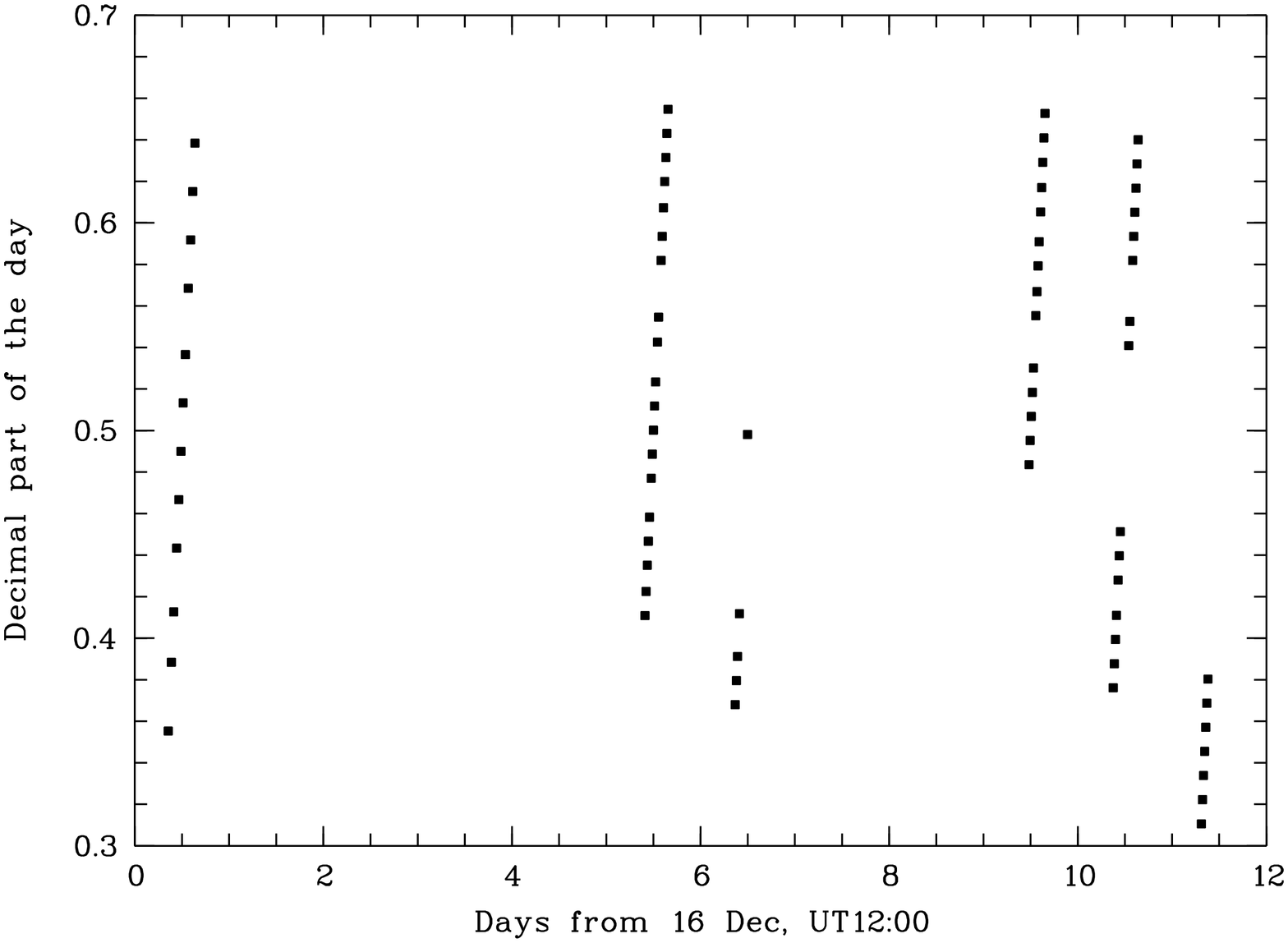}}
    \caption{Datapoint distribution for the TLS (upper panel) and the 
    CA run (lower panel). Plotted is the non-integer fraction of the observing 
    times against the observing times.
    \label{sampling}}
\end{figure}

Both observing runs span at least ten days. The distribution of the datapoints 
over the observing run, however, is irregular because of variable weather conditions.
The number of datapoints per night varies from 0 to 27. Figure\,\ref{sampling} shows the 
datapoint distribution for both runs.

Image reduction and photometry of both campaigns was done following the recipes
explained in the previous section. An object catalogue was created
for the deepest time series image of both campaigns. For each time series image, we
determined the spatial offset between the object catalogue and the positions in the
actual image. Applying this offset to the object catalogue and centering 
these new positions on the respective image generates an object catalogue
for each image. We used a fixed sample of PSF stars for each night of the TLS 
time series. After the photometry procedure, we obtained lightcurves for all objects 
in the respective object catalogue. 

\subsection{Relative calibration}

The lightcurves of both campaigns were calibrated using an average lightcurve
from a set of non-variable reference stars in the observed field. This is necessary
to correct for changing atmospheric conditions as well as the variable airmass
over the time series. In the following, we describe the selection 
process for the reference stars.

First, a sample of stars was selected, whose photometry errors were below 
0.1\,mag in all images. To reject variable objects from this initial reference star list, we 
examined the variability of each star with respect to the average lightcurve of all other stars. 
We used the following procedure based on the routine by Allain (\cite{a94}). 
The number of a reference star is indicated with $i=1\ldots N_R$, the number of an image 
with $j=1\ldots N_B$.

\begin{enumerate}
\item{averaging the instrumental magnitudes for each star
\begin{equation}
\overline{m_i}= \frac{1}{N_B} \sum_{j=1}^{N_B} m_i(t_j)
\end{equation}}
\item{subtraction of this average from every value of the time series for every star
\begin{equation}
m_i^0(t_j)=m_i(t_j)-\overline{m_i}
\end{equation}}
\item{calculation of average $\overline{m_j^0}$ and standard deviation $\sigma _j$ 
for the differences from Eq.\,(4) 
\begin{equation}
\overline{m_j^0} = \frac{1}{N_R} \sum_{i=1}^{N_R} m_i^0(t_j)
\end{equation}
\begin{equation}
\sigma _j=\sqrt{\frac{1}{N_R-1} \sum_{i=1}^{N_R} (m_i^0(t_j)-\overline{m_j^0})^2}
\end{equation}}
\item{rejection of obvious 'bad' images by examination of $\sigma _j$ ($N_B'$ is the
number of the remaining images)}
\item{repeated execution of steps 1--3 without the rejected images}
\item{calculation of a quality number $test_i$, which allows for every star
an examination of its intrinsic variability compared to all other reference stars;
if $|m_i^0(t_j)-\overline{m_j^0}| \geq \sigma _j$ is $test_{i,j}=1$, 
elsewhere 0
\begin{equation}
test_i=\sum_{j=1}^{N_B'} test_{i,j}
\end{equation}}
\end{enumerate}

After the described procedure, the stars with $test_i<3$ were selected
as reference stars. We obtained $\approx 170$ reference stars for the TLS and
$\approx 80$ for the CA campaign. For all images, the average brightness of the 
reference stars was determined.
\begin{equation}
\overline{m^{ref}}(t_j)=\frac{1}{N_\mathrm{ref}}\sum_{i=1}^{N_\mathrm{ref}} m_i^{ref}(t_j)
\end{equation}
This mean lightcurve was subtracted from all time series. Thus, we obtained extinction corrected 
relative magnitudes $m^{rel}(t_j)$ from the instrumental magnitudes for all objects.
$m(t_j)$.
\begin{equation}
m^{rel}(t_j)=m(t_j)-\overline{m^{ref}}(t_j)
\end{equation}

\subsection{Optimal Image Subtraction}

For CCD images from the 1.23-m telescope on the Calar Alto, a
dedicated image analysis package has been developed by the Wendelstein Calar Alto
Pixellensing Project (WeCAPP, Riffeser et al. \cite{rfg01}) team. This software is based on 
'Optimal Image Subtraction' (OIS, Alard \& Lupton \cite{al98}). After astrometric and photometric 
alignment, it performs an image convolution with respect to a reference frame (G{\"o}ssl \& Riffeser
\cite{gr02}). The reference image is obtained by coaddition of the time series images with the best 
seeing. For the long exposure time series from CA, we used this pipeline to check the 
quality of our lightcurves from our already described 'classical' photometry. Because of
bad image quality, six images were not used for this procedure.

The basic idea of difference image analysis algorithms like OIS is to measure {\it only} the
variability and {\it not} the constant portion of the star's flux. 
The result of the pipeline are frames which contain only variable 
sources. On these frames, we performed aperture photometry for the previously determined 
object catalogue. The resulting fluxes were transformed to relative fluxes by
division by the flux of the respective object in the reference image. 
Relative fluxes were transformed to relative magnitudes with 
$m_{rel} = -2.5\log_{10}(1+f_\mathrm{rel})$.

To compare the results from OIS with those from {\it daophot}, we determined the 
photometric precision of both measurements: After excluding 3$\sigma$\,outliers, we calculated
average and rms for all lightcurves. The lightcurves of the brightest objects scatter with
4\,mmag in the OIS data, 3\,mmag less than with classical photometry.
Throughout the usable magnitude range, OIS improves the precision by several mmag. 
Therefore, we used the lightcurves from the WeCAPP software for the further analysis. 
The short exposure CA images, however, cannot be reduced properly with the OIS 
pipeline, since they do not contain enough stars for the fitting process.

Three obstacles prevented a successful application of OIS to the TLS data:
\begin{itemize}
\item{The pixel scale of the TLS Schmidt camera is roughly two times that of the
CCD at the 1.23-m telescope on CA. This leads to an undersampled PSF, which makes
exact image centering and image folding difficult.}
\item{The sky background in the TLS images was in most cases $>25000$\,ADU, because of
snow cover around the observatory. Therefore, moderately 
crowded open cluster fields with many bright stars lead to many saturated pixels 
which are useless for difference image analysis and decrease the precision of the
fitting process.}
\item{Every bright star causes a so-called Schmidt ghost in the image, disturbing 
the fitting process. However, if the ghosts were masked so that the fit can be done 
without them, we would loose a considerable fraction of the targets.}
\end{itemize}
For all these reasons, we returned to 'classical' PSF fitting photometry for the
TLS images. High background caused by light reflection from snow, bad seeing and
variable atmospheric conditions limited the precision of the TLS lightcurves. Here, we
reach 0.015\,mag for the brightest stars.

\section{Time Series Analysis}
\label{tsa}

\subsection{Generic variability test}
\label{generic}

\begin{figure}[htbd]
    \centering
    \resizebox{\hsize}{!}{\includegraphics[angle=-90]{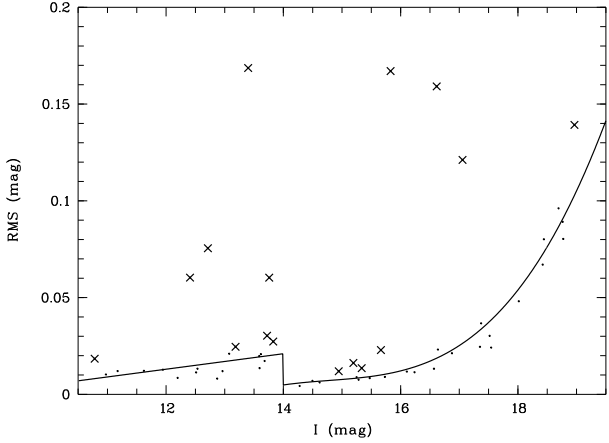}}
    \caption{Root mean square of the time series photometry vs. I-band magnitude for all 
    CA candidates. The solid line is the mean RMS of the relative
    photometry, determined by fitting the RMS of {\it all} objects in the
    field. Below I=14\,mag the long exposure photometry is 
    influenced by saturation, brighter targets were analysed with the 
    short exposure lightcurves. All objects marked with a cross are variable at 
    the 99\% level.}
    \label{rmsca}
\end{figure}

\begin{figure}[htbd]
    \centering
    \resizebox{\hsize}{!}{\includegraphics[angle=-90]{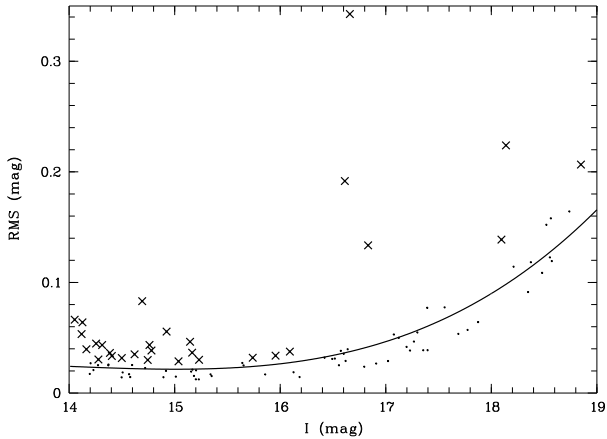}}
    \caption{Root mean square of the time series photometry vs. I-band magnitude for all 
    TLS candidates. The solid line is the mean RMS of the relative
    photometry, determined by fitting the RMS of {\it all} objects in the
    field. All objects marked with a cross are variable at the 99\% level.
    At the bright end, the variability test is contaminated by extinction
    residuals and beginning saturation.}
    \label{rmstls}
\end{figure}

The candidate lightcurves of both campaigns were analysed with a unique procedure.
In a first step, we examined all lightcurves visually to register obvious signs of rapid
variability. In particular, we looked for sudden brightness eruptions (e.g. flares),
eclipses or totally irregular variability. This search was not successful.
It turned out that some TLS objects are either too bright or too faint to be
detected in most of the images; time series analysis is not possible for these targets.
Subsequently, the lightcurves were filtered: In three iterations, we excluded 
3$\sigma$\,outliers, but only if they were framed by other datapoints to avoid the exclusion 
of intrinsic short-term variability. The root mean squares of the filtered lightcurves are
listed in Table\,7 and plotted in Figs.\,\ref{rmsca} and \ref{rmstls}.  
The solid lines in these Figures indicate our photometric precision determined by fitting the rms of 
{\it all} lightcurves with low degree polynomials. This average rms was compared to the
rms of the candidates with a statistical F-test, which is particularly well-suited for the comparison
of scatter in data. All variable objects at a significance level of 99\% are marked with a cross 
(and with a 'v' in Table\,7). 

Of the 52 CA candidates, 16 (30\%) show significant variability. In contrast, {\it all} 
objects in our images, i.e. mostly field stars, have a much smaller fraction of variable objects 
($\approx$10\%). It is obvious from Fig.\,\ref{rmsca} that the fraction of variable objects as well 
as the variability amplitude does not change significantly over the entire magnitude range, which 
corresponds to masses down to $0.02\,M_{\odot}$. Figure\,\ref{rmstls} shows the same data for the TLS run. 
Unfortunately, strongly variable conditions and high background lead to strongly scattered rms 
for bright objects. Additionally, the high image background prevents the detection of bright {\it and}
highly variable targets, since they would be saturated around their lightcurve maxima.
Therefore, the data for targets with $I<15$\,mag are of limited use. 
However, the plot again shows no decrease of variability amplitude with I-magnitude (and hence mass).
Interestingly, one variable substellar target (no. 108) was identified with an X-ray source by 
Mokler \& Stelzer (\cite{ms02}), confirming that this object is an active Brown Dwarf.

Summarizing, these are the two results from this test: a) The young VLM objects in our observed
field contain a large fraction of variable objects. b) The fraction of variable objects and the 
variability amplitude is independent of mass from 0.5 down to 0.02\,$M_{\odot}$.

\subsection{Period search}
\label{perser}

Our period search is based on the widely used Scargle periodogram (Scargle \cite{s82}). 
Compared to the classical periodogram, this modified version has one advantage: The 
probability that a periodogram peak at a given frequency has the height $z$ should be 
simply $P(z)=\exp{(-z)}$, even for non-uniformly sampled datasets. This makes the Scargle 
periodogram particularly well-suited for astronomical time series with their characteristic 
gaps between the observing nights. The False Alarm Probability $\mathrm{FAP_{Scargle}}$ for a 
periodic signal is the probability that a peak at {\it any} frequency has the height z:
$\mathrm{FAP_{Scargle}}=1-[1-\exp(-z)]^{N_i}$. $N_i$ is the number of independent frequencies.
For evenly spaced data, $N_i$ can be estimated as $N_i = -6.362 + 1.193N + 0.00098N^2$,
where N is the number of datapoints (Horne \& Baliunas \cite{hb86}). For 'clumped' data, however,
$N_i$ (and with it the $\mathrm{FAP_{Scargle}}$) will decrease drastically. For a first estimate 
of the $\mathrm{FAP_{Scargle}}$, we use $N_i=N/2$. Later, we will determine reliable empirical 
False Alarm Probabilities for the detected periods.

\begin{table}[h]
\caption{Candidates with significant periodic variability in the CA campaign. Periods (P) are determined
by fitting the CLEANed periodogram peak with a Gaussian. Period errors ($\Delta$P) are based on the half width
at half maximum of the periodogram peak, transformed in time space. The amplitudes (A) correspond to the
peak-to-peak-range of the binned lightcurve. N is the number of datapoints used for the period
search.}
\begin{tabular}{rcccccc} \hline
No. & M ($M_{\odot}$) & P (h) & $\Delta$P (h) & A (mag) & $\mathrm{FAP_{E}}$ (\%) & N\\
\hline
2  & 0.07 & 14.7 & 0.26 & 0.385 & $<$0.01 & 71 \\ 
8  & 0.10 & 79.1 & 12.0 & 0.045 & $<$0.01 & 71 \\ 
9  & 0.17 & 246~ & 102~ & 0.017 & $<$0.01 & 71 \\ 
13 & 0.77 & 5.63 & 0.05 & 0.024 & 0.13    & 76 \\ 
14 & 0.17 & 7.57 & 0.09 & 0.040 & $<$0.01 & 72 \\ 
15 & 0.20 & 52.9 & 4.24 & 0.034 & $<$0.01 & 71 \\ 
19 & 0.65 & 39.3 & 2.34 & 0.397 & $<$0.01 & 77 \\ 
21 & 0.67 & 201~ & 74.4 & 0.147 & $<$0.01 & 76 \\ 
22 & 0.06 & 14.4 & 0.40 & 0.035 & $<$0.01 & 69 \\ 
28 & 0.03 & 21.1 & 0.65 & 0.057 & $<$0.01 & 71 \\ 
33 & 0.17 & 228~ & 86.4 & 0.454 & $<$0.01 & 72 \\ 
37 & 0.70 & 193~ & 66.3 & 0.192 & $<$0.01 & 78 \\ 
43 & 0.06 & 74.4 & 8.25 & 0.318 & $<$0.01 & 72 \\ 
48 & 0.11 & 28.4 & 1.25 & 0.028 & $<$0.01 & 71 \\ 
\hline				     
\end{tabular}			     
\label{resca}			     
\end{table}		

\begin{table}[htbd]
\caption{Candidates with significant periodic variability in the TLS campaign. Columns
as in Table\,\ref{resca}.}
\begin{tabular}{rcccccc} \hline
No. & M ($M_{\odot}$) & P (h) & $\Delta$P (h) & A (mag) & $\mathrm{FAP_{E}}$ (\%) & N\\
\hline
2   &  0.07 & 63.6   & 8.14   & 0.547 & $<$0.01 & 75 \\ 
16  &  0.03 & 57.9   & 7.41   & 0.523 & $<$0.01 & 74 \\ 
22  &  0.06 & 8.36   & 0.13   & 0.068 & $<$0.01 & 76 \\ 
23  &  0.19 & 3.64   & 0.03   & 0.027 & $<$0.01 & 74 \\ 
32  &  0.04 & 9.09   & 0.16   & 0.071 & $<$0.01 & 76 \\ 
33  &  0.17 & 44.6   & 3.79   & 1.114 & $<$0.01 & 77 \\ 
43  &  0.06 & 62.1   & 7.33   & 0.314 & $<$0.01 & 77 \\ 
80  &  0.17 & 15.1   & 0.44   & 0.173 & $<$0.01 & 76 \\      
85  &  0.02 & 37.9   & 2.66   & 0.468 & $<$0.01 & 74 \\       
95  &  0.02 & 8.03   & 0.11   & 0.283 & $<$0.01 & 76 \\    
101 &  0.26 & 62.5   & 7.37   & 0.049 & $<$0.01 & 75 \\    
107 &  0.27 & 34.0   & 2.20   & 0.097 & 0.07 & 75 \\    
128 &  0.02 & 5.78   & 0.07   & 0.066 & 0.39 & 73 \\    
\hline				     	      
\end{tabular}			     
\label{restls}			     
\end{table}		

Periodograms are always contaminated with 'false' peaks, namely sidelobes and aliases,
caused by the windowing of the data. The recorded time series is the convolution 
of the signal with this window function. Thus, any period search based on the
raw periodogram alone will deliver spurious detections. To circumvent this problem, we
used the CLEAN algorithm by Roberts et al. (\cite{rld87}), which deconvolves 'dirty' 
spectrum and window function and thus 'cleans' the periodogram. This algorithm is therefore
able to distinguish between real peaks and spurious features in the periodogram. Since the 
CLEAN algorithm is based on the classical periodogram, it does not allow an {\it a priori} 
estimation of the FAP. The combination of Scargle periodogram and CLEAN algorithm, however,
delivers reliable period detections {\it and} a FAP estimate. Therefore, the period search
procedures of numerous recent variability studies rely exclusively on these two
techniques (e.g., Bailer-Jones \& Mundt \cite{bm99}, \cite{bm01}, Terndrup et al. \cite{tkp99},
O'Dell et al. \cite{ohc97}, Patten \& Simon \cite{ps96}).

\begin{figure*}[htbd]
\resizebox{5.5cm}{!}{\includegraphics[angle=-90,width=6.5cm]{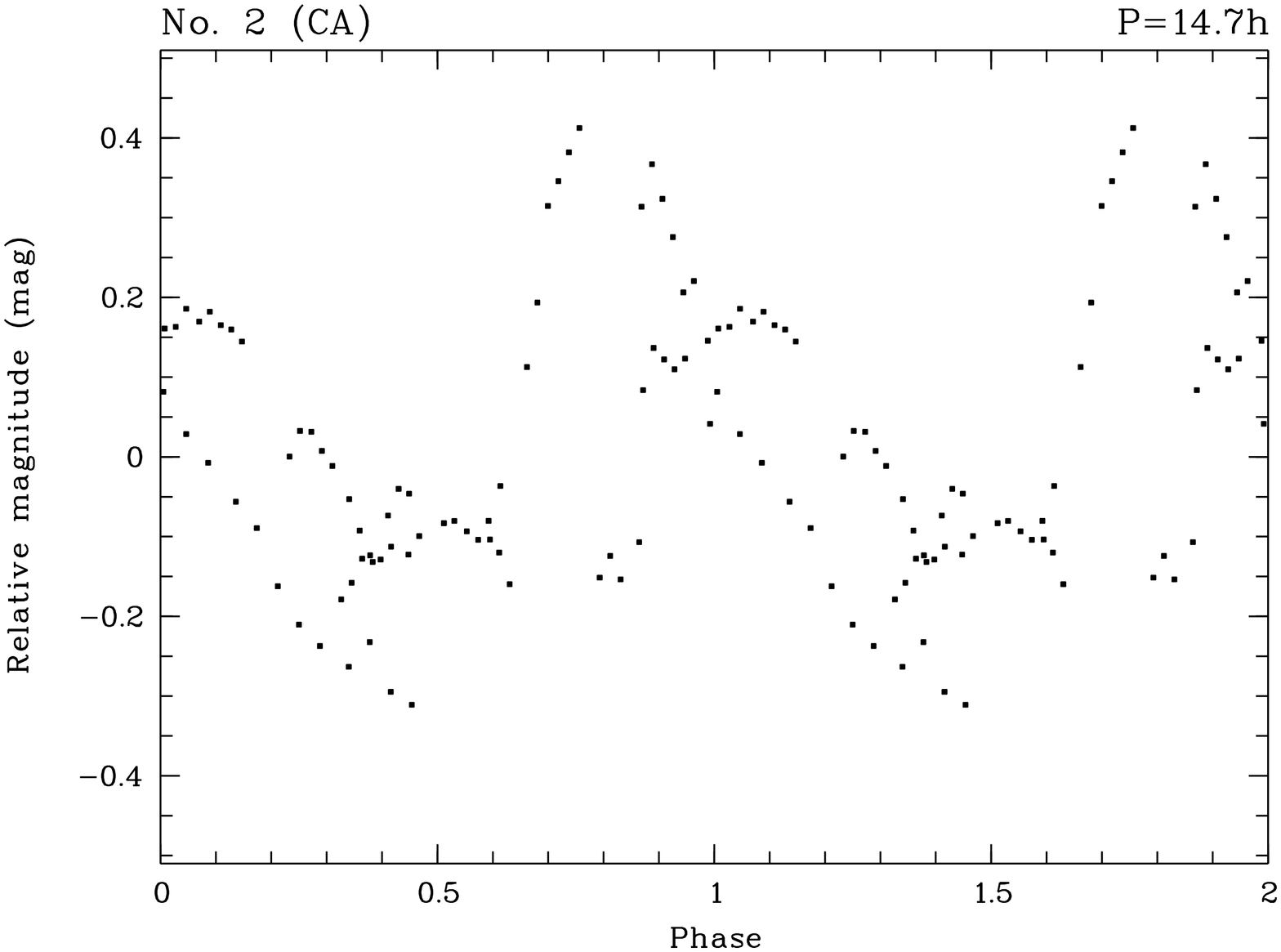}} \hfill
\resizebox{5.5cm}{!}{\includegraphics[angle=-90,width=6.5cm]{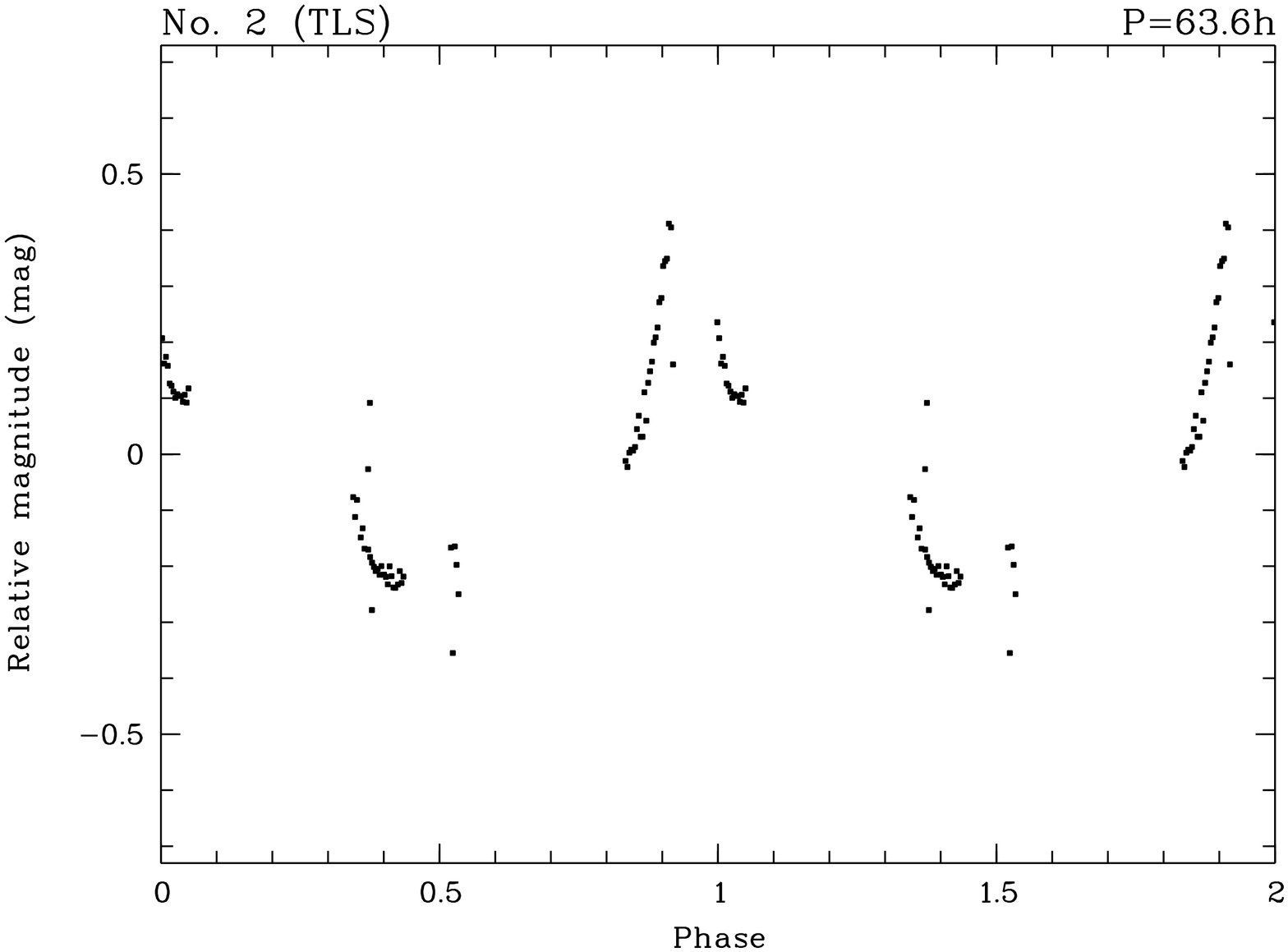}} \hfill
\resizebox{5.5cm}{!}{\includegraphics[angle=-90,width=6.5cm]{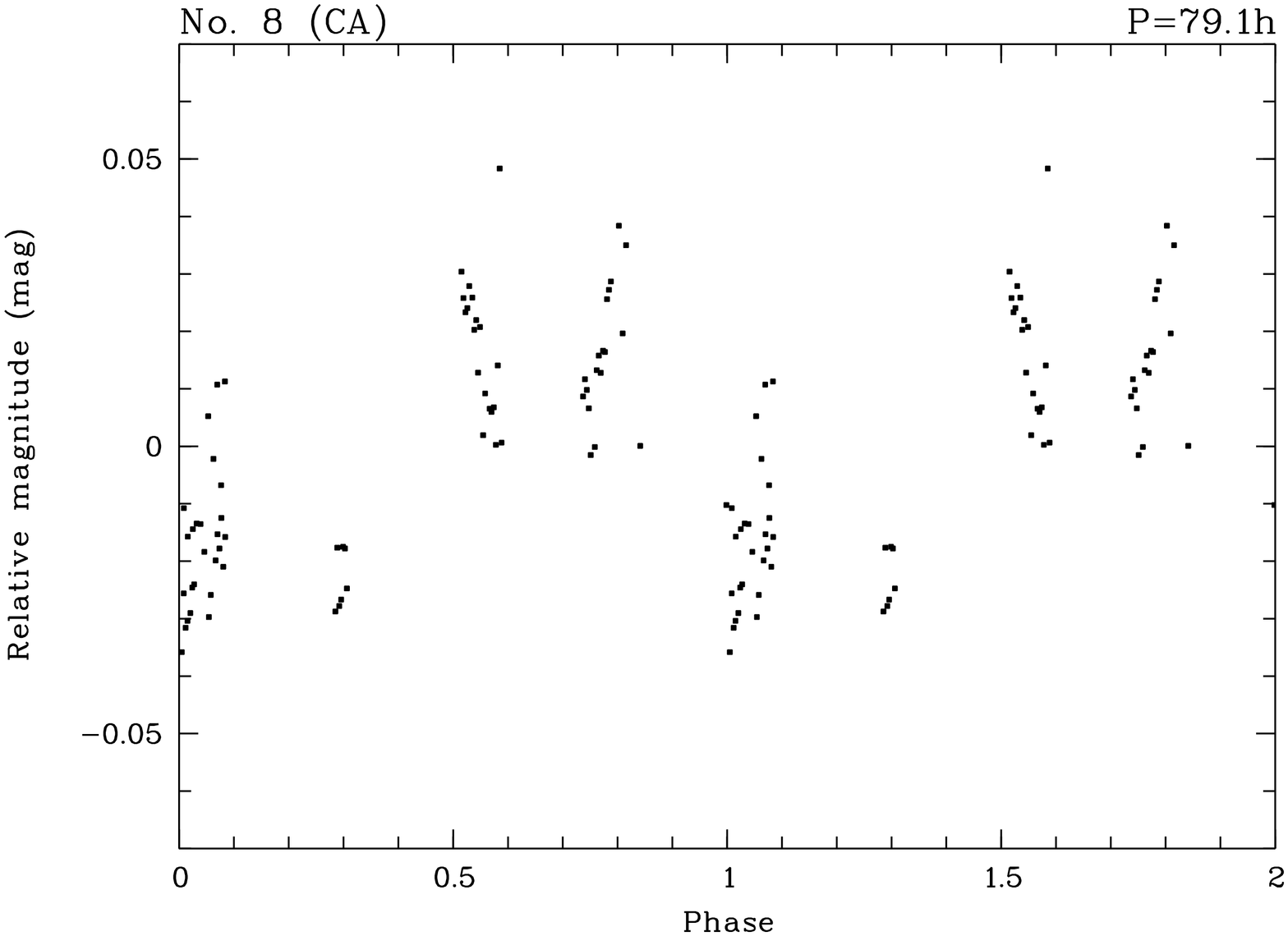}} \\
\resizebox{5.5cm}{!}{\includegraphics[angle=-90,width=6.5cm]{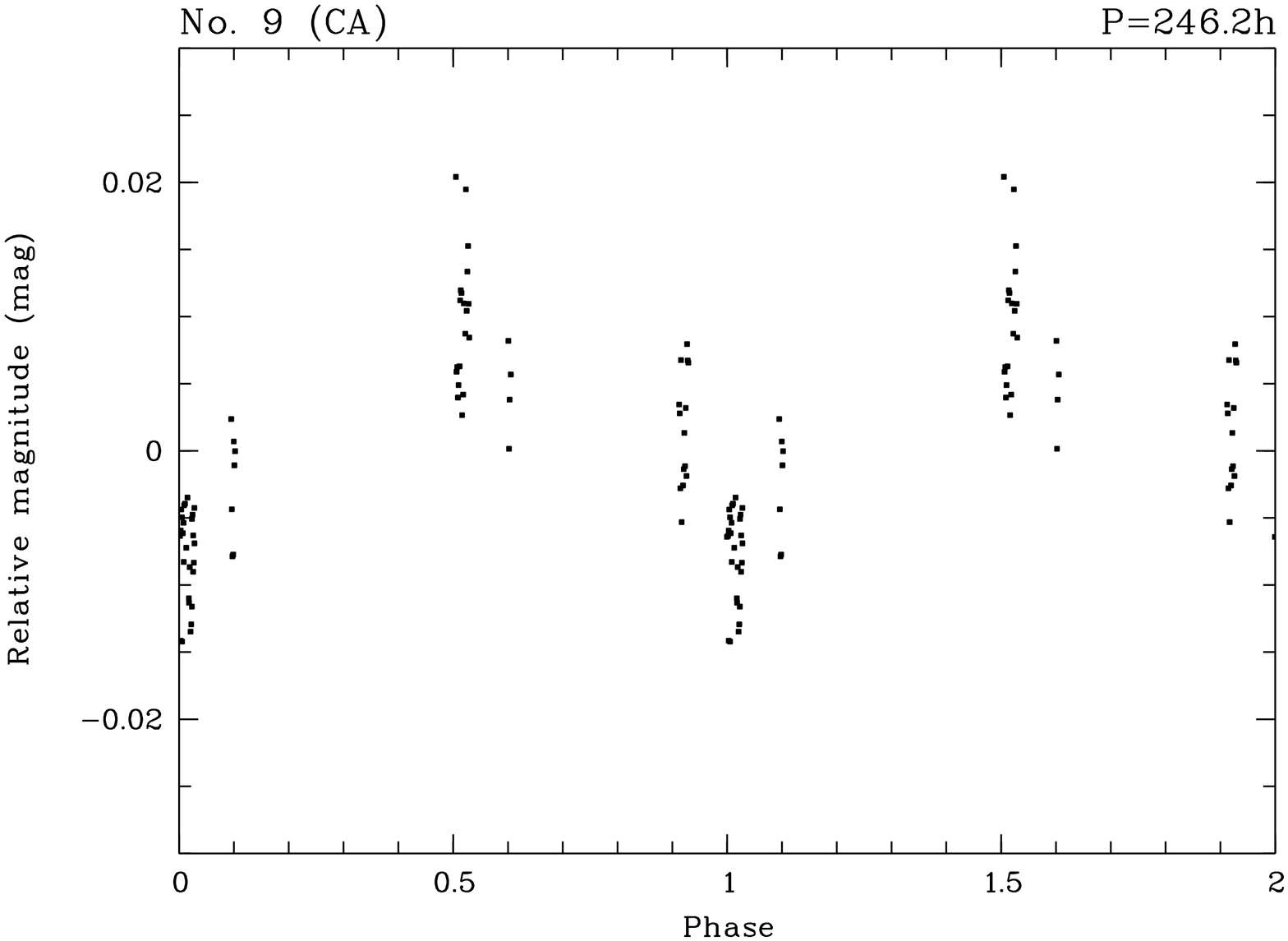}} \hfill
\resizebox{5.5cm}{!}{\includegraphics[angle=-90,width=6.5cm]{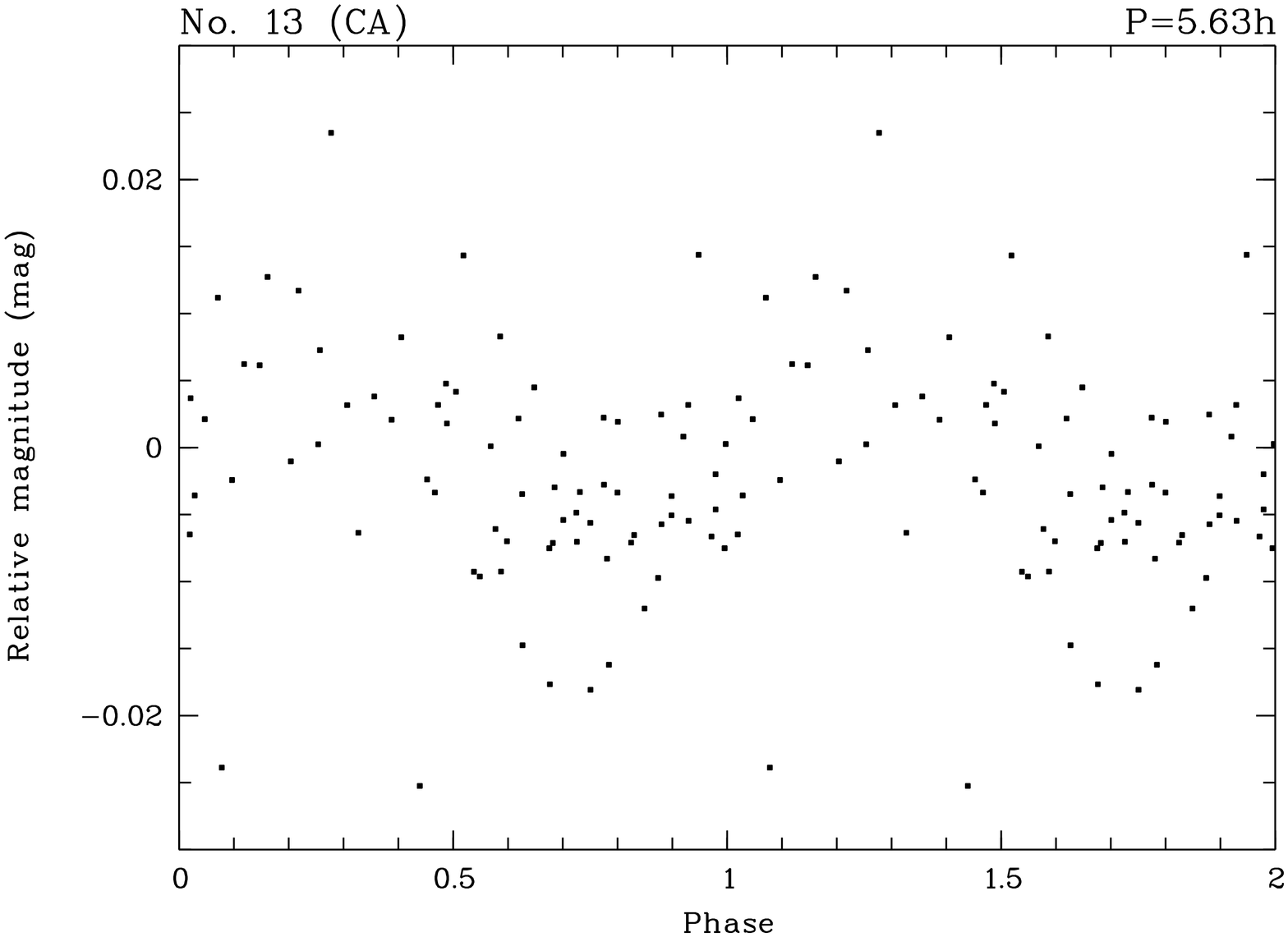}} \hfill
\resizebox{5.5cm}{!}{\includegraphics[angle=-90,width=6.5cm]{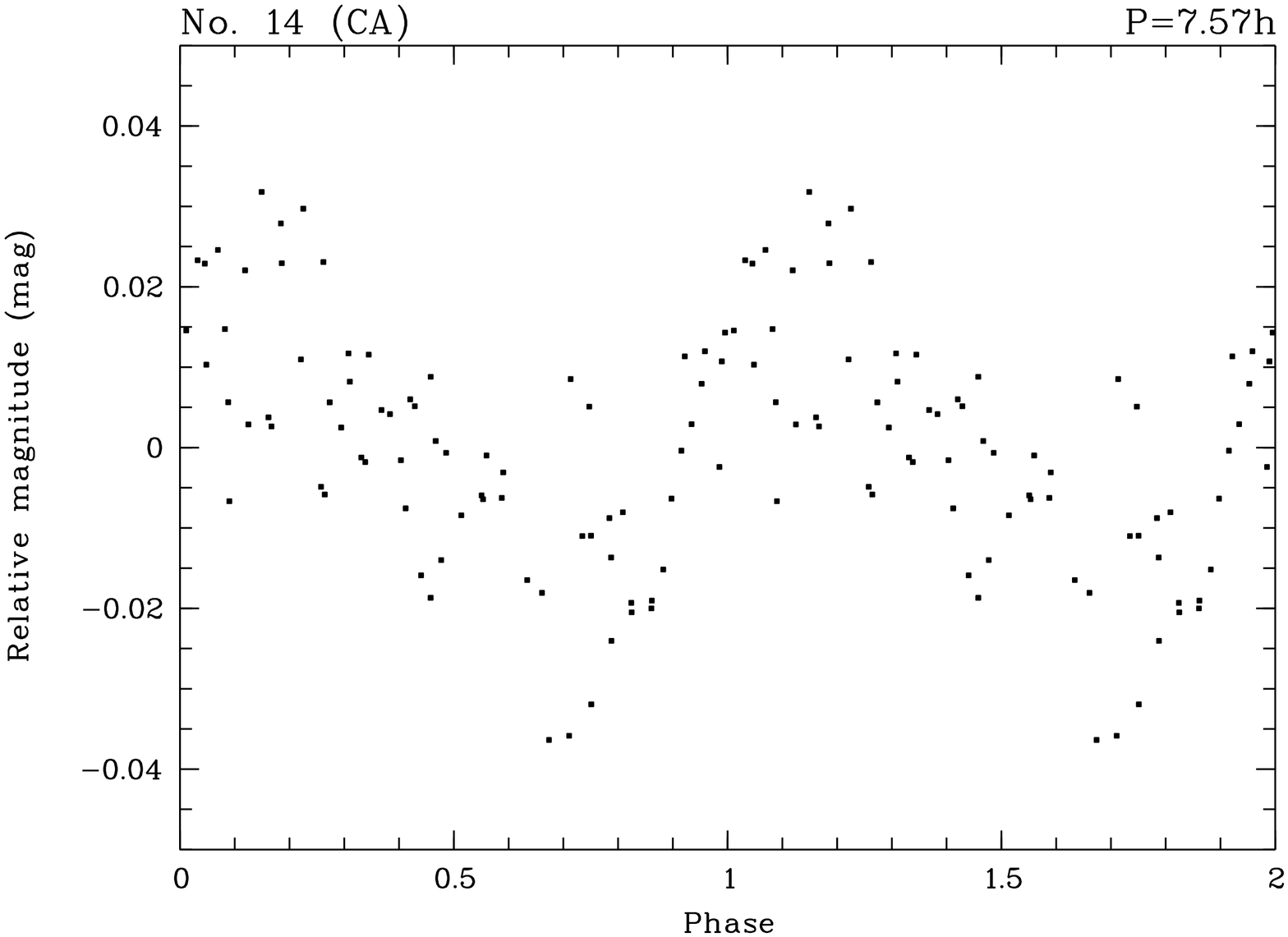}} \\
\resizebox{5.5cm}{!}{\includegraphics[angle=-90,width=6.5cm]{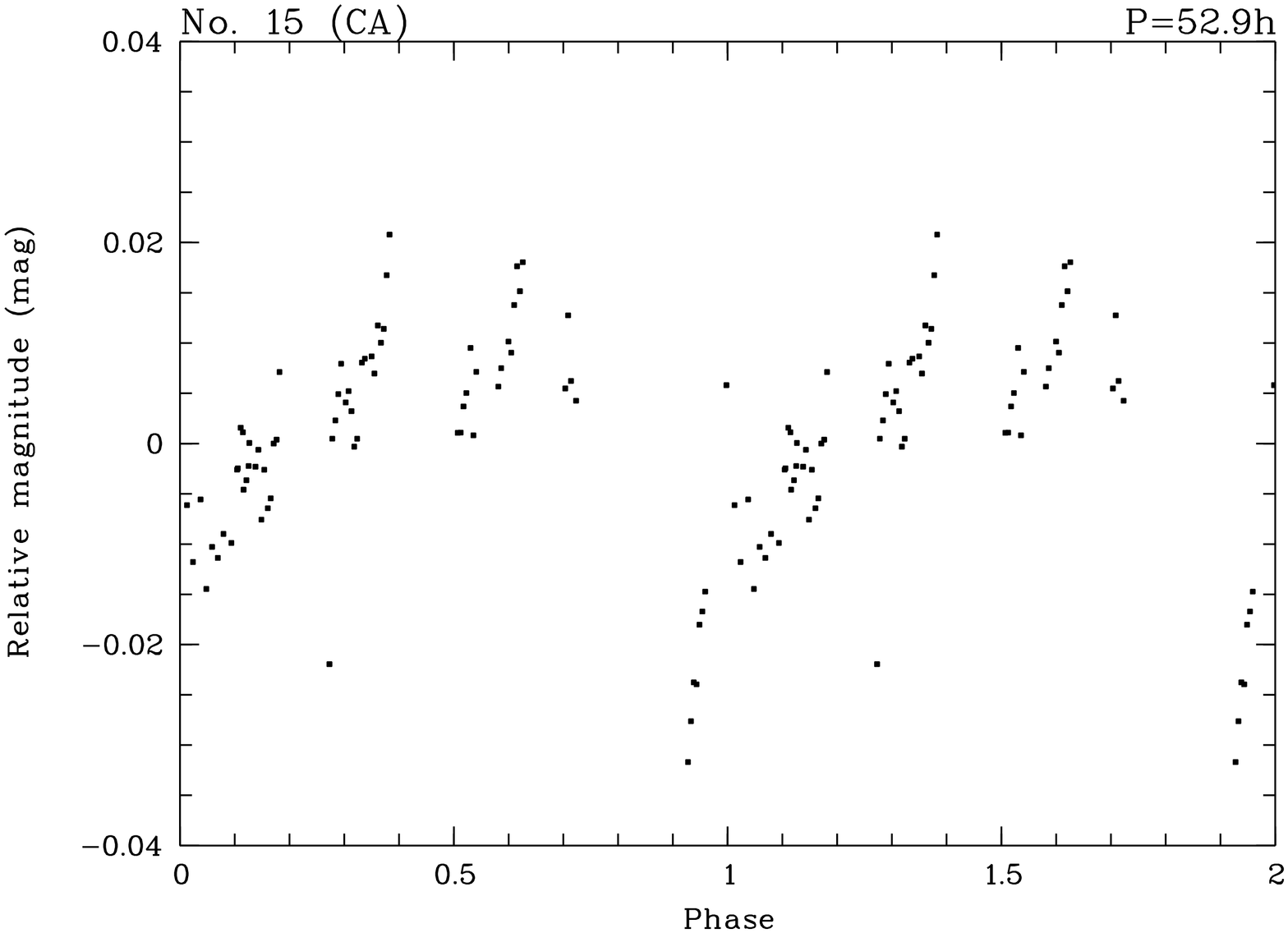}} \hfill
\resizebox{5.5cm}{!}{\includegraphics[angle=-90,width=6.5cm]{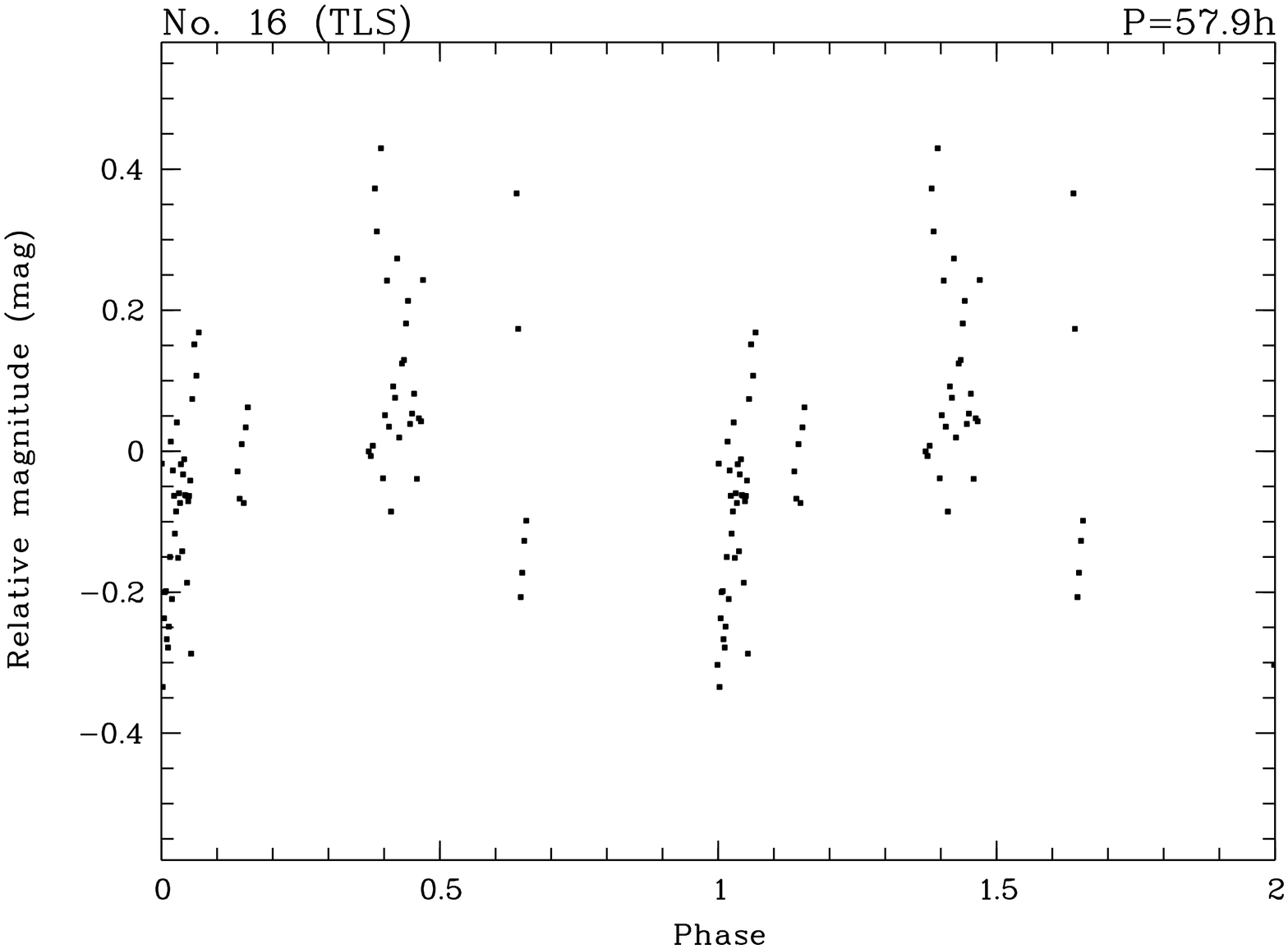}} \hfill
\resizebox{5.5cm}{!}{\includegraphics[angle=-90,width=6.5cm]{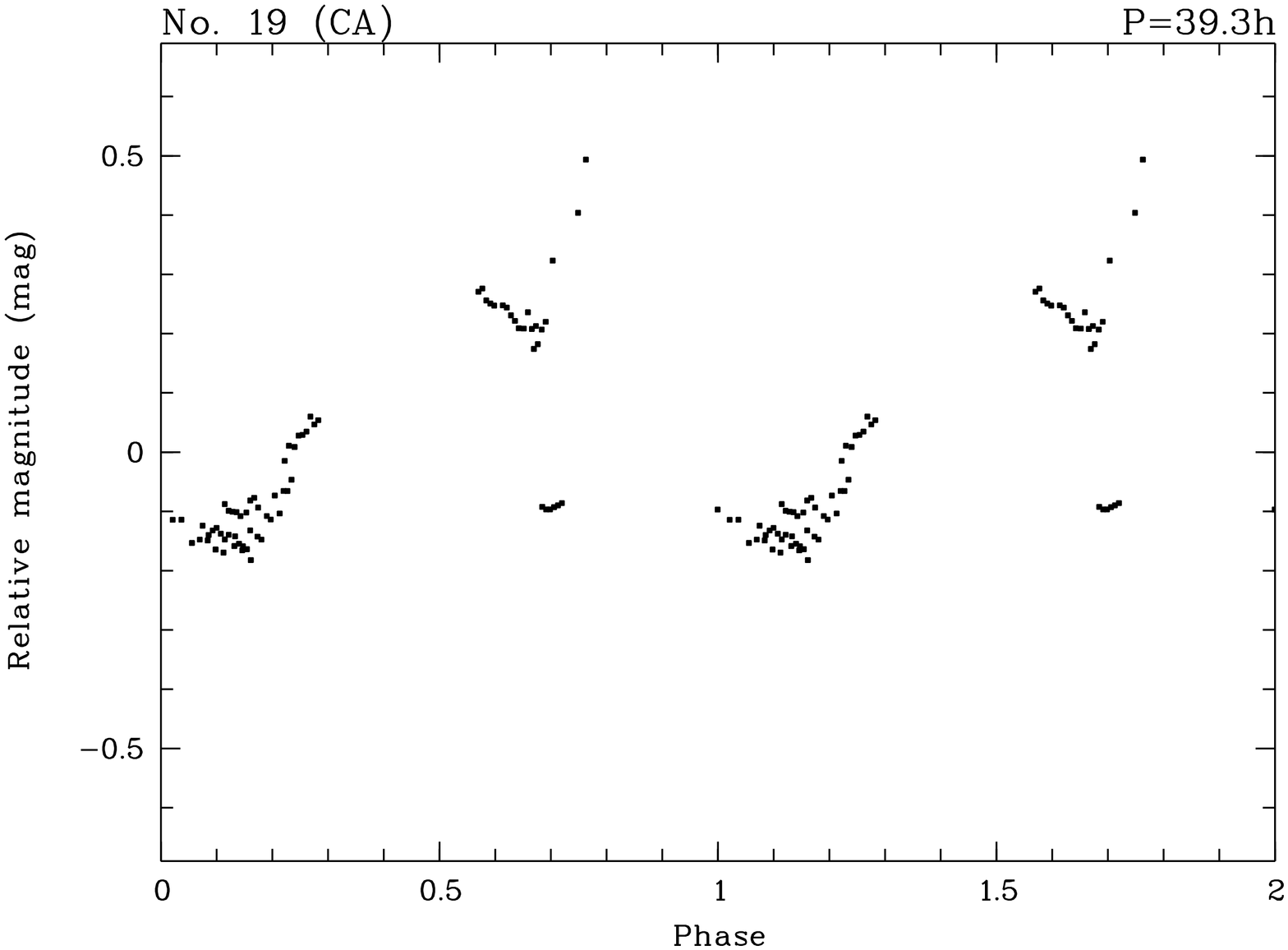}} \\
\resizebox{5.5cm}{!}{\includegraphics[angle=-90,width=6.5cm]{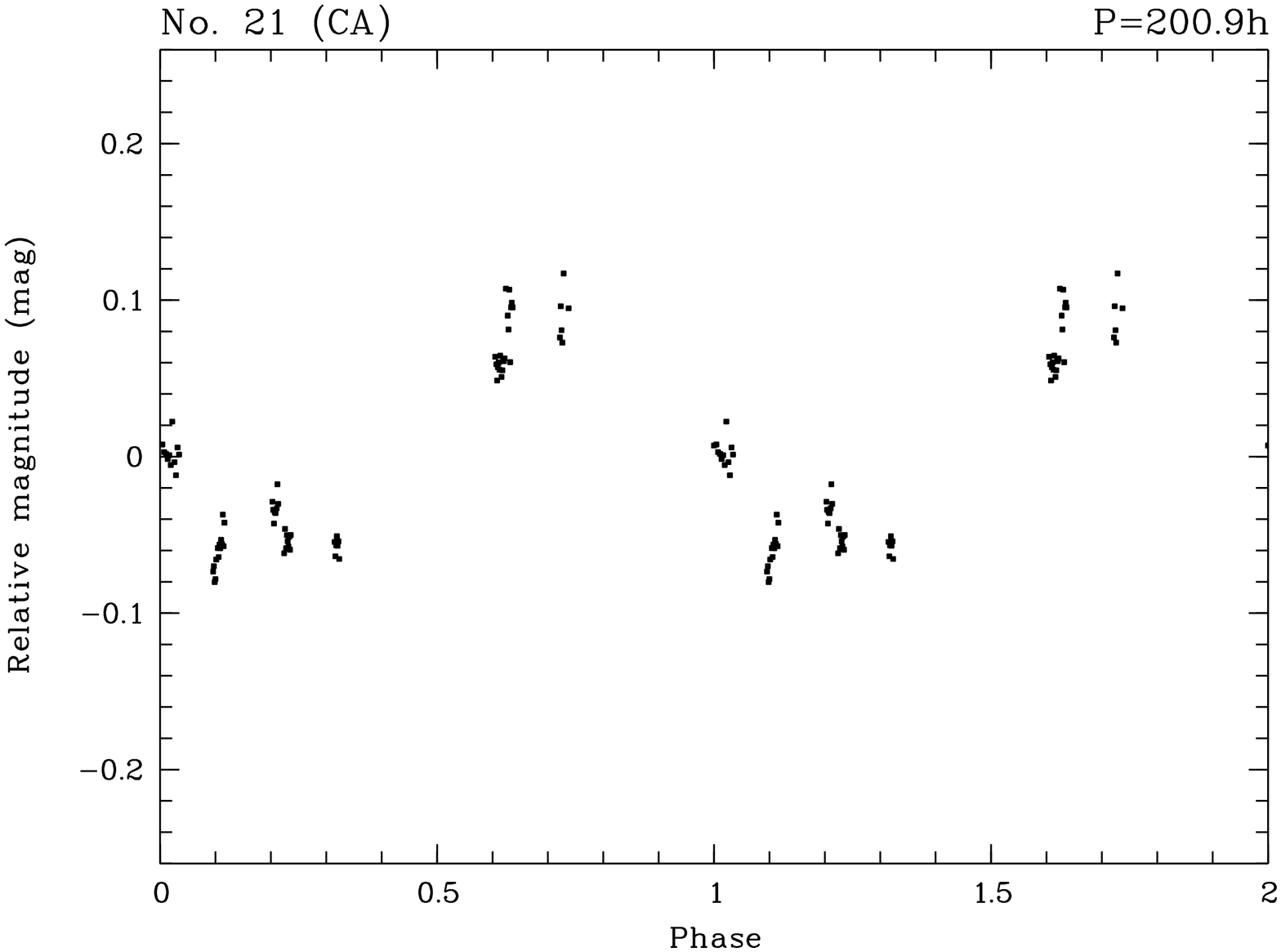}} \hfill
\resizebox{5.5cm}{!}{\includegraphics[angle=-90,width=6.5cm]{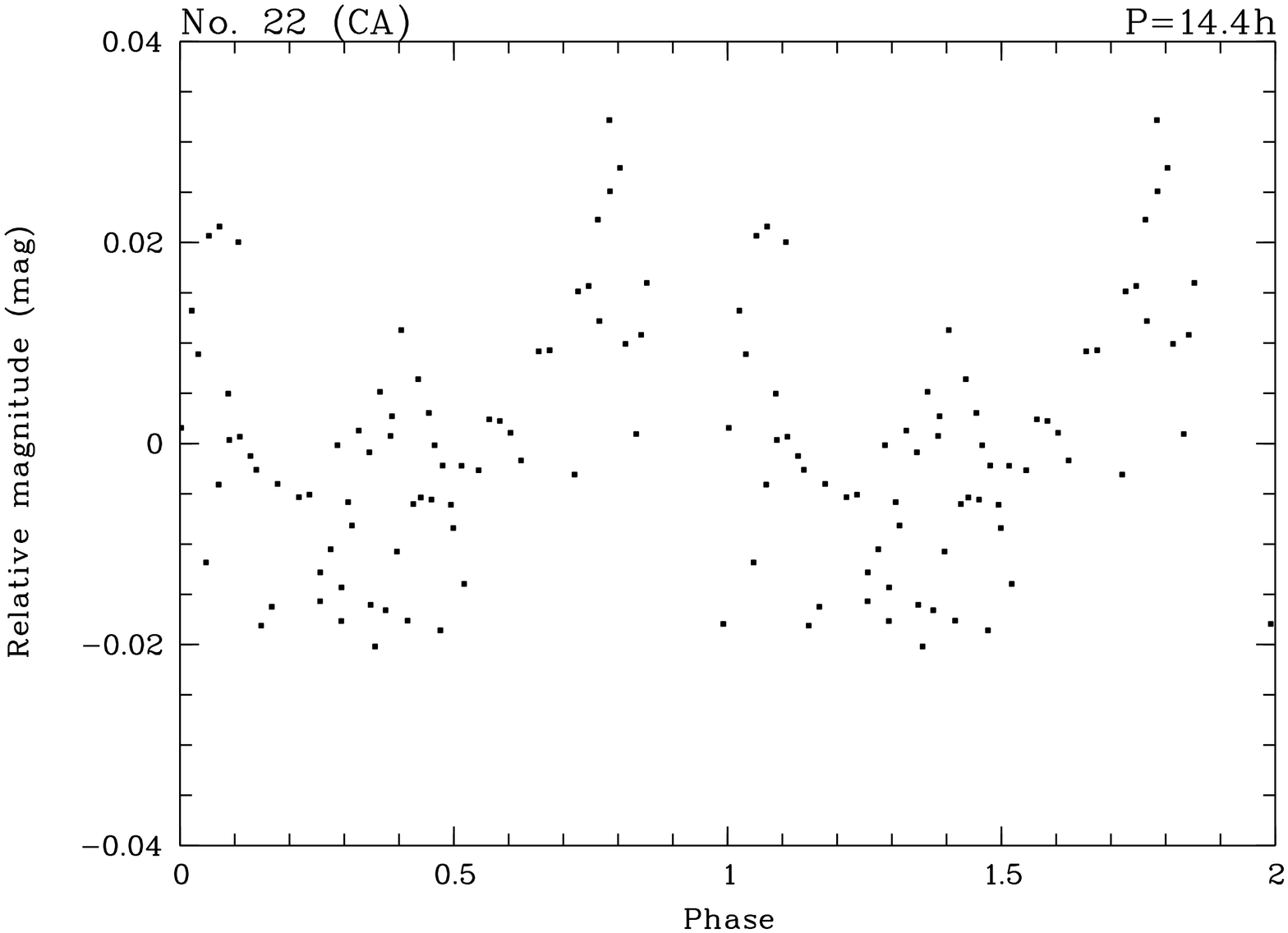}} \hfill
\resizebox{5.5cm}{!}{\includegraphics[angle=-90,width=6.5cm]{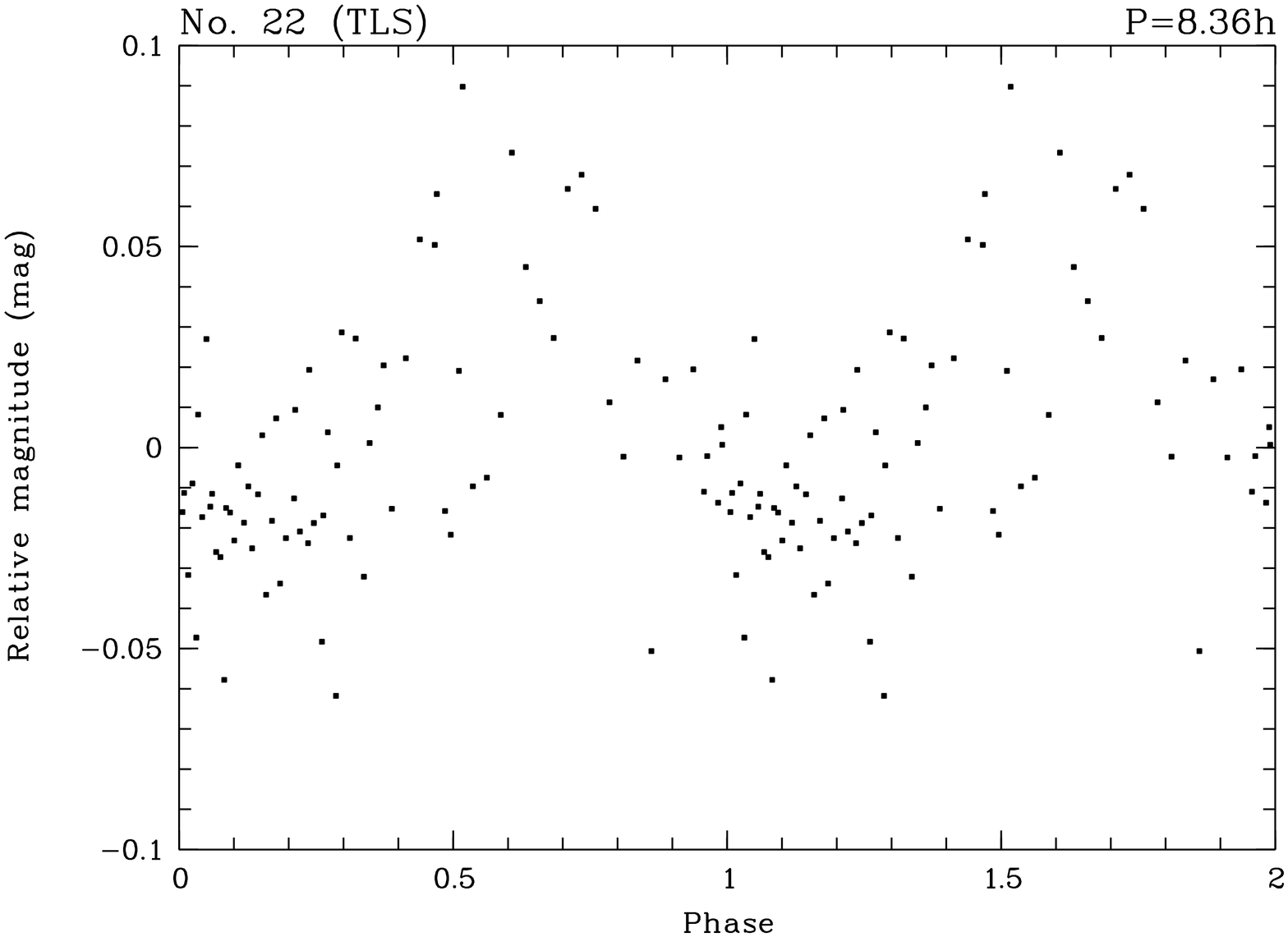}} \\
\resizebox{5.5cm}{!}{\includegraphics[angle=-90,width=6.5cm]{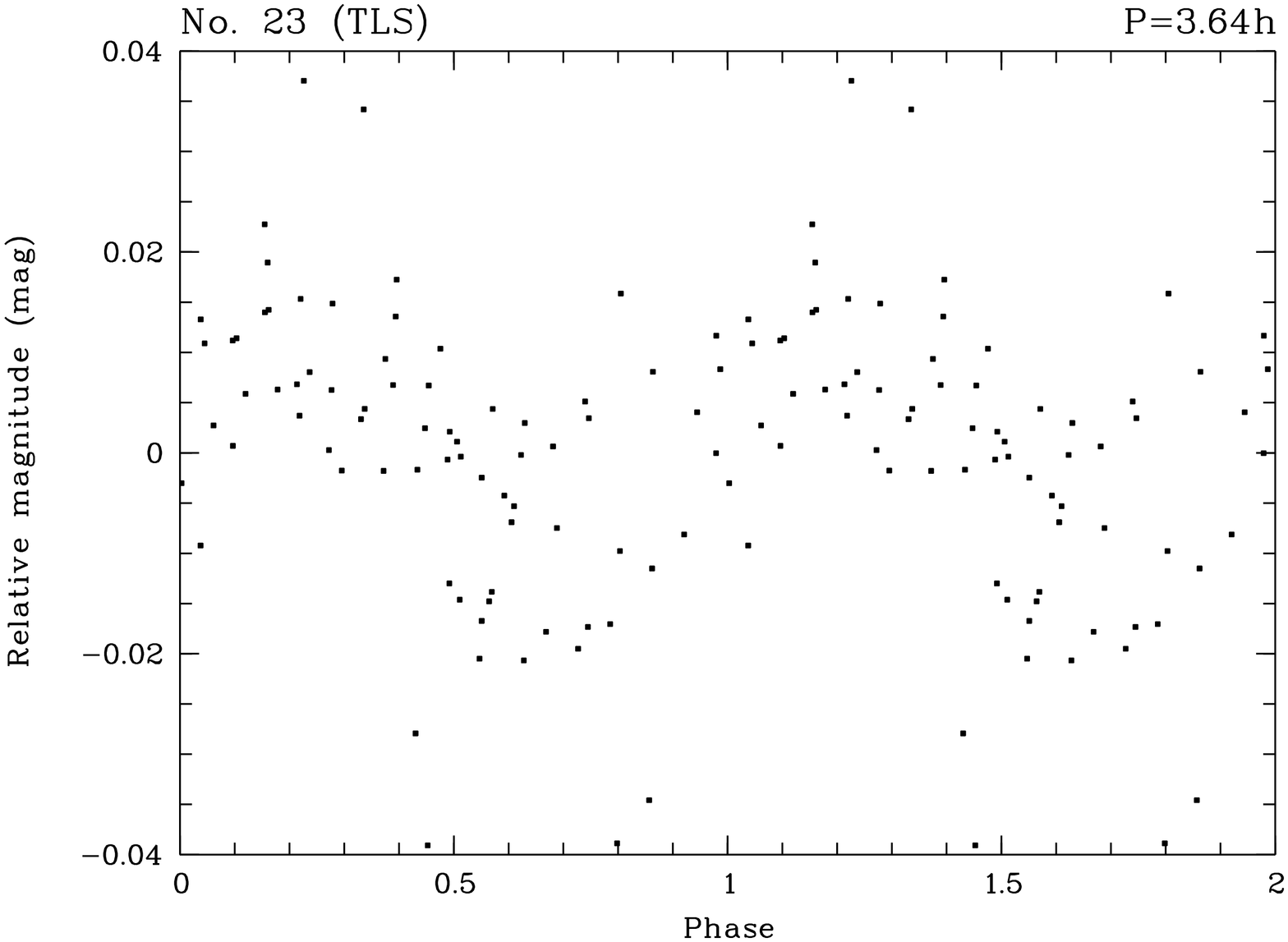}} \hfill
\resizebox{5.5cm}{!}{\includegraphics[angle=-90,width=6.5cm]{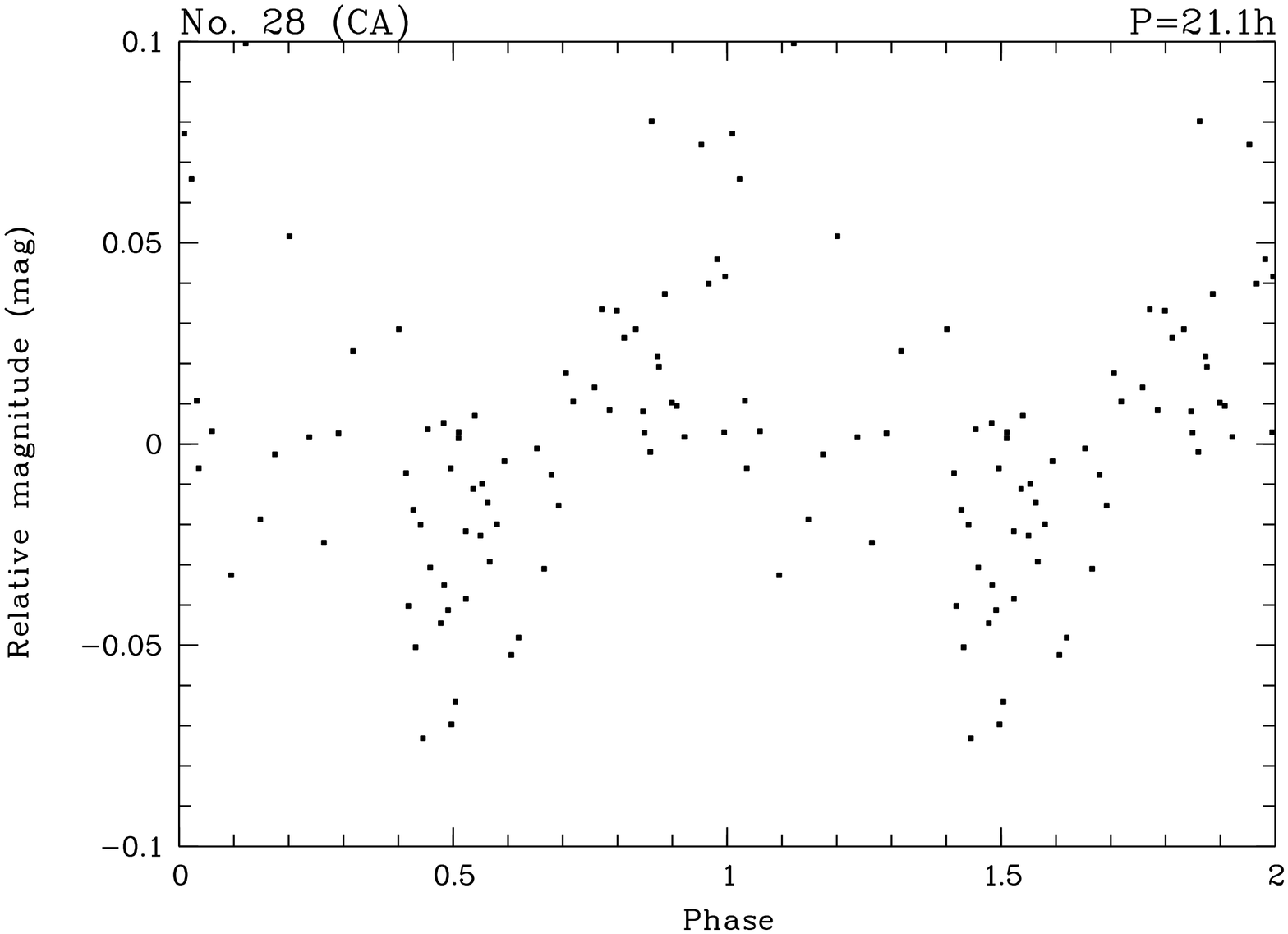}} \hfill
\resizebox{5.5cm}{!}{\includegraphics[angle=-90,width=6.5cm]{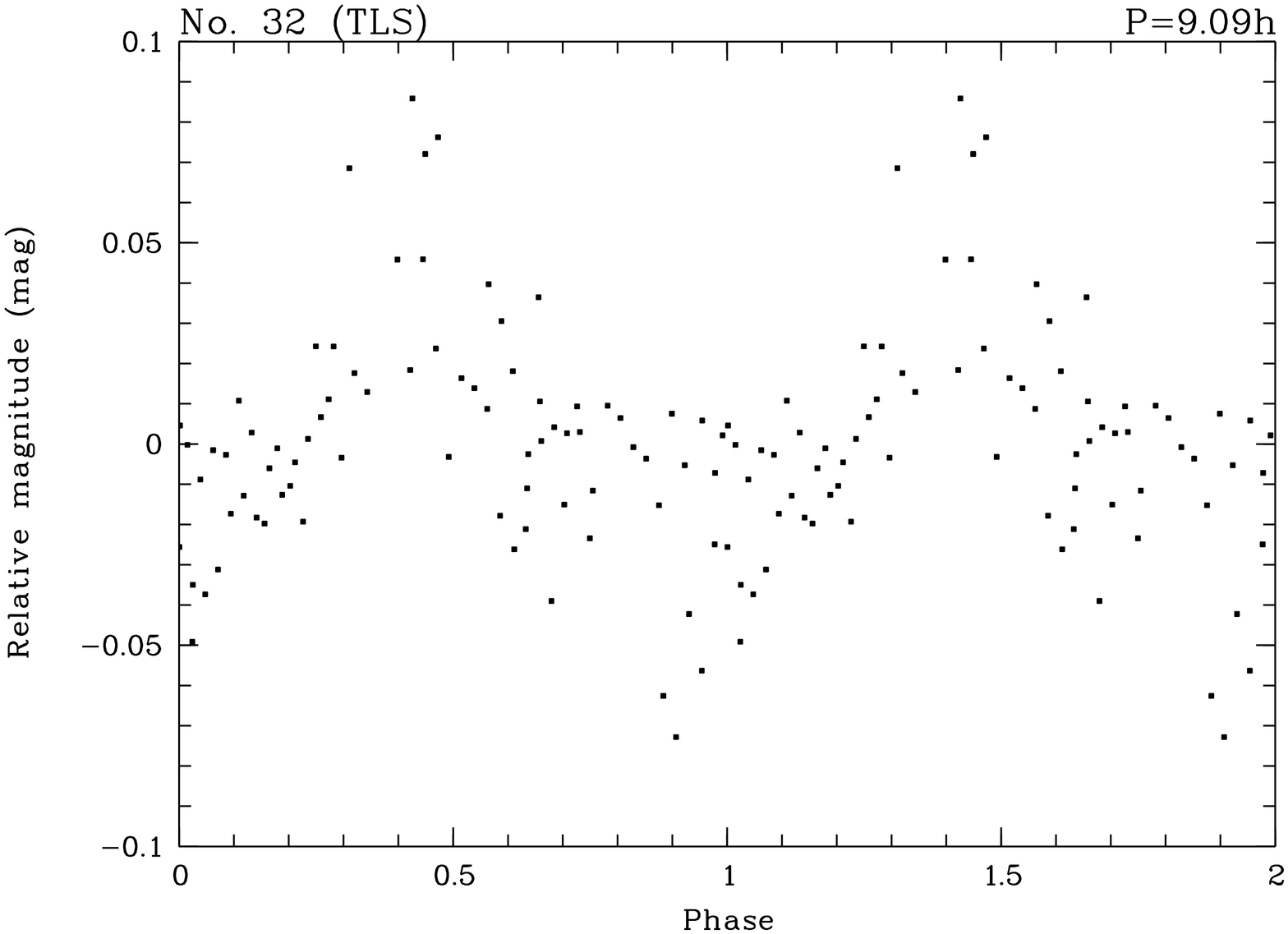}} \\
\caption{Phased lightcurves for the detected periodicities. No. and period from Table\,\ref{resca} 
and \ref{restls} are indicated.}
\label{phase1}
\end{figure*}

\begin{figure*}[htbd]
\resizebox{5.5cm}{!}{\includegraphics[angle=-90,width=6.5cm]{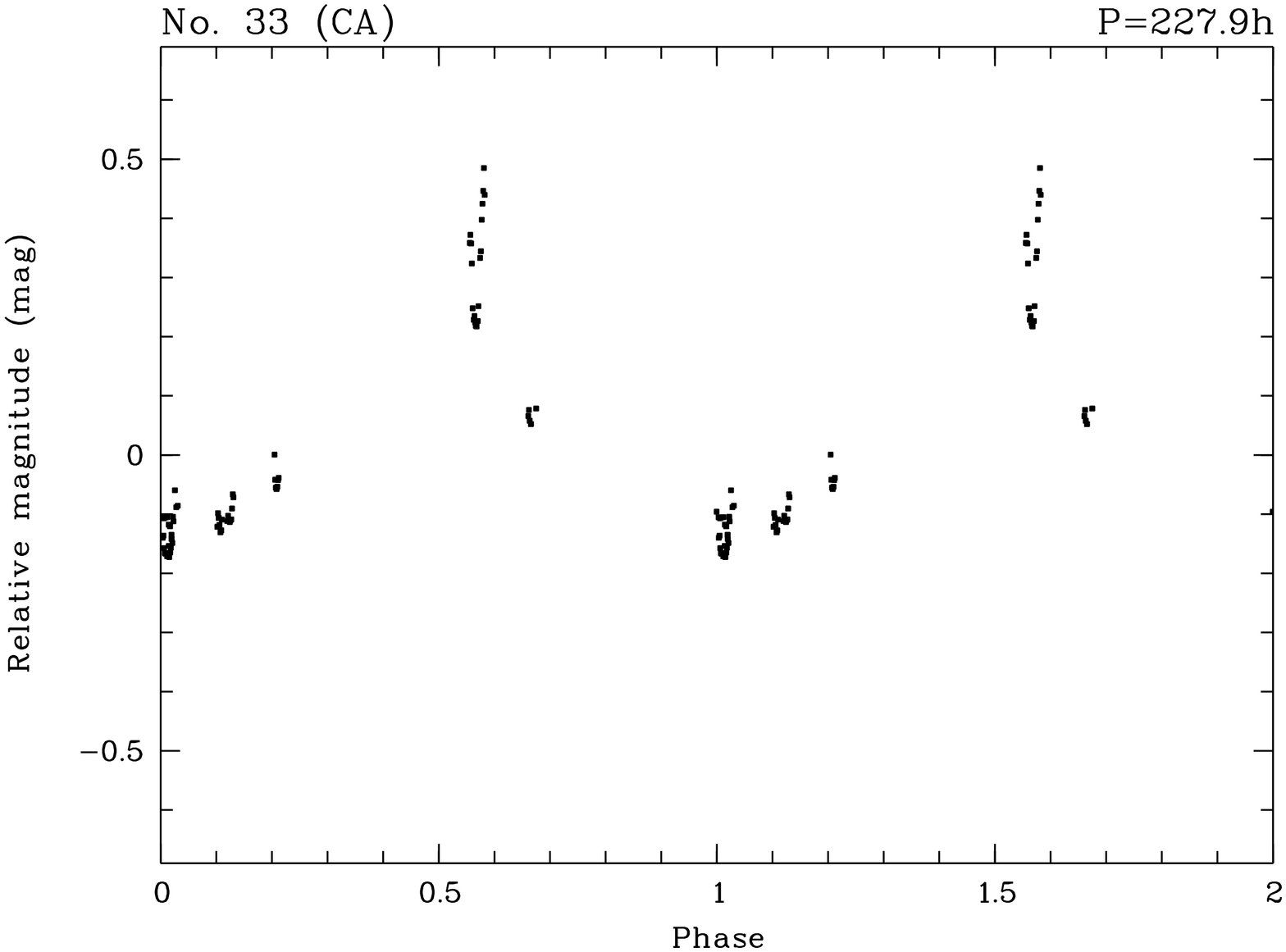}} \hfill
\resizebox{5.5cm}{!}{\includegraphics[angle=-90,width=6.5cm]{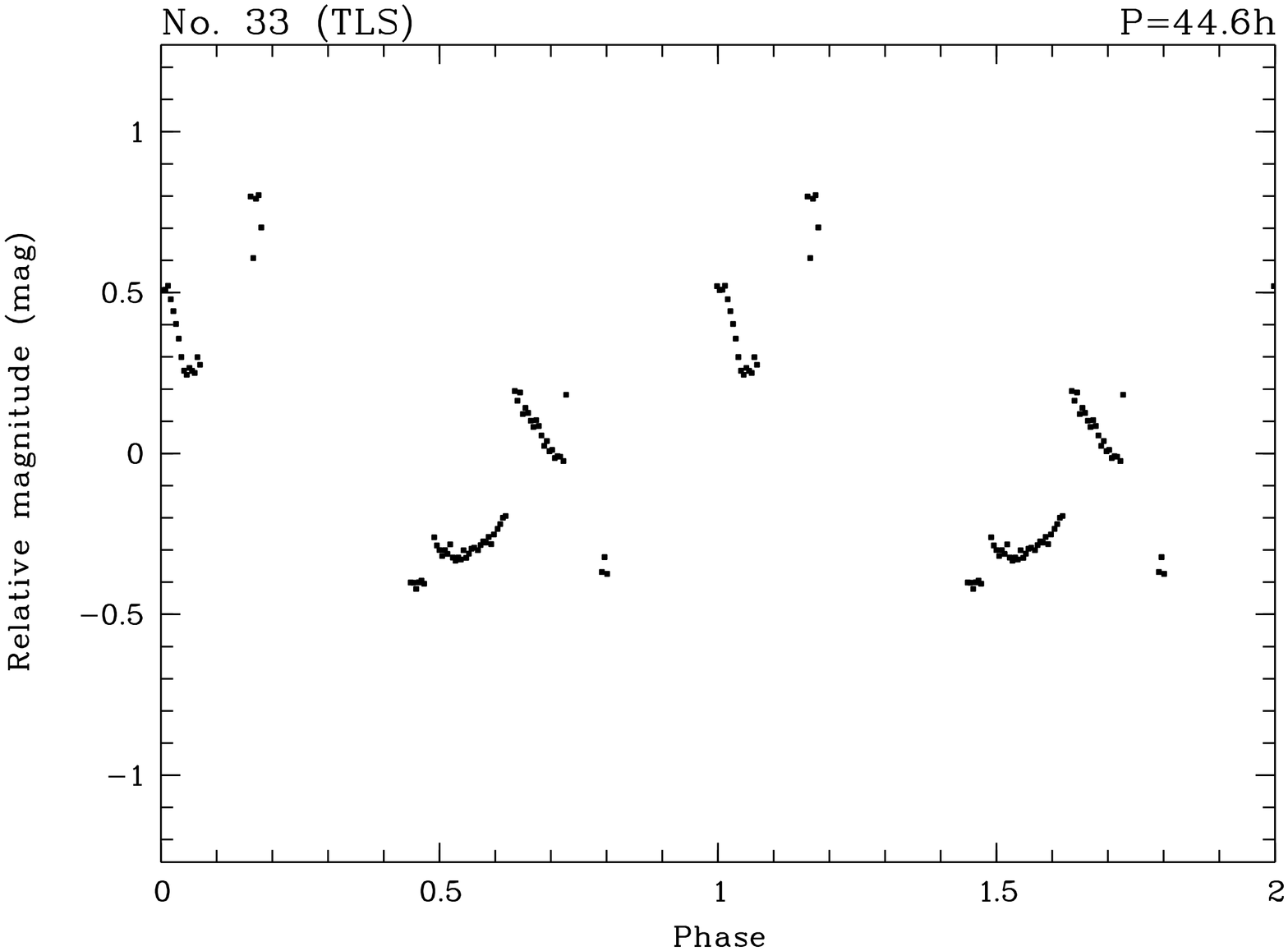}} \hfill
\resizebox{5.5cm}{!}{\includegraphics[angle=-90,width=6.5cm]{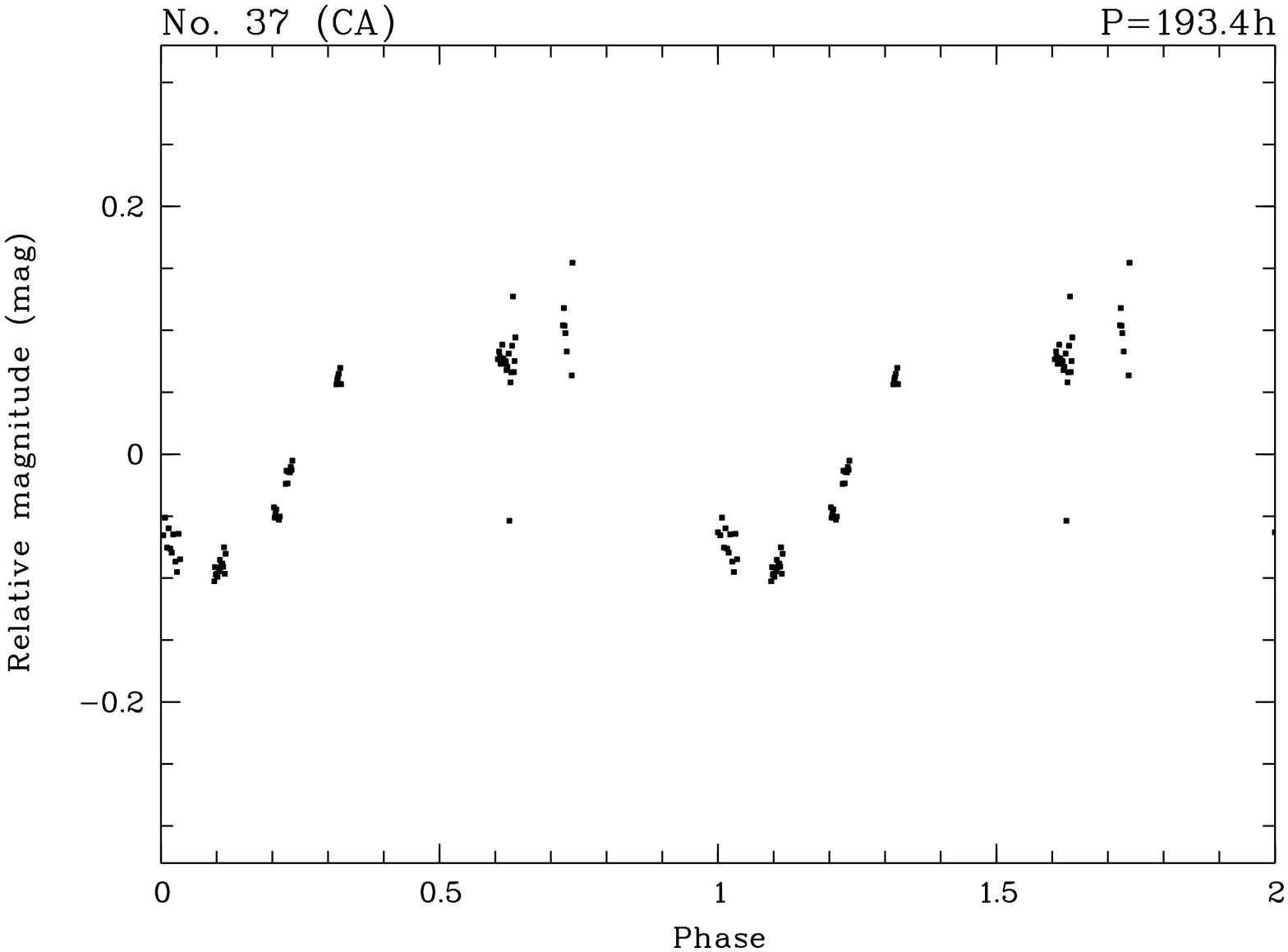}} \\
\resizebox{5.5cm}{!}{\includegraphics[angle=-90,width=6.5cm]{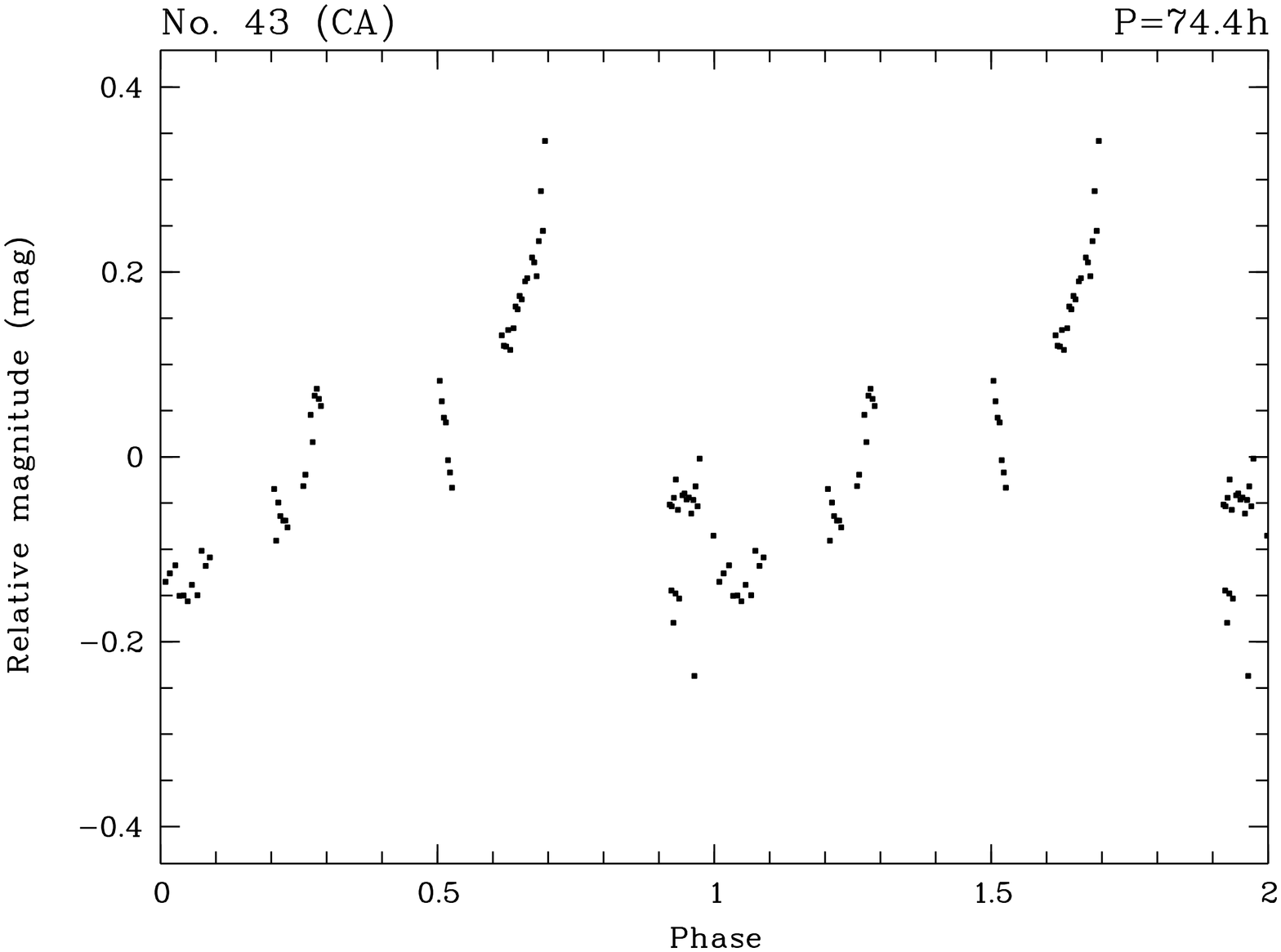}} \hfill
\resizebox{5.5cm}{!}{\includegraphics[angle=-90,width=6.5cm]{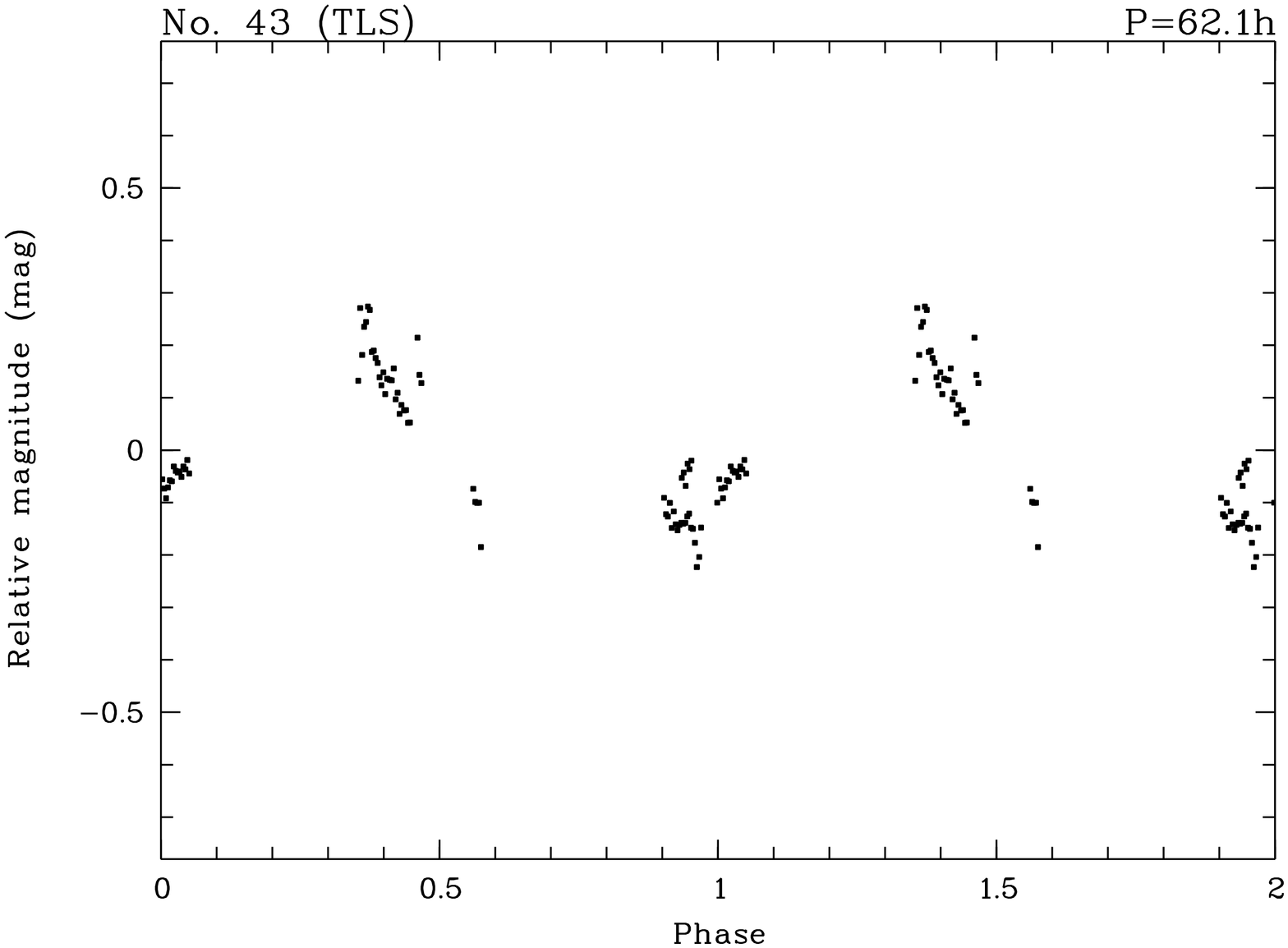}} \hfill
\resizebox{5.5cm}{!}{\includegraphics[angle=-90,width=6.5cm]{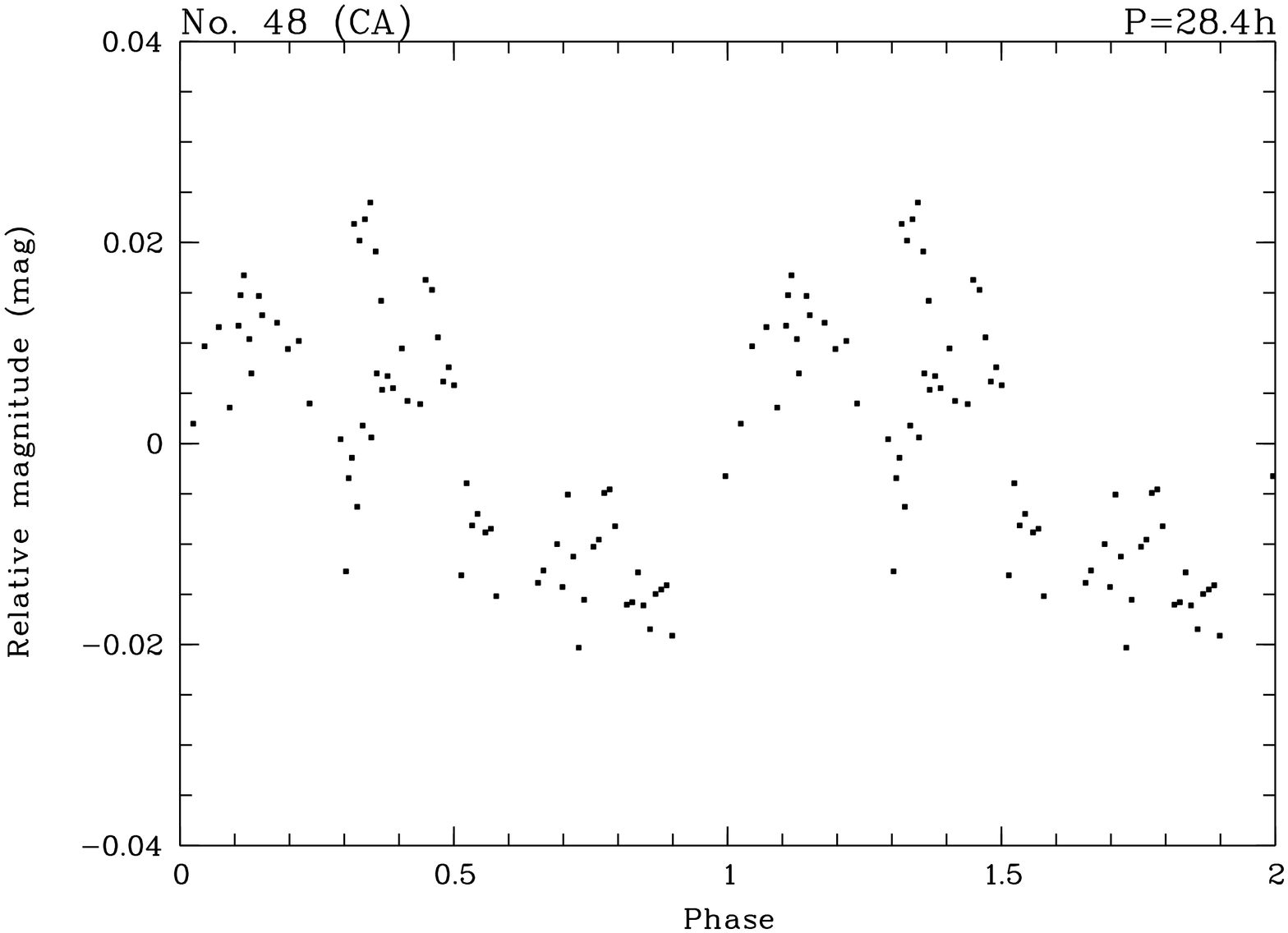}} \\
\resizebox{5.5cm}{!}{\includegraphics[angle=-90,width=6.5cm]{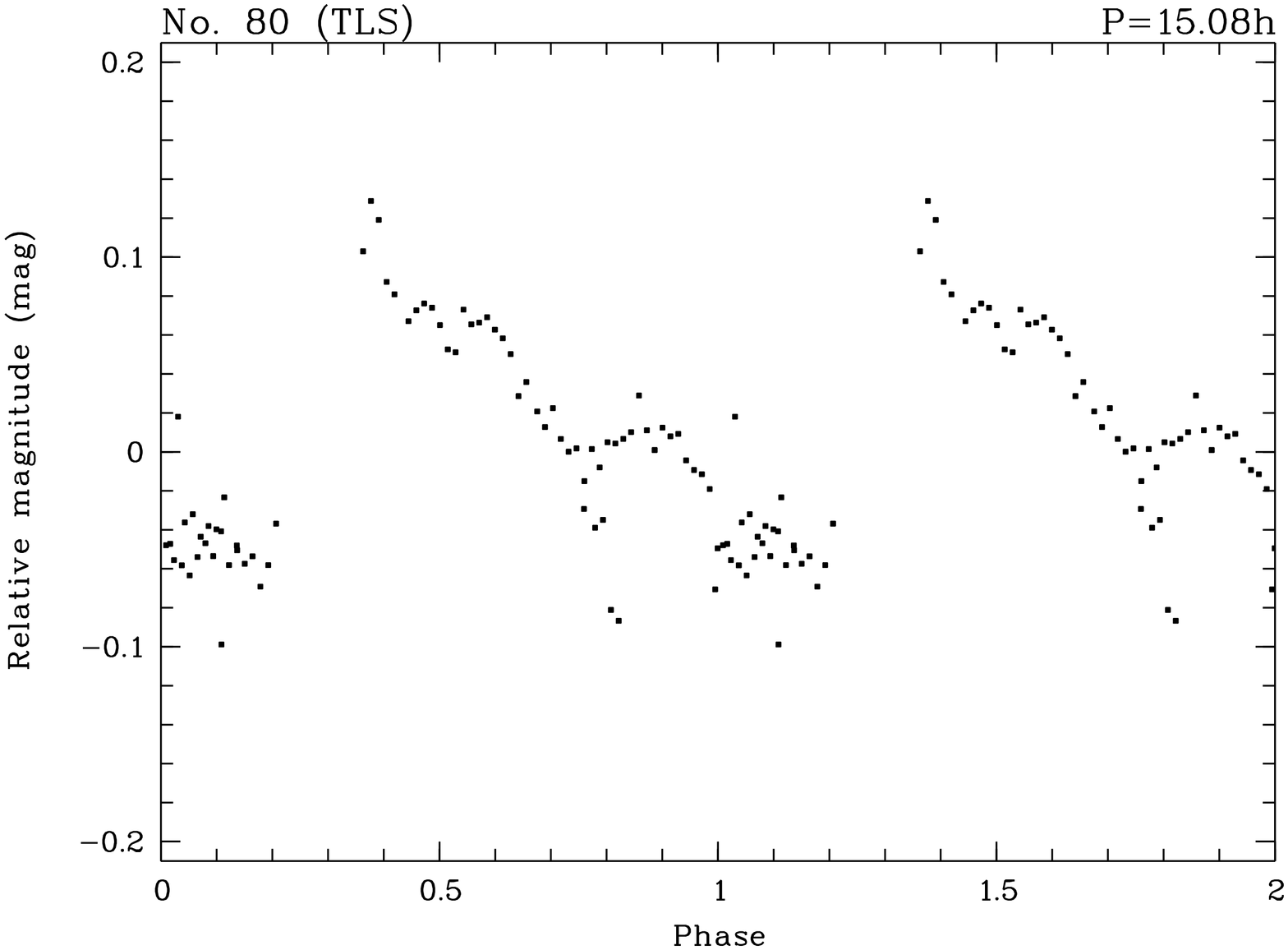}} \hfill
\resizebox{5.5cm}{!}{\includegraphics[angle=-90,width=6.5cm]{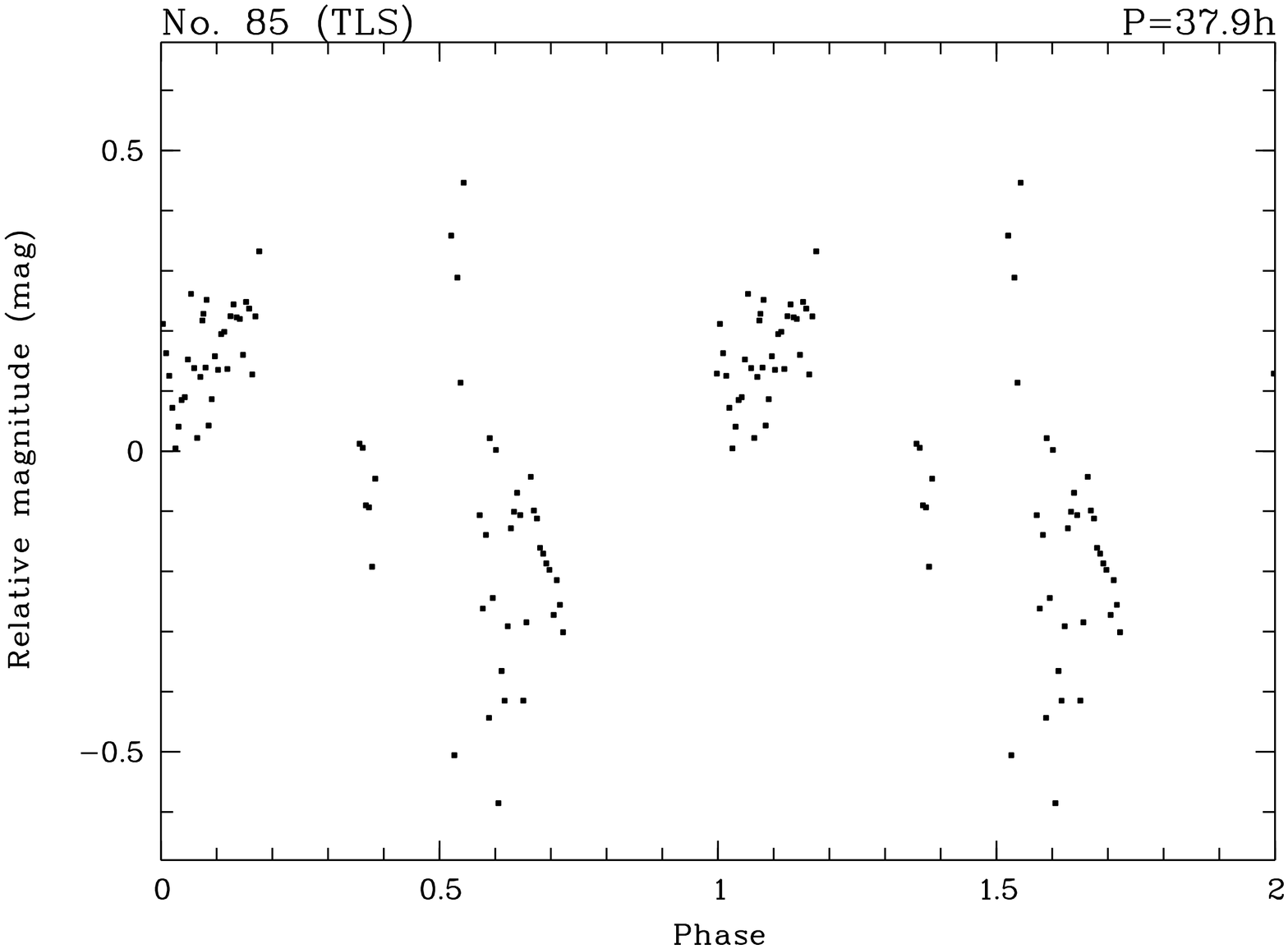}} \hfill
\resizebox{5.5cm}{!}{\includegraphics[angle=-90,width=6.5cm]{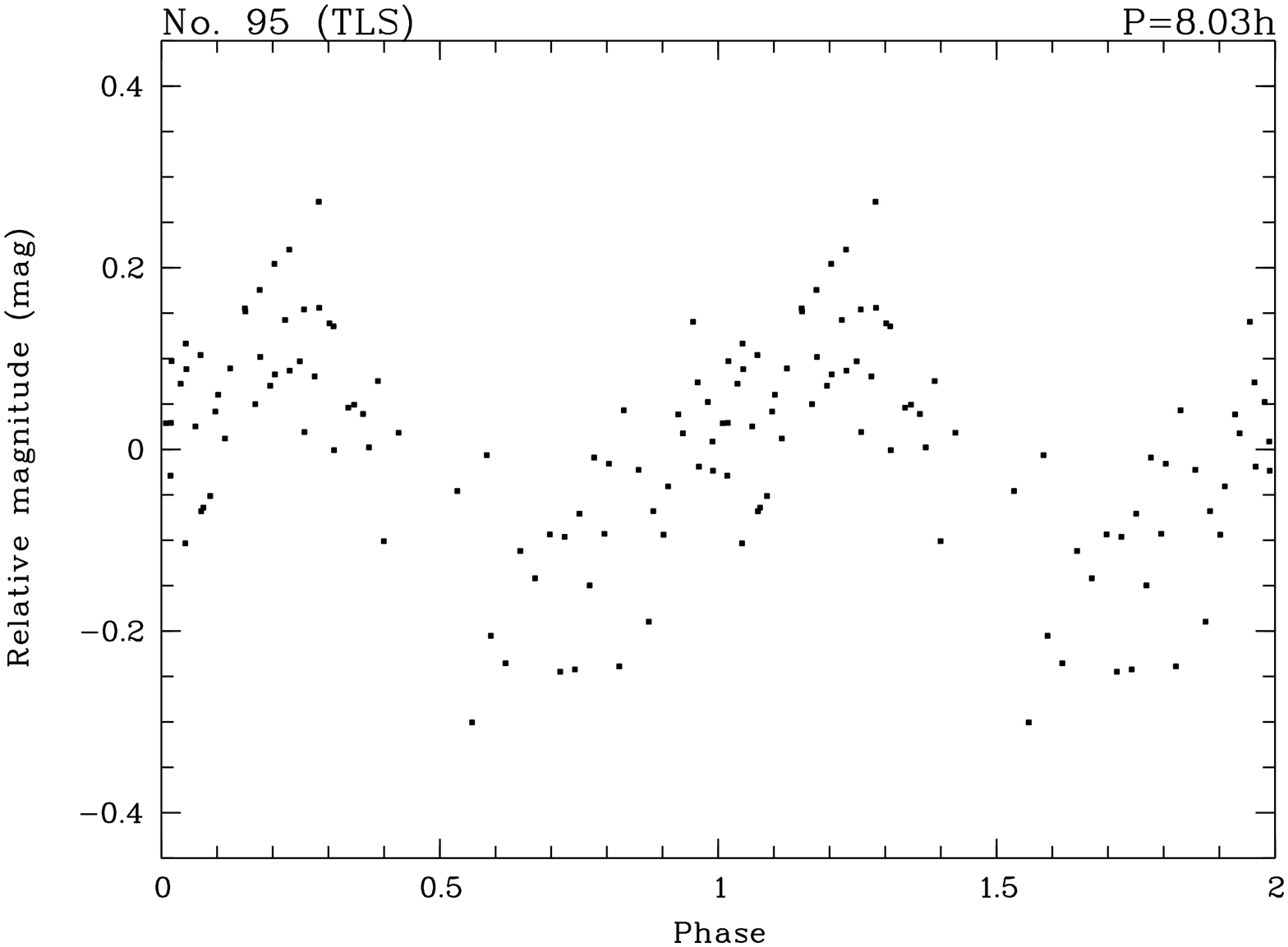}} \\
\resizebox{5.5cm}{!}{\includegraphics[angle=-90,width=6.5cm]{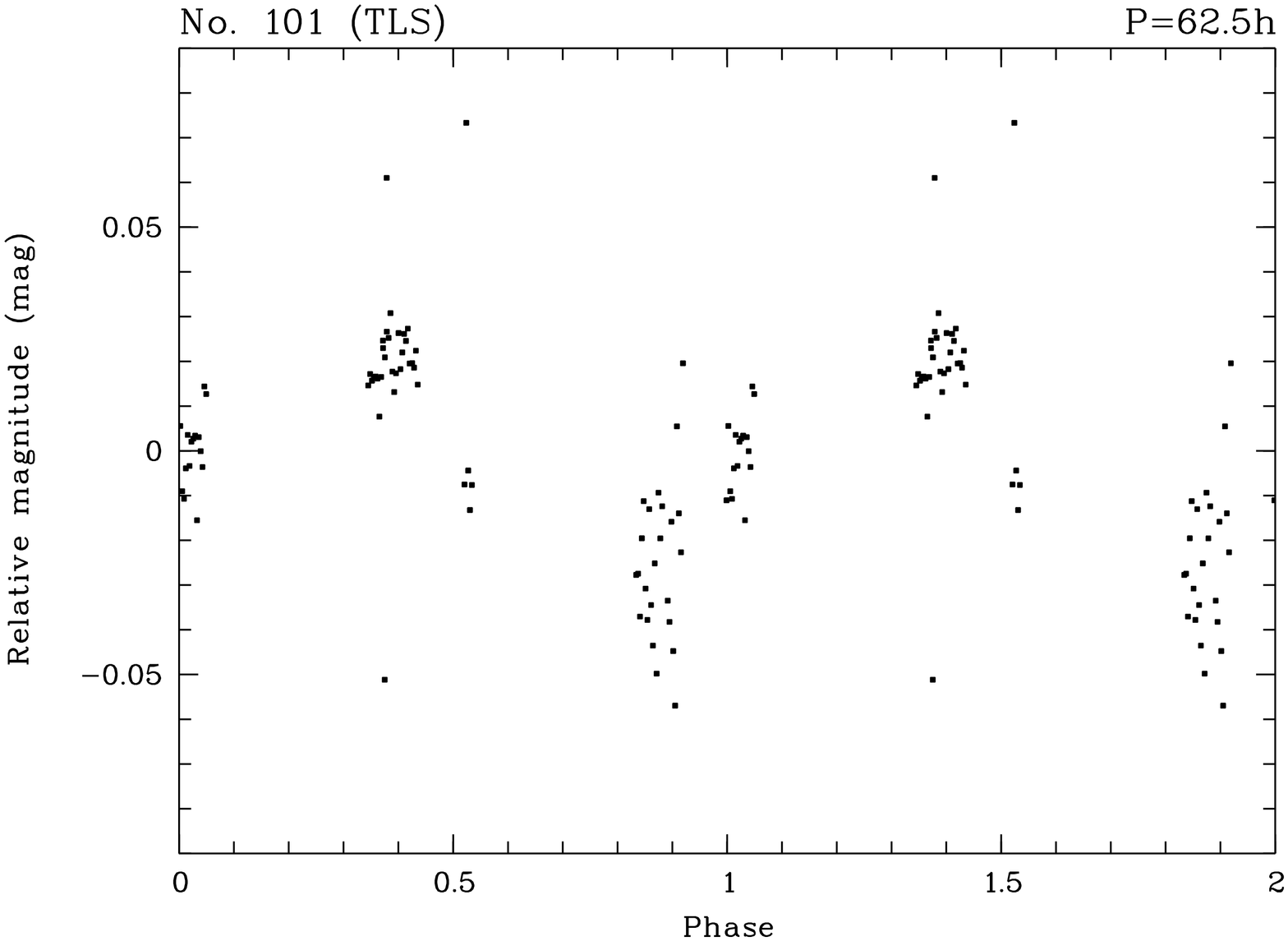}} \hfill
\resizebox{5.5cm}{!}{\includegraphics[angle=-90,width=6.5cm]{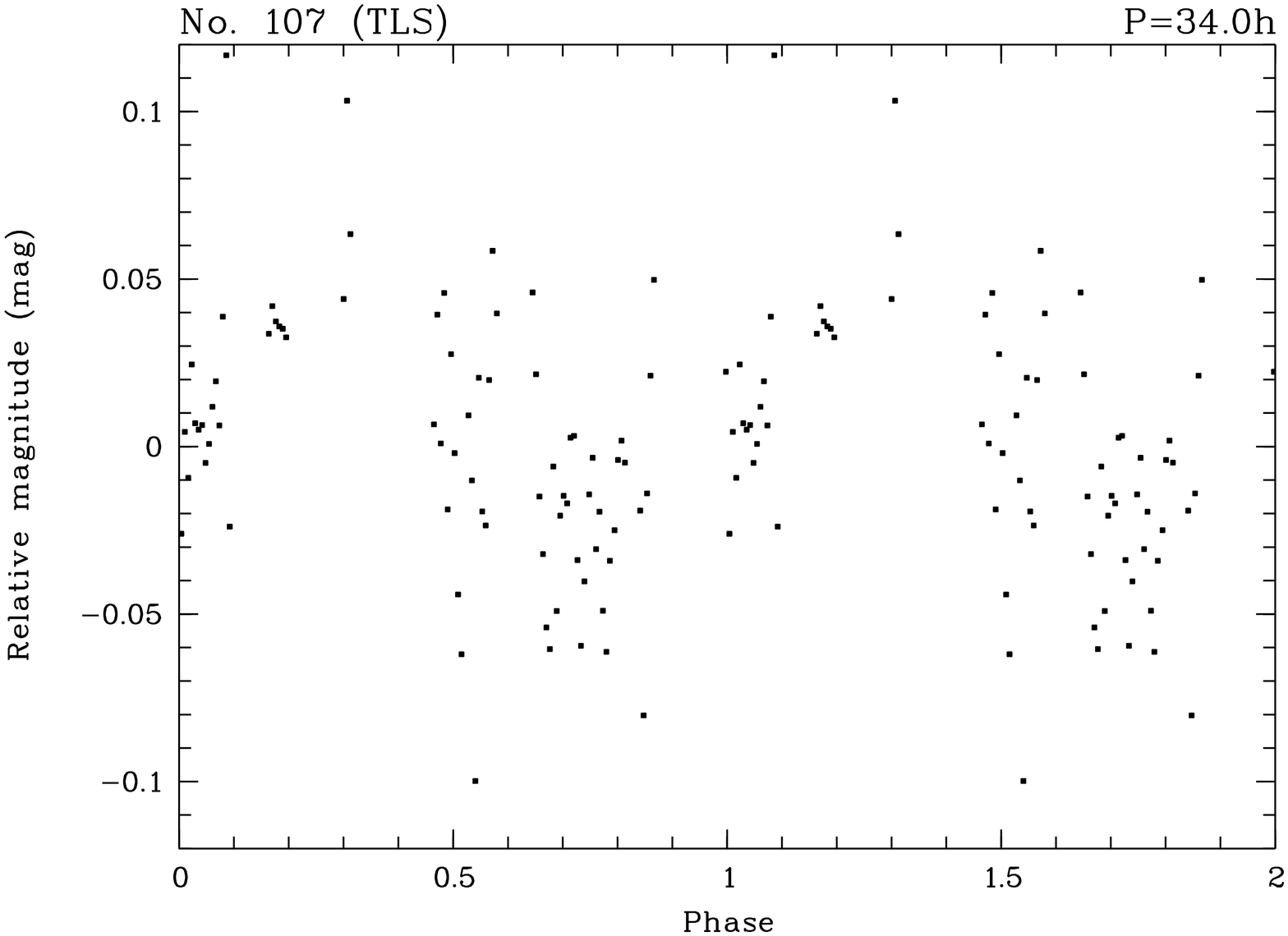}} \hfill
\resizebox{5.5cm}{!}{\includegraphics[angle=-90,width=6.5cm]{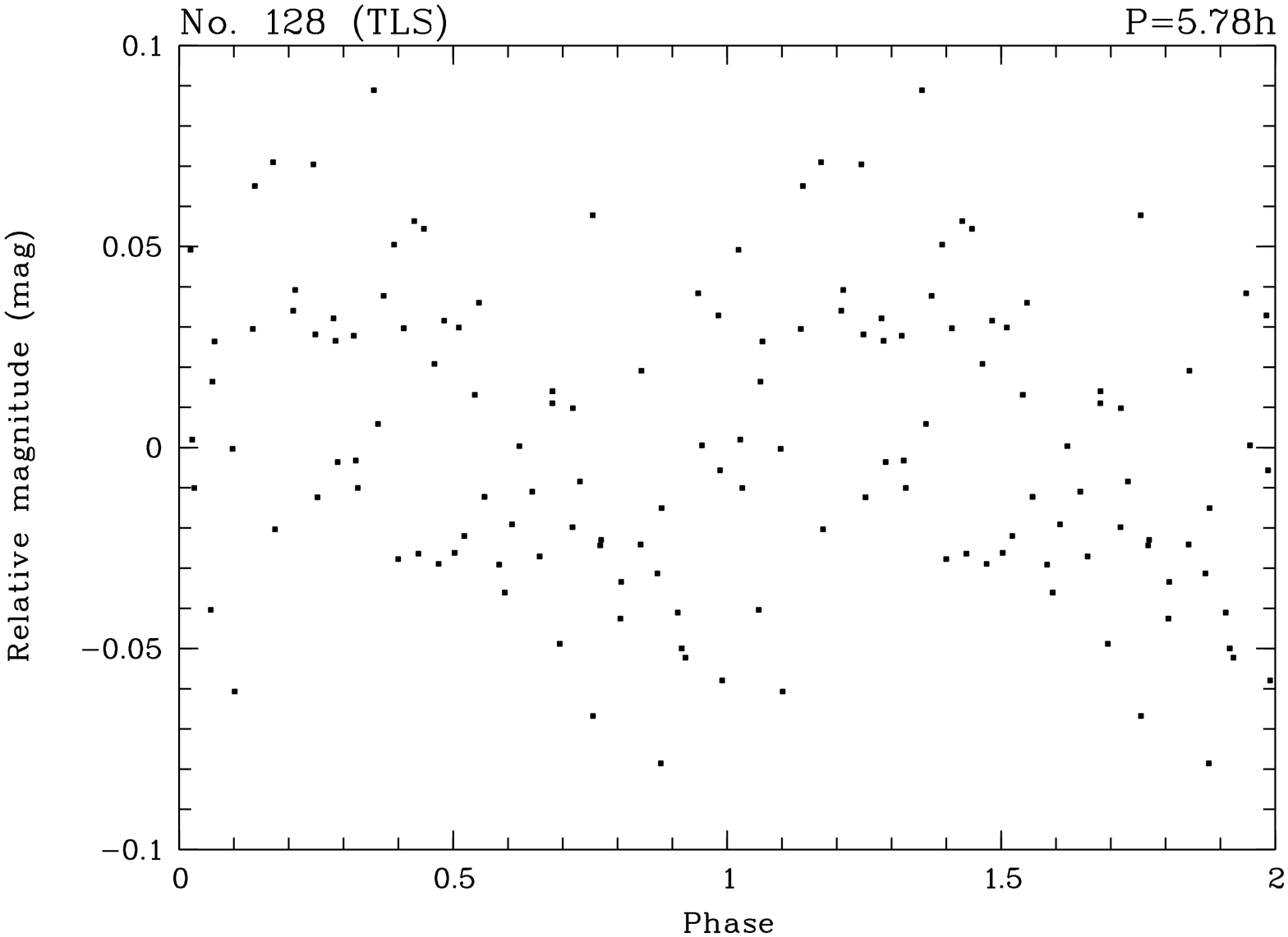}} \\
\caption{Phased lightcurves for the detected periodicities (continued).}
\label{phase2}
\end{figure*}

As an additional test to control the significance of periods independently from periodograms, 
we propose here a method based on the F-test. If a detected period is not significant, then 
the variance of the time series without the period should be not significantly different from the
variance with that period. We fitted each detected period with a sine wave and compared the
scattering of the original time series $\sigma$ with that of the residuals of the fit
$\sigma_\mathrm{sub}$. From $F=\sigma^2 / \sigma_\mathrm{sub}^2$ follows the probability that the 
two variances are equivalent. Thus, this $\mathrm{FAP_{F-test}}$ indicates, how probable it is that 
the found period is caused by chance variations in the noise of the photometry (and not in the 
periodogram). Our results (see Fig.\,\ref{phase1} and \ref{phase2}) show that the sine wave 
approximation is reasonable at least for low signal-to-noise, i.e. for objects for which reliable 
FAP determination is critical. It turns out that the $\mathrm{FAP_{F-test}}$ is in all cases larger 
than the $\mathrm{FAP_{Scargle}}$.

As a first step in our final period search procedure, we filtered the lightcurves
to exclude obvious outliers. The following criteria must be fulfilled to 
accept a period.

\begin{itemize}
\item{The Scargle periodogram has a significant peak, i.e. with $\mathrm{FAP_{Scargle}}$ 
$<1\%$.}
\item{The Scargle periodograms of at least ten near neighbour stars show no significant 
peak at a similar frequency, i.e. no corresponding period is within $\pm10\%$ of the 
candidate period.}
\item{Phased lightcurve and the original lightcurve itself show the period clearly.}
\item{The variance of the original lightcurve is significantly different from the
variance of the lightcurve {\it without} the (sine wave approximated) period, i.e. 
$\mathrm{FAP_{F-test}}<10\%$.}
\item{The lightcurves of nearby stars phased to the candidate period show no 
significant periodicity.}
\item{The CLEAN algorithm does not reject the period as spurious feature in the
periodogram.}
\end{itemize}

We established periodic variability for 14 CA and 13 TLS targets. Periods range from
4 to 240\,hours, amplitudes from 0.02 to 1.1\,mag. Tables\,\ref{resca} and \ref{restls} 
list the relevant data for all objects with periodic variability. Figures\,\ref{phase1} and 
\ref{phase2} show the lightcurves phased to the measured period.

To determine reliable FAP for the periodicities, we used a bootstrap approach as proposed e.g.
by K{\"u}rster et al. (\cite{ksc97}). The resulting empirical FAP will be called $\mathrm{FAP_{E}}$ 
in the following. We obtained 10000 randomized data sets with the same sampling 
as the original lightcurve by retaining the observing times and randomly redistributing the relative 
magnitudes amongst the observing times. We calculated the Scargle periodogram for each of these 
data sets and recorded the power of the highest peak. The $\mathrm{FAP_{E}}$ is the fraction of data sets for 
which the power of the highest peak exceeds the power of our detected periodicity. This time-expensive 
simulation must be accomplished for every object on its own to account for the slightly different window 
function caused by our filtering. Therefore, we did not include these simulations in the period search
criteria outlined above, and used instead the $\mathrm{FAP_{Scargle}}$ and $\mathrm{FAP_{F-test}}$ as 
preliminary significance test. It turned out, however, that all detected periods were indeed highly 
significant with $\mathrm{FAP_{E}}<1$\%. Although the estimated $\mathrm{FAP_{Scargle}}$ tend to be 
somewhat higher than the $\mathrm{FAP_{E}}$, the two values are consistent within the statistical 
uncertainties of the simulation. 

It should be noted that the described bootstrap simulation possibly delivers optimistic values
for the FAP. Young stars frequently show irregular variability on timescales of hours to days, and 
it is likely that some of our targets exhibit a similar behaviour. Therefore, Herbst \& Wittenmyer 
(\cite{hw96}) discussed that datapoints collected within one night may not be independent. On the
other hand, our bootstrap algorithm works on uncorrelated data sets and makes no attempt to simulate
possible irregular variations. Since the data sets are produced by 'shuffling' the original 
lightcurve, such variations are taken into account as increased noise. 
 
Kearns et al. (\cite{keh97}) and Stassun et al. (\cite{smm99}) circumvent the problem by determining
the FAP with Monte-Carlo simulations based on synthetic lightcurves of noise with two dispersions,
one representing the variations during one night and the other representing the night-to-night
variability, calculated as the standard deviation of nightly means. This simulation, 
however, does not adequatly reproduce intrinsic irregular variability, because even
periodic variations will cause an increased night-to-night scatter. Hence, it is not guaranteed
that the synthetic lightcurves contain no periodic component. Therefore, the method will
overestimate the FAP. Moreover, the method assumes {\it a priori} that the datapoints 
are normally distributed and that the timescale for intrinsic fluctuations is {\it one} day. 
A third drawback of the method emerges when the observing conditions are highly variable
and there are nights with very few datapoints, like in our case. Under these circumstances,
the night-to-night scatter (as defined above) is determined with high uncertainty.
For all these reasons, we consider this approach not applicable to our data.

Another possibility to mimic intrinsic night-to-night variability in simulated lightcurves is to 
redistribute only the integer part of the observing dates randomly (Herbst et al. \cite{hbm02}). 
This approach, however, delivers test lightcurves with significantly different sampling
than the original time series, a serious drawback, because the FAP depends critically on the 
window function. Moreover, this procedure only makes sense for time series with datapoints from 
many observing nights, and is therefore not applicable to our data. Lamm et al. (\cite{lbm03}) 
compare the FAP from bootstrap simulations with and without taking into account the correlations
between the datapoints. They found that the Scargle power for a FAP of 1\% decreases only marginally when 
the simulation relies on uncorrelated datapoints. We conclude that our $\mathrm{FAP_{E}}$ might be 
optimistic values, but since the $\mathrm{FAP_{E}}$ for all our periods is below 0.4\% (and for 89\% 
of our periods even below 0.01\%), we are nevertheless confident about their significance. Finally,
we note that every FAP can only be treated as a relative value, only comparable with values derived 
with exactly the same procedure.

\subsection{Sensitivity and completeness}
\label{sensper}

The sensitivity range of the period search is determined by our time sampling.
For regularly spaced lightcurves, the upper frequency limit -- and thus the
lower period limit $P_{min}$ -- is given by the Nyquist frequency $\nu_\mathrm{max}=\frac{1}{2\Delta}$, 
where $\Delta$ is the (constant) distance between two datapoints. The upper period limit $P_{max}$, 
on the contrary, corresponds to the overall length of the time series. For irregularly spaced data, 
these relations are only a rough approximation. For a reliable determination of $P_{min}$ and 
$P_{max}$, we carried out a simulation: For both campaigns, we selected a number of field stars which are 
non-variable according to the generic variability test described in Sect.\,\ref{generic}. Periods from 0.1 to 
300\,h were added as pure sinewaves to the lightcurves of these objects. The amplitudes of the test 
periods were chosen so that the signal-to-noise ratio (SNR, defined as ratio between amplitude 
of periodicity and scatter in the original lightcurve) is larger than 2. For each period, the frequency 
of the highest peak in the Scargle periodogram was recorded. The difference between imposed and detected 
period delivers an estimate for the reliability of the period search for this period. 

\begin{figure}[t]
  \centering
  \resizebox{\hsize}{!}{\includegraphics[angle=-90,width=8cm]{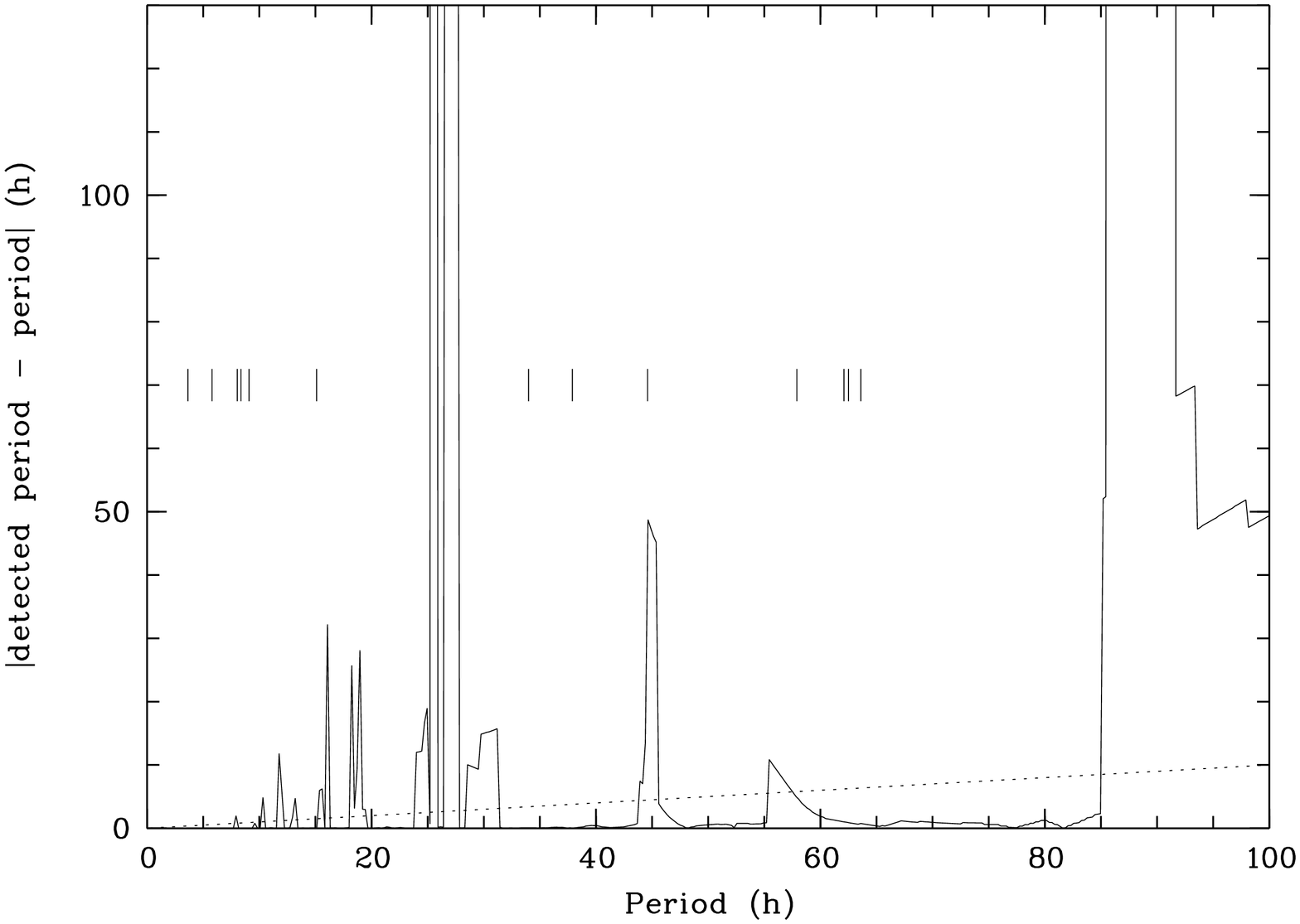}}
  \resizebox{\hsize}{!}{\includegraphics[angle=-90,width=8cm]{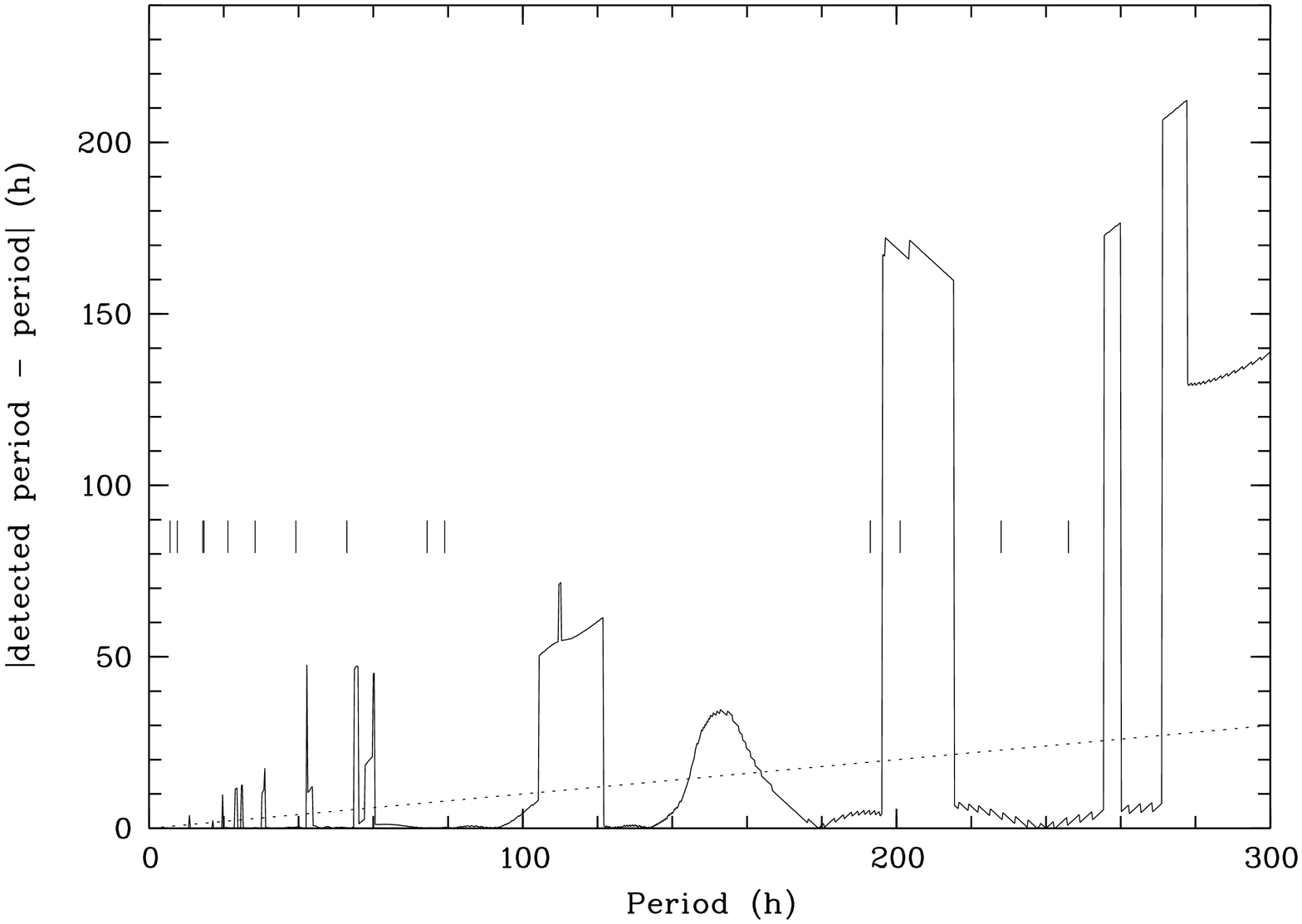}}
  \caption{Sensitivity of the period search: The absolute difference between detected period
  and true period vs. true period for the TLS (upper panel) and the CA run (lower panel). The 
  signal-to-noise ratio of the shown periodicity is 5; the dotted line corresponds to a period 
  error of 10\%. The vertical bars indicate the periods found with our time series analysis.
  \label{senssim}}
\end{figure}

Figure\,\ref{senssim} shows the absolute difference between true and detected period for both
campaigns. The dotted line corresponds to a period error of 10\%. The simulation shows that the 
sensitivity varies with the period. 
We define $P_{max}$ as the longest period for which a detection with error below 10\% is 
possible and obtain 90\,h for the TLS and 270\,h for the CA dataset. Below this limit, there exist 
several, usually narrow windows where the period search with the Scargle periodogram is not reliable. 
These windows are caused by gaps in the datapoint distribution shown in Fig.\,\ref{sampling}. Also 
for short periods the reliability of the period search fluctuates somewhat. This illustrates the 
commonplace that for non-continuous time series the period search will never be complete. 
In Fig.\,\ref{senssim}, we also show the periods found with our time series analysis (see 
Tables\,\ref{resca} and \ref{restls}) as vertical bars. With one exception, all periods fall 
in regions where the period determination is reliable. The only exception, the 200\,h CA period 
for object no. 21, will be discussed below.

On the other hand, the simulation shows that it is easily possible to detect periods down to the Nyquist 
limit (which is $\approx 0.5$\,h for both runs) and below. This is in agreement with Eyer \& Bartholdi 
(\cite{eb99}) who show that for time series with irregular spacing the Nyquist limit is only an upper limit for 
$P_{min}$. These results are more or less independent of the SNR, at least for values 
$>2$, in agreement with similar simulations by Wolk (\cite{w96}). For the simulation shown in 
Fig.\,\ref{senssim}, we used a periodicity with SNR of 5, a value typical for our 
periodicities. For a particular target, this plot can look slightly different. When
the SNR of its lightcurve is very high, the width and heigth of the low-sensitivity windows 
will be slightly decreased. The 200\,h period for CA target no. 21 (see Fig.\,\ref{phase1}) falls 
into such a window. Since the lightcurve of this object has a SNR of 10, much 
higher than the SNR of 5 for which Fig.\,\ref{senssim} was computed, we consider this detection 
to be reliable within the given errors (see Table\,\ref{resca}).

\subsection{Pooled variance diagrams}
\label{pool}

\begin{figure*}[t]
  \centering
  \resizebox{8cm}{!}{\includegraphics[angle=-90,width=7cm]{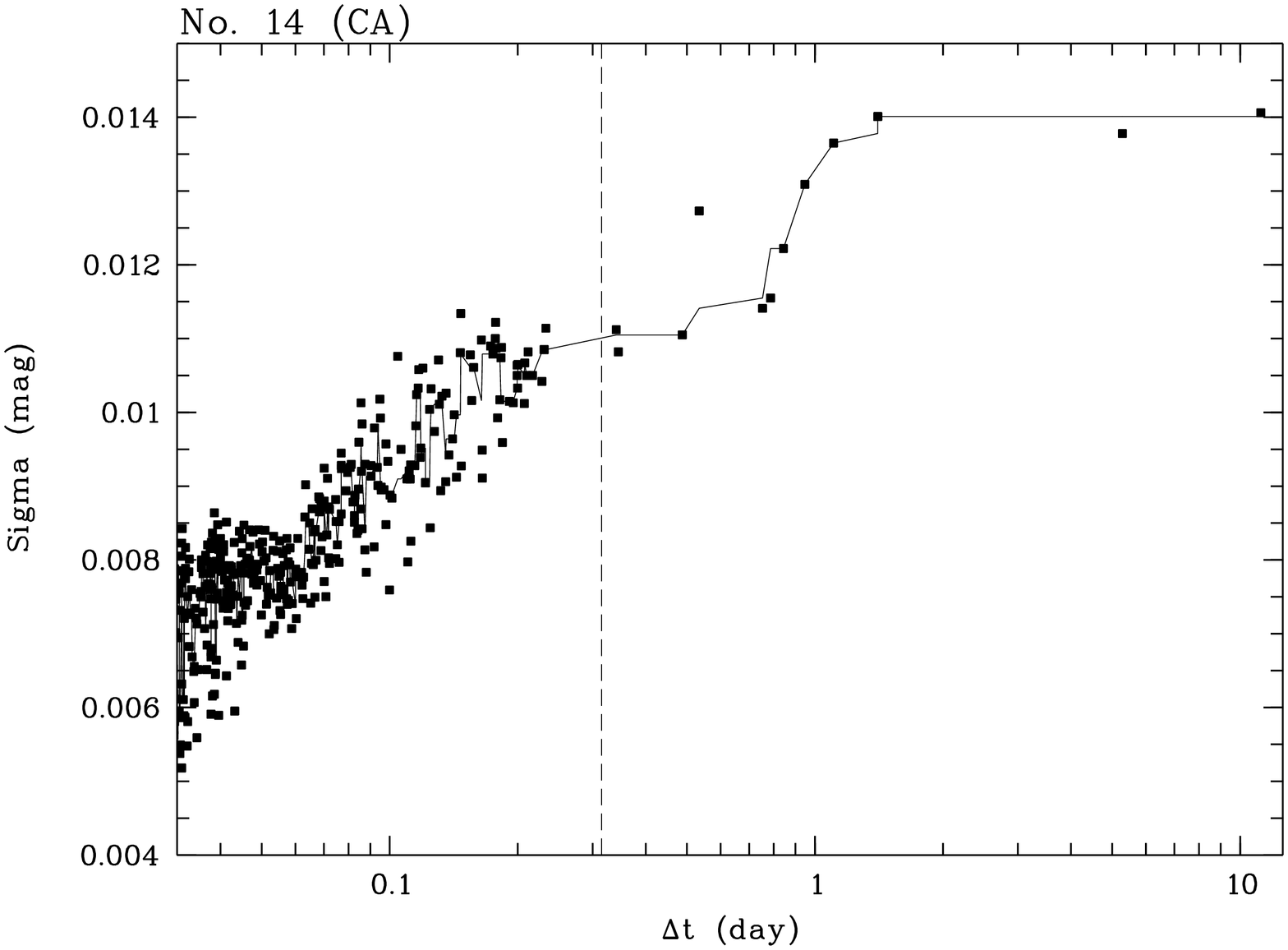}}
  \resizebox{8cm}{!}{\includegraphics[angle=-90,width=7cm]{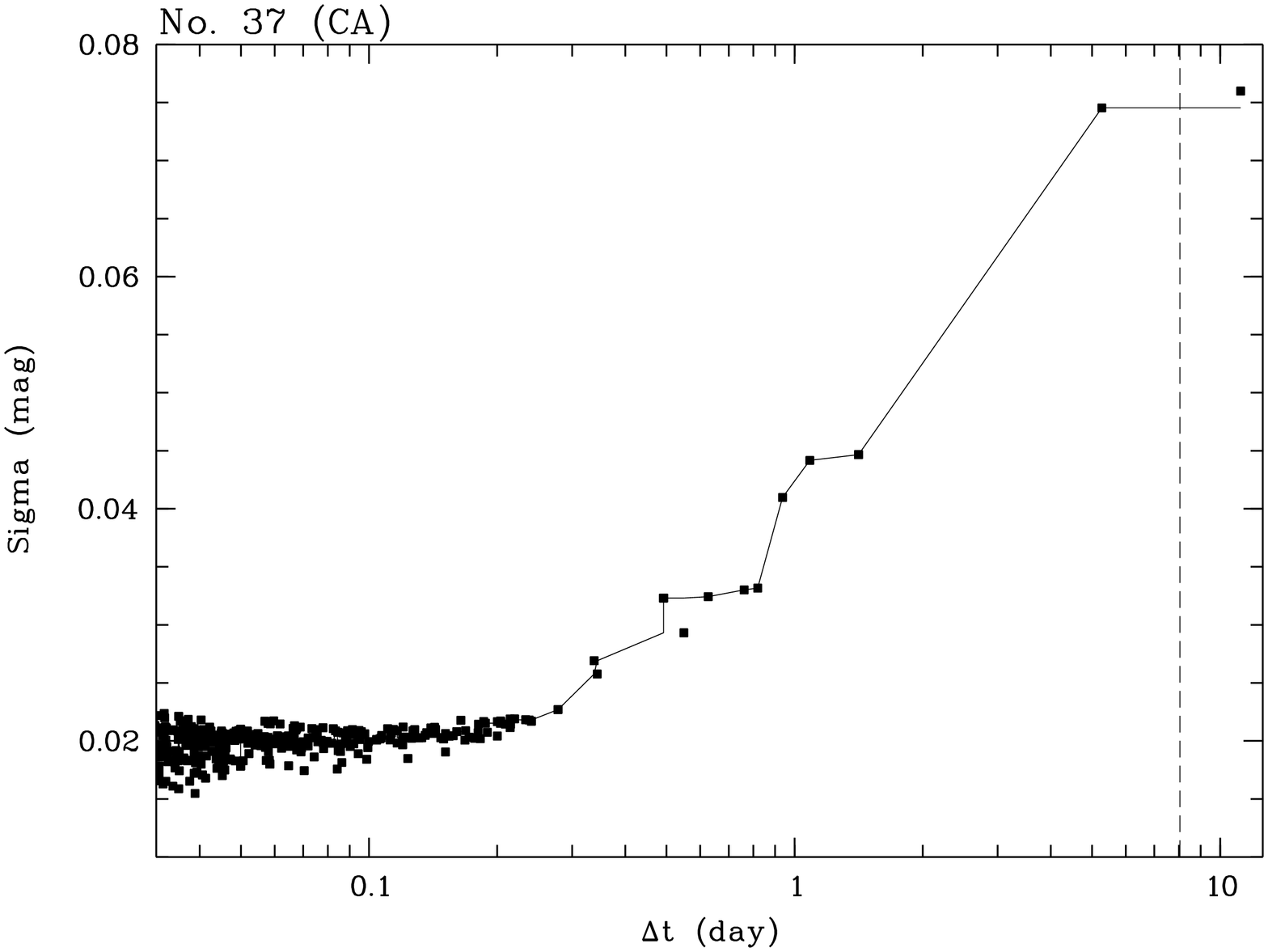}}\\
  \resizebox{8cm}{!}{\includegraphics[angle=-90,width=7cm]{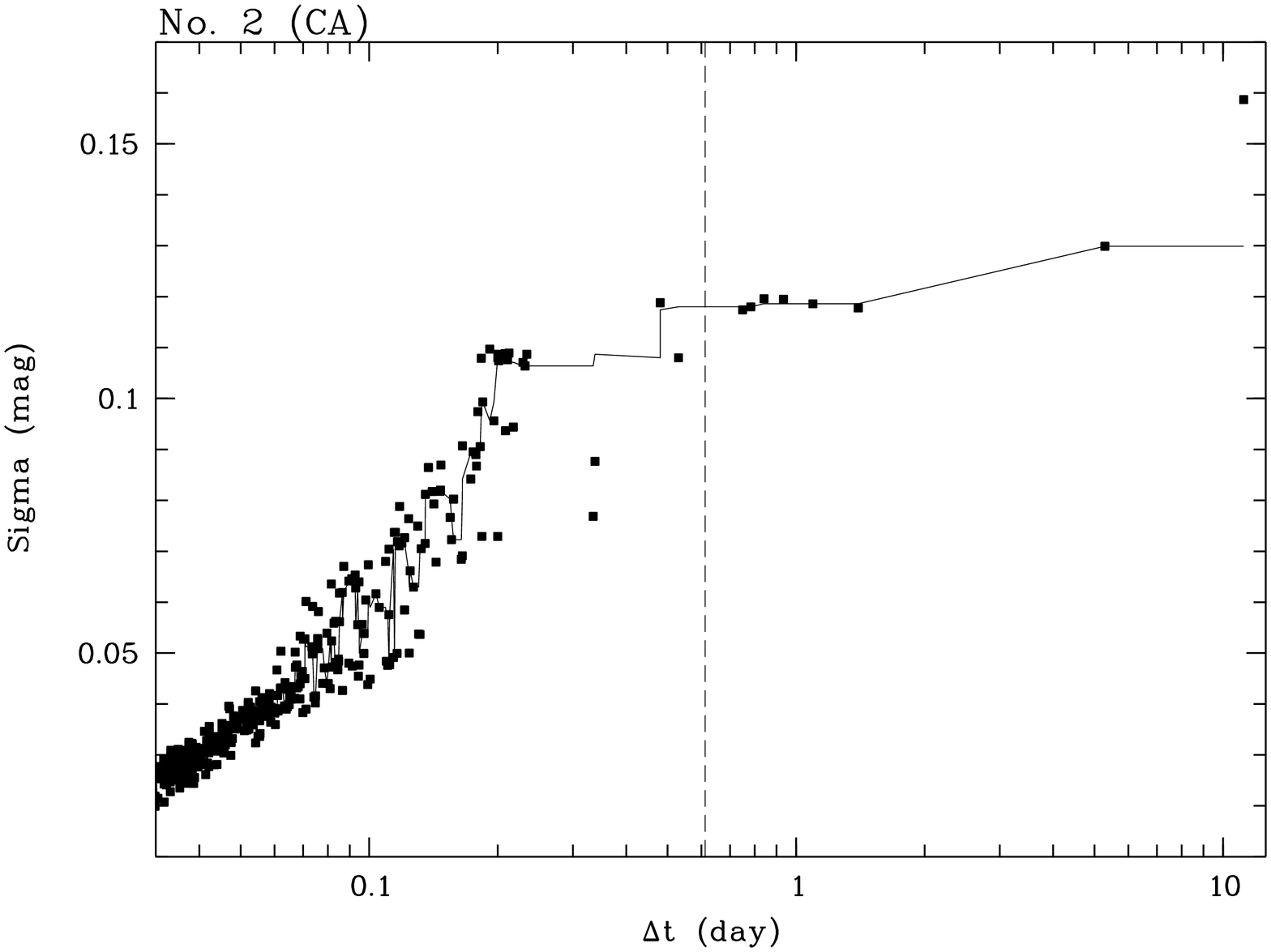}}
  \resizebox{8cm}{!}{\includegraphics[angle=-90,width=7cm]{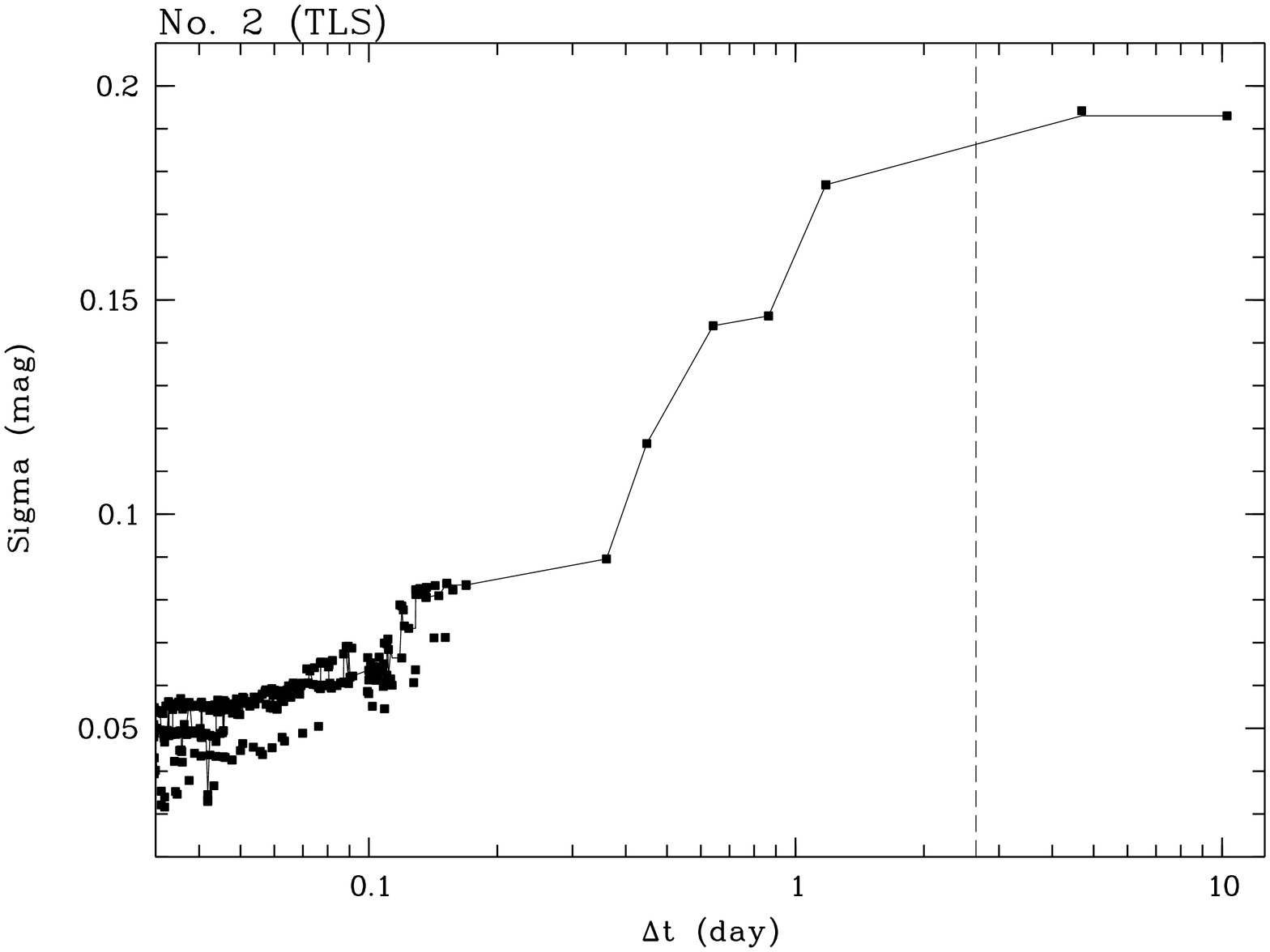}}
  \caption{Pooled variance diagrams for selected lightcurves.
  The solid lines show a running median over five datapoints. 
  The periods adopted in Tables\,\ref{resca} and \ref{restls} are indicated as dashed line.
  \label{pooldia}}
\end{figure*}

The period analysis described in Sect.\,\ref{perser} detected periodic variability in 27 lightcurves.
Some of these lightcurves, however, clearly show variability on timescales different from the adopted
rotation period. Hence, the corresponding phase diagrams show additional variability superimposed
on the adopted period. Moreover, some of the targets also show different periods in both runs, as we will 
outline in Sect.\,\ref{longterm}. To verify the period search results, in particular for the objects 
where the phase plots are less convincing, we therefore investigated the pooled-variance diagram (PVD) 
algorithm, which was introduced by Dobson et al. (\cite{ddr90}). The basic approach of this method is to 
divide the time series into equal bins of specified length and to calculate the variance for each bin. 
The mean of these values (the pooled variance) measures the variability on the specified timescale. 
If the time series contains a period, the pooled variance should show a plateau around this period, starting
roughly at $P/2$. Although the PVD is surely not suitable for precise period determination, it allows a 
robust verification of the found periods. Moreover, it will reveal if there is substantial evidence for 
variability on timescales different from the period. We used a PVD implementation from M. K{\"u}rster, 
described in detail in K{\"u}rster et al. (\cite{kee00}). The program was applied to all lightcurves with 
detected periodic variability. Especially interesting examples for the resulting PVD are shown in 
Fig.\,\ref{pooldia}. 

The most important result of our PVD analysis is the confirmation of the periods given in Tables\,\ref{resca} 
and \ref{restls}. There is no case where the PVD shows no plateau around the adopted period.
According to the PVD, the objects with detected periodicity clearly fall into two groups. The PVD for
targets with low-amplitude variability reveal a broad plateau around the period,
as shown in the upper left panel of Fig.\,\ref{pooldia}. Sometimes there is an additional plateau at timescales
comparable with the overall length of our lightcurves, which can be attributed to long-term variations.
On the other hand, objects with high photometric amplitude in most cases show several plateaus (see the upper 
right panel of Fig.\,\ref{pooldia} for an example). Apart from a clear flat area around the adopted period, 
there are often plateaus between 0.2 and 1.0 days, which cannot be attributed to periodicities. Thus, for
these objects the PVD indicate the existence of short-term variability superimposed to the period. 

The objects no. 2, 33, and 43 show high amplitude variations in both campaigns. In the following,
we discuss exemplarily the PVD for object no. 2. In the lower part of Fig.\,\ref{pooldia},
we show the PVD for the CA (left panel) and the TLS (right panel) lightcurve. The 
CA lightcurve clearly shows a plateau around the adopted period of 14.7\,h. There is also evidence for
variability on even shorter timescales, but no signs for a period of 63.6\,h, as detected in the TLS
lightcurve. The PVD for the TLS data, however, shows several plateaus, including one around the 
63.6\,h period and a second around 16\,h, in agreement with the period from the CA lightcurve.
Hence, only the CA period is confirmed by both diagrams. The same result was obtained for objects no. 
33 and 43. Since the TLS periodograms deliver ambiguous results, offering two possible periods, of which 
the CA periodograms confirm only one, which is also in agreement with the PVD analysis, we therefore 
consider the CA periods of these three targets more likely to be the rotation period.

\subsection{Long-term variability} 
\label{longterm}

The variability analysis was accomplished separately for both campaigns. 
As mentioned above, 27 objects in the CA field were also contained in the TLS catalogue. 
Since both time series campaigns were separated by 10 months, it is possible to investigate 
the long-term variability for those targets that were observed twice. The strongly
different noise characteristics of both runs, however, complicates the comparison of the 
analysis results.

First, we compare the generic variability test results from Sect.\,\ref{generic}.
Five objects (the high-amplitude objects no. 2, 33, 43 and the low-level 
variables no. 14, 51) are variable in both runs. The remaining 
targets with variability detection in the CA lightcurve are either too bright or have
too low amplitudes to be detected also in the TLS data. On the other hand, there are no
objects which are {\it only} variable in the noisier TLS time series. Thus, we conclude that
variability on young VLM objects can persist over one year.

We examined whether a detected period is present in the alternative campaign as well. 
For nine of the doubly-detected targets, we find periodic variability in the CA time series. 
\begin{itemize}
\item{The objects no. 2, 33 and 43 show high amplitude variations in both campaigns.
As discussed in Sect.\,\ref{pool}, the CA periods can be recovered also from the
TLS lightcurves, but not vice versa. Therefore, the CA periods are more likely to be the
true rotation periods.}
\item{Targets no. 8, 9, 15, 28 show low amplitude periods in the CA lightcurves. None of the
four objects exhibits significant variability in the TLS data, neither periodic nor
non-periodic. For two of these targets, the CA amplitudes are too low to be detected in the 
more scattered TLS lightcurves, i.e. the average rms of the TLS lightcurves for the respective 
I-magnitude is higher than the CA amplitude. However, for targets 8 and 15 the CA amplitude 
would be detectable even in the TLS lightcurves. Thus, the flux modulation must be 
significantly reduced compared to the CA data.}
\item{Target no. 14 has an 8\,h period in the CA lightcurve and was significantly variable 
during the TLS run. However, the lightcurve is slightly contaminated by extinction residuals,
possibly preventing the detection of a period.}
\item{Target no. 22 exhibits a significant 8.4\,h period in the TLS lightcurve and a 14.4\,h period
in the CA data, i.e. the CA period is roughly twice the TLS period. In the next section, we will 
attribute the periodic modulations to co-rotating surface features. It is easily possible that the 
distribution of surface features produces a flux modulation at half of the rotation period, e.g. 
through two opposite spots. Therefore, the 14.4\,h period is more probably the true rotation period.}
\end{itemize}

Additionally, three targets with period detection in the TLS run (no. 16, 23, 32) are not variable
during the CA campaign. This is particularly surprising in the case of no. 16, since it shows 
high-amplitude variations during the TLS run. The remaining two cases, no. 23 and 32, show the 
inverse behaviour of no. 8 and 15, i.e. their low-amplitude variability is only present in one run 
and at least diminished in the other. 

\section{Interpretation of the variability}
\label{inter}

\begin{figure}[htbd]
    \centering
    \resizebox{\hsize}{!}{\includegraphics[angle=-90]{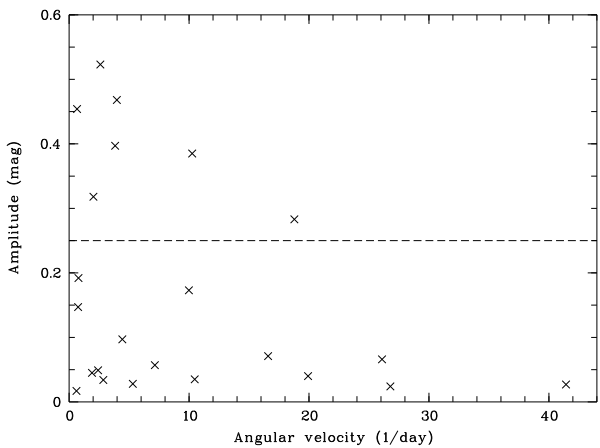}}
    \caption{Angular velocity versus amplitude. The dashed line delineates the separation 
    between low-amplitude and high-amplitude objects.}
    \label{peramp}
\end{figure}

In this section, we argue that the observed variability has its origin in 
two fundamentally different kinds of surface activity. We discriminate the two
types of variability by means of the photometric amplitude, as given
in Tables\,\ref{resca} and \ref{restls}. For periodicities detected in
the TLS and the CA lightcurve, we use the CA amplitude in the following.
In Fig.\,\ref{peramp} we plot amplitude vs. angular velocity ($\Omega=2\pi/P$) 
for all variable objects. Clearly, there is a strong cumulation of targets with 
low-level variability. The remaining objects show variability with amplitudes 
ranging from 0.28 to 0.55\,mag. There is a clear gap of around 0.1\,mag width
around amplitudes of 0.25\,mag, indicated by a dashed line in Fig.\,\ref{peramp}. 
Furthermore, the high-amplitude lightcurves deviate visibly from a strict sine shape. 
To quantify these deviations, we fitted all lightcurves with a sine function and 
measured the rms of the residuals. If the variability is well-approximated with a 
sine fit, the rms should not significantly exceed the mean rms of field stars, as 
given in Figs.\,\ref{rmsca} and \ref{rmstls} (solid lines). It turns out that most 
of the highly variable objects with amplitudes $>$0.25\,mag still show considerable
variability after the sine wave has been subtracted, discriminating them clearly 
from the low-amplitude objects. The classification in 'low-amplitude' and 
'high-amplitude' variables is also confirmed by the pooled variance diagrams 
(Sect.\,\ref{pool}), since only the highly variable objects show clear signs 
of variability on timescales different from the rotation period.
In the following, we discuss possible origins for the low-level and the high-level
variability.

\subsection{Low-level variability}
\label{low}

As shown above, we established periodic variability with amplitudes $<$0.2\,mag for 16 
objects. Similar periodicities are well-known for stars in young open clusters (e.g., 
Krishnamurthi et al. \cite{ktp98}, Patten \& Simon \cite{ps96}). Such variability can be 
explained by the existence of asymmetrically distributed surface features co-rotating with
the objects.

Our variable targets have effective temperatures of $>$2700\,K, estimated from the 3-Myr-isochrone
of Baraffe et al. (\cite{bca98}). Hence, they are probably too hot to harbour condensated
dust clouds in their atmospheres (see Allard et al. \cite{aha01}).
The origin of the low-level variability are therefore most probably magnetically induced
spots, slightly cooler or hotter than their photospheric environment. The properties
of these spots, i.e. temperature and filling factor, can only be determined from
multi-filter/spectroscopic monitoring or future Doppler imaging campaigns. Nonetheless, our
results give first clues about photospheric surface activity on very young VLM objects. Only 
17\% of the targets show periodic variability -- a significantly lower 
rate than in similar studies for solar mass stars, indicating low activity levels. This is
contrary to the results of the X-ray study by Mokler \& Stelzer (\cite{ms02}), who find
no decrease of X-ray activity with mass. Their sample of X-ray detections
could, however, be biased against objects with ongoing accretion. Thus, at least part of the 
X-ray flux is possibly not caused by photospheric surface activity.

There are only two objects which show low-level variability in both campaigns (no. 14 and 22).
For these two objects no unambiguous period could be found. For object no. 22, this
can be explained by evolving spot distributions (see Sect\,\ref{longterm}). On the other hand, 
there are four variable objects whose flux modulation must be significantly reduced in one of the 
campaigns compared to the other, i.e. we find that the activity level of 50\% of our objects
has significantly changed in the course of 10 months. Thus, these results give first evidence 
for the evolution of spot activity on young VLM objects. 

\subsection{High-level variability}
\label{high}

Apart from the low amplitude periodic objects, there are 7 additional objects whose lightcurve
amplitudes are $>$0.25\,mag and reach up to 1.1\,mag. The periodograms of their lightcurves show 
several highly significant peaks. After fitting these periods with a sine wave and subtracting the 
fit, significant residuals with timescales of hours remain in most cases. The amplitude of these irregular 
residuals is strongest for targets no. 2, 33 and 43. We conclude that the signal of these targets 
consists of high-amplitude periodic variations, probably corresponding to the rotation period, and 
superimposed, sporadic, short-term features. The pooled variance diagrams for highly variable
objects imply that the timescale for these irregular variations is $0.2\ldots1.0$\,days.
(Sect.\,\ref{pool}).
 
It is not possible to explain this high-level variability by the usual scenario of photospheric 
spots alone: To cause a flux variation of 0.75\,mag, we would need a zero-emission spot with half the size of 
the visible hemisphere -- this is physically impossible. Eclipse events deliver no plausible interpretation 
as well, since they should cause strictly periodic light changes. Thus, we are confronted with a variability 
characteristic not known before for VLM objects. 

As outlined in Sect.\,\ref{intro}, however, the observed variability characteristic is very similar
to the high-level photometric variations of classical T Tauri stars (CTTS). Several groups obtained
lightcurves of young solar-mass stars which show a) periodic variability on timescales of days and b)
irregular variations on timescales of hours (e.g., Herbst et al. \cite{hmw00}, Fern\'andez \& Eiroa 
\cite{fe96}, Bouvier et al. \cite{bck95}). This behaviour is usually explained by the existence of hot spots 
formed by matter flowing from an accretion disk onto the stellar surface. Short-term variations can then be 
understood as a consequence of accretion rate variations and disk instabilities. Our photometry shows that 
the photometric behaviour of CTTS extends far down into the substellar regime, without obvious mass 
dependence. This implies the existence of T Tauri-like accretion disks around VLM objects. 

\begin{figure}[htbd]
    \centering
    \resizebox{\hsize}{!}{\includegraphics[angle=-90]{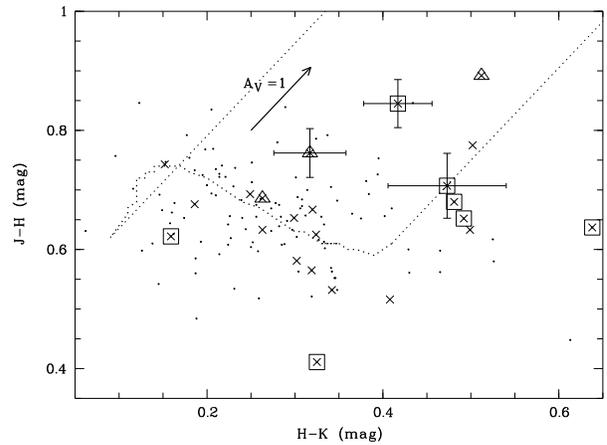}}
    \caption{(H-K,J-K) colour-colour diagram for the cluster member candidates in
    our field. Crosses are periodically variable objects. Overplotted symbols flag
    objects with high photometric amplitude: Squares mean that the lightcurve amplitude 
    exceeds 0.25\,mag, and thus mark the objects defined as high-amplitude targets in 
    Sect.\,\ref{inter}. Triangles are assigned to transition objects with lightcurve 
    amplitudes between 0.1 and 0.25\,mag. The solid line indicates the 3 Myr isochrone 
    from the Baraffe et al. evolutionary tracks. Dotted lines show the interstellar
    extinction vector calculated from Mathis (\cite{m90}). A reddening vector for
    $A_V=1$\,mag is given. For some reddened objects, we overplot the errorbars for the
    photometry.
    \label{jk}}
\end{figure}

A widely used method to check for the existence of circumstellar disks is the comparison
of the near-infrared colours with the intrinsic colours of the targets (e.g., Edwards et al. 
\cite{esh93}). A disk would manifest itself by a significant near-infrared excess. Figure\,\ref{jk} 
shows the (H-K,J-K) colour-colour diagram for the cluster member candidates, constructed from
2MASS data (see Table\,7). The solid line indicates the 3\,Myr isochrone from the Baraffe 
et al. (\cite{bca98}) evolutionary tracks. Periodically variable objects are marked with a cross. Overplotted 
squares are assigned to highly variable objects with amplitudes $>$0.25\,mag, triangles to transition objects
with amplitudes between 0.1 and 0.2\,mag. Most of the non-variable cluster members are distributed 
symmetrically around the isochrone, where the scatter is determined by the photometric errors, reconfirming
that the extinction towards $\sigma$\,Ori is negligible and uniform over the cluster area (see also B\'ejar 
et al. \cite{bzr99}, \cite{bmz01}). The periodic variable objects with low amplitude populate (with 
two exceptions) the same region in the diagram. 

The highly variable objects, in contrast, clearly tend to lie on the red side of the isochrone, down
or even outside the reddening band (dotted lines). Spectroscopy for three of these objects
assures their membership of the VLM population around $\sigma$\,Ori (see Sect.\,\ref{class}).
The fact, however, that the vast majority of the $\sigma$\,Ori members only scatter around the 
isochrone shows that the reddened objects must suffer from intrinsic reddening. The fraction 
of objects with near-infrared excess  (i.e. more than 1$\sigma$ reddening) is 71\% for high amplitude 
and 66\% for transition objects, but 15\% for low amplitude targets. Thus, we conclude that most of 
the highly variable objects exhibit a near-infrared excess, indicating that they are surrounded by 
an accretion disk. This may also apply to the transition objects. Figure\,\ref{jk} shows that some 
accreting objects might exist among low amplitude and non-variable targets as well.

The results from Sect.\,\ref{longterm} show that the high-level variations are persistent
over nearly one year, although with significantly changed amplitude. The only exception, 
object no. 16, with high-level variability in the TLS and {\it not} in the CA lightcurve, can be 
understood if it had a reduced amplitude, since this object has $I=18.8$\,mag and the mean rms 
at this brightness is 0.15\,mag in the CA time series. Thus having shown the permanence of 
high-amplitude variability, these targets are the first photometrically selected sample of strongly 
accreting VLM objects. 

Based on this conclusion, we can estimate disk frequencies and lifetimes. Out of
135 cluster member candidates, 7 show high-amplitude variations and thus probably possess an
accretion disk. This corresponds to a fraction of 5 to 7\% (taking into account a field
star contamination of up to 30\% of our VLM sample). Since we might have missed accreting objects 
without strong photometric variability, this should be considered as a lower limit.
For comparison, Barrado y Navascu\'es et al. (\cite{bbm03}), Oliveira et al. (\cite{ojk02}), 
and Barrado y Navascu\'es \& Mart\'{\i}n (\cite{bm03}) estimate disk frequencies of 5--14\% for 
VLM objects in the $\sigma$\,Ori cluster. Based on a very small sample of six Brown Dwarfs, 
Jayawardhana et al. (\cite{jas03}) derived a disk frequency of $33\pm24$\%. Given
the statistical uncertainties, all these results are in good agreement with our estimate. 
Thus, it seems that disks are rare in the $\sigma$\,Ori cluster, but they exist for 
VLM stars as well as for Brown Dwarfs. The fact that we do not observe totally irregular 
variability with high amplitudes is a further indication for the scarcity of strong accretors 
among VLM objects in the $\sigma$\,Ori cluster, since such variations are believed to be typical 
for the most actively accreting T Tauri stars in younger clusters (e.g., Herbst et al. 
\cite{hmw00}). 

The low disk frequency for VLM objects in the $\sigma$\,Ori cluster implies that most
of these objects dissipate their disks before they reach the age of this cluster. Since the most
probable age for the cluster is 3\,Myr (Zapatero Osorio et al. \cite{zbp02}), the disk lifetime
for VLM objects, defined as the timescale for essentially all the stars to lose
their disk (Haisch et al. \cite{hll01}), is roughly 3-4\,Myr. In comparison, solar-mass stars 
retain their disks significantly longer (Haisch et al. \cite{hll01}), their disk lifetime is 
6\,Myr. Thus, if the age estimate for the $\sigma$\,Ori cluster is valid, disk dissipation 
timescales are shorter in the VLM regime. This is confirmed by a recent study of Oliveira 
et al. (\cite{ojv03}), who investigated near-infrared colours of a sample of $\sigma$\,Ori 
cluster members with $J<13.3$\,mag (corresponding to $M>0.2\,M_{\odot}$). They derived an 
overall disk frequency of 46-54\%, but the near-infrared excess shows a strong decline 
for $K>11$\,mag, corresponding to $M<0.5\,M_{\odot}$ (see their Fig. 5). Thus, most of the 
stars with disks have solar-like masses, and the disk frequency is decreasing with mass.

\section{Rotation Periods}
\label{rot}

As discussed above, we attribute the detected periodicities to a rotational modulation of the
emitted flux. In this section, we will analyse the implied rotation periods. 
We will compare the periods from high-level variable with those for low-level variable objects.
We will then discuss the mass dependence of rotation. The period sample for all analyses comprises
14 CA and 9 TLS periods, excluding those TLS periodicities also detected in the CA lightcurves
(see also the discussion of these objects in Sect.\,\ref{longterm}).
The two rotation periods for $\sigma$\,Ori objects (SOri 31 and SOri 33) measured by Bailer-Jones 
\& Mundt (\cite{bm01}) will be used to underline the results.

\subsection{High amplitude vs. low amplitude}
\label{highlow}

In Sect.\,\ref{inter} we argue that we observe two kinds of variability, high-level and
low-level variations, where the high-level variability is interpreted as a consequence of
ongoing accretion processes, as confirmed by near-infrared photometry (Sect.\,\ref{high})
and spectroscopy (see Sect.\,\ref{accr}). Thus, highly variable objects are very likely
surrounded by circumstellar disks. This could have consequences for their rotational
behaviour, since magnetic coupling between star and circumstellar disk is believed 
to be the mechanism responsible to brake the rotation of T Tauri stars. For solar mass stars, 
there is indeed observational evidence for a correlation between rotation and disk indicators, 
in the sense that stars with disks tend to be slow rotators (e.g. Edwards et al. \cite{esh93}, 
Herbst et al. \cite{hbm02}). However, there exist controversial investigations which find no 
correlation between rotation and disk presence (e.g., Stassun et al. \cite{smv01}), indicating 
that disk-locking may not be the full solution to angular momentum distribution in young stars 
(Rebull et al. \cite{rws02}). Nevertheless, assuming that disk-locking plays a major role 
as an angular momentum regulation mechanism, we should expect that accreting VLM objects show a 
rotation period distribution different from the non-accreting ones. In particular, 
objects with disks should on average rotate more slowly. 

In Fig.\,\ref{peramp}, we plot photometric amplitude vs. angular velocity for all our targets.
As argued in Sect.\,\ref{high}, the plot shows two populations of datapoints, separated by a 
gap around amplitudes of $0.2\ldots0.25$\,mag, delineated by a dashed line. The objects above 
this line are probably active accretors. For these objects, the average angular velocity is 
$\Omega = 6.0\,d^{-1}$. For amplitudes $<0.25$\,mag, however, $\Omega$ is on average $11.1\,d^{-1}$. 
Thus, low-
amplitude objects tend to rotate faster on average, as we would expect for a disk-locking 
scenario. The periods from Bailer-Jones \& Mundt (2001) are also consistent with this picture, since 
they are shorter than 10\,h (corresponding to $\Omega>15$) and have amplitudes $<0.05$\,mag.
Hence, the available data may indicate disk-locking processes on VLM objects. Since this
result could be affected by low number statistics, we postpone its further discussion until
more rotational data are on hand.

It is instructive to compare our rotation periods with those recently published by Joergens et al.
(\cite{jfc03}). They measured 4 periods for Brown Dwarfs and VLM stars in the ChaI star forming region, 
which is 1\,Myr old and thus considerable younger than the $\sigma$\,Ori cluster. All their periods are 
longer than 2 days. In the context of the disk-locking scenario, these objects could be still 
disk-locked. This would be in agreement with the large disk frequencies (65\%) among Brown Dwarfs 
in the similarly-aged Trapezium cluster (Muench et al. \cite{mal01}). 

\subsection{Mass dependence}

Several groups have reported a bimodal distribution of the rotation periods for very young 
solar mass stars (e.g. Herbst et al. \cite{hbm02}, Edwards et al. \cite{esh93}). Since longer periods 
are significantly correlated with infrared excess attributed to the presence of a circumstellar 
disk, this is a direct sign for a disk-locking mechanism braking the rotation in the T Tauri phase.
There is observational evidence, however, that the bimodality becomes less obvious for low-mass 
objects and finally vanishes for objects with masses below $0.2\,M_{\odot}$ (Herbst et al. 
\cite{hbm01}). Our periods show no bimodal distribution throughout the whole mass regime. Since 
we observed only few targets with masses comparable to the solar mass, the detection of the
slow rotator peak in the histogram could be prevented by our low-number statistics.
Hence, our data allow no verification of the evolution of bimodality with decreasing mass.
Nevertheless, we note that there is definitely no bimodality within our statistical
uncertainties for objects with masses in the range from 0.03 to 0.4\,$M_{\odot}$.

\begin{figure}[htbd]
    \centering
    \resizebox{\hsize}{!}{\includegraphics[angle=-90]{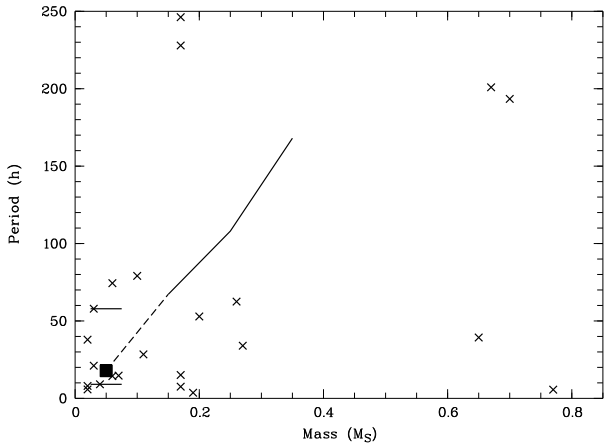}}
    \caption{Mass dependence of the rotation period: Plotted are all measured periods
    from TLS and CA vs. the object mass. The solid line shows the median period published
    by Herbst et al. (\cite{hbm01}). The median for our objects with $M<0.1\,M_{\odot}$ is shown
    as filled square, vertical lines indicate the quartiles (giving an estimate of the
    statistical uncertainties). This median found for our objects extends the line of Herbst 
    et al. (\cite{hbm01}) into the Brown Dwarf regime (dashed line).}
    \label{periods}
\end{figure}

To quantify the mass dependence of the periods, Herbst et al. (\cite{hbm01}) calculate a moving
median for their periods in the mass regime $0.1\ldots0.4\,M_{\odot}$, measured for objects
in the ONC. They find that lower-mass objects tend to rotate gradually faster. Our results confirm 
this result. Figure\,\ref{periods} shows masses vs. rotation periods for our $\sigma$\,Ori targets.
The median period is 46.1\,h for VLM stars, but only 14.7\,h for Brown Dwarfs. Our periods
make it possible for the first time to extend the analysis of Herbst et al. (\cite{hbm01}) to masses 
below their detection limit. Figure\,\ref{periods} shows the mass-period relation of Herbst et al. 
combined with our datapoint for masses $<0.1\,M_{\odot}$. Obviously, our periods nicely 
extend the mass-rotation relation down into the substellar regime. Because
of the statistical uncertainties, however, this needs further confirmation.

Recapitulating, we observe two effects: no bimodality and the tendency 
of faster rotation towards lower masses. Both could be explained by a decrease of the 
effectiveness of rotational braking through magnetic coupling to a circumstellar disk. 
Indeed, there is evidence for decreasing disk lifetimes with decreasing mass, as pointed
out in Sect.\,\ref{highlow}. Moreover, efficient angular momentum removal, either through
magnetic coupling or stellar winds, could also be prevented by low magnetic field strengths,
as implied by the the low activity level inferred from the small rate of variable objects
(see Sect.\,\ref{low}).

\section{Accreting VLM objects}
\label{accr}

\begin{figure}[htbd]
    \centering
    \resizebox{\hsize}{!}{\includegraphics[angle=-90]{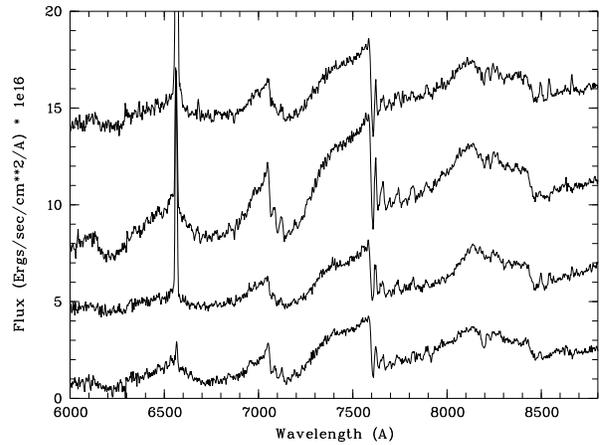}}
    \caption{Optical spectra for three highly variable targets (from top to bottom: no. 2, 33,
    43) and one non-variable target (no. 90, beneath). The spectra are shifted by 12, 6, 4, 0 
    (from top to bottom) units for clarity.}
    \label{spectra}
\end{figure}

In this section, we concentrate again on the discussion of the highly variable
targets, which were identified as actively accreting VLM objects in Sect.\,\ref{inter}.
To substantiate this interpretation, we obtained low-resolution spectroscopy with MOSCA
at the 3.5-m telescope on Calar Alto and grism {\it red500}, covering a wavelength
range of $5400\ldots10000$\,\AA\,with a dispersion of 2.9\,\AA/pixel. We observed two 
Brown Dwarfs and one VLM star with high-level variability in both time series (no. 2, 33, 43) 
and, for comparison, one non-variable Brown Dwarf (no. 90). For relative flux calibration, we 
additionally obtained a spectrum of the spectrophotometric standard star GD71 (Bohlin et al. \cite{bcf95}). 
Spectra of Ar/ArHg/Ne arc lamps were used for wavelength calibration. Reduction and calibration was 
done with the IRAF standard routines within the package {\it onedspec}. We note that the flux is 
considerably damped for $\lambda>8000\,$\AA, because of second order overlap in the spectra. All four 
spectra are shown in Fig.\,\ref{spectra}.

\subsection{Spectral classification}
\label{class}

The spectra show all the characteristics expected from late-type objects, 
in particular deep TiO and VO bands (Kirkpatrick et al. \cite{khm91}). 
We determined $T_\mathrm{eff}$ and $\log{g}$ from the spectra by comparing them with the 
STARdusty2000 models of Allard et al. (\cite{ahs00}). For this comparison, we used only the
spectral range from 6700 to 7700\,\AA, to exclude regions with strong emission lines and
avoid the long wavelength end, where our spectra are not properly flux calibrated. 
The model grid spans $T_\mathrm{eff}=2000\ldots4000$\,K in 100\,K steps and 
$\log{g}=3.5\ldots5.5$ in steps of 0.5. To account for varying atmospheric conditions during 
the observations and for possible continuum excesses as expected for accreting objects, we 
allowed for a wavelength-independent scaling and shifting of the models. 

We calculated the root mean square between each model and the data. It turns out that the 
rms values strongly depend on $T_\mathrm{eff}$, with a clear minimum. On the other hand, the 
dependency on $\log{g}$ is much weaker. Therefore, we first determined the effective temperature 
at which the rms is smallest. From the set of models with this $T_\mathrm{eff}$, we then selected
the model with the best fitting $\log{g}$. Visual comparison of the selected model with 
the data shows good agreement. The results, i.e. $T_\mathrm{eff}$ and $\log{g}$, are listed 
in Table\,\ref{spectype}. Spectral types are estimated with the effective temperature scale 
of Leggett et al. (\cite{lad00}), which is based on essentially the same models as our 
analysis.

\begin{table}[htbd]
\caption[]{Comparison of the spectra with models of Allard et al. (\cite{ahs00}). The first two columns
are the parameters of the best fitting model spectrum. The spectral types follow with the effective 
temperature scale of Leggett et al. (\cite{lad00}). For comparison, we additionally give the $T_\mathrm{eff}$ 
estimated from photometry alone, presuming cluster membership.
  $^*$\,Non-variable target observed for comparison}
  \label{spectype} 
  \begin{tabular}{l|rrrr}
  \hline
  no. & 2~~ & 33~ & 43~ & 90$^*$~\\
  \hline
  $T_\mathrm{eff}$\,(K)        & 2800 & 3100 & 2900 & 3000\\
  $\log{g}$                    & 4.0  & 3.5  & 4.0  & 3.5 \\
  Spectral type                & M5.5 & M4.0 & M5.0 & M4.5\\
  $T_\mathrm{eff}$\,(K) -- phot & 2930 & 3030 & 2890 & 2850\\
  \hline
  \end{tabular}
\end{table}

Assuming that the object is indeed a cluster member, a second estimate for $T_\mathrm{eff}$
can be obtained by comparing the photometry with the cluster isochrone of Baraffe et al. 
(\cite{bca98}). This value of $T_\mathrm{eff}$ will be either too high or too low, if the target 
resides in the fore- or background of the cluster. An estimate in agreement with the true 
$T_\mathrm{eff}$, as measured from the spectra, confirms the cluster membership. 
We cannot expect perfect agreement, however, since this comparison is affected by photometry errors and the
grid size of the models. A further complication is of course the large variability of targets no. 
2, 33 and 43. It might be that photometry and spectroscopy caught the object in a totally different 
accretion state. With these considerations in mind, all objects show good agreement between spectroscopic 
and photometric effective temperatures, confirming them als VLM cluster members of $\sigma$\,Ori.
The youth of all four targets its also confirmed by the low surface gravity values. Evolved objects
with comparable masses exhibit $\log{g}>5.3$ (Baraffe et al. \cite{bca98}), whereas we derived 
$\log{g}=3.5\ldots4.0$, in agreement with the values expected
for a 3\,Myr object with $T_\mathrm{eff}$ around 3000\,K. 

\subsection{Emission-line spectrum}
\label{emiss}

The dominant spectral feature in all three highly variable objects is the large H$\alpha$ emission 
line. The non-variable object has H$\alpha$ in emission as well, but with considerably lower intensity.
A variety of other emission features is present in the spectra, similar to emission spectra
of CTTS. We measured equivalent widths for all prominent emission 
lines by subtracting a linearly fitted continuum from the line flux (Table\,\ref{ew}). Obviously, 
the highly variable targets, in particular targets no. 2 and 43, show much more intense 
emission features than the comparison object. Target no. 33 seems to be more active than 
no. 90, judged from the H$\alpha$ equivalent widths. However, compared to no. 2 and 43,
H$\alpha$ and Ca emission lines are relatively faint. It might be that we caught target no. 
33 at the minimum of its lightcurve and thus in a relatively quiet state.  

Forbidden lines such as [OI] and [SII] can be attributed to low-density regions, e.g. in stellar
winds, and are thus tracers of a mass loss process. On the other hand, features like the
H$\alpha$ emission line and the Ca triplet are clear signs of the accretion process. These lines
are also common in outflows. For both features, however, we  attribute most of the
emission to accretion processes: On the one hand, the ratio between H$\alpha$ and [SII] in 
our spectra is much larger than 4, which is the upper limit given for (outflow-dominated) 
Herbig Haro objects (e.g. B{\"o}hm \& Goodson \cite{bg97}). On the other hand, the ratio of the 
Ca triplet equivalent widths is close to 1:1:1, whereas the predicted ratio for optically thin emission
would be 1:9:5. Similar values are common among strongly accreting CTTS (e.g., Reipurth et al.
\cite{rbg86}) and clearly differ from those of Herbig Haro flows. The presence of the He$\lambda$6678 line 
is a third indicator of ongoing accretion. We conclude that the emission spectrum of those highly 
variable objects is dominated by the signature of strong accretion processes. Comparing H$\alpha$
linewidths and spectral types with the classification scheme of White \& Basri (\cite{wb03}), 
the highly variable objects belong to the VLM analogues of classical T Tauri stars, while 
the non-variable object is similar to a weak-line T Tauri star.

\begin{table}[htbd]
\caption[]{Equivalent widths for prominent emission lines. The first column contains name and
laboratory wavelength (in \AA) of the emission line, then we list the equivalent widths for the
four targets. Non-detection of the feature is indicated with a dash, 'b' means that the line is 
present, but strongly blended by nearby features.\\ 
  $^*$ Non-variable target observed for comparison}
  \label{ew} 
  \begin{tabular}{l|rrrr}
  \hline
  no. & 2~~ & 33~ & 43~ & 90$^*$~\\
  \hline
  ~$[OI]$ 6300    ~~~  & ~b    ~ &~ 1.1  ~& ~b     ~& ~-   ~\\
  ~$[OI]$ 6364    ~~~  & ~0.9  ~ &~ -	 ~& ~-     ~& ~-   ~\\
  ~H$\alpha$ 6563 ~~~  & ~104.5~ &~ 21.3 ~& ~103.7 ~& ~10.7~\\
  ~HeI 6678	  ~~~  & ~1.8  ~ &~ 2.4  ~& ~-     ~& ~-   ~\\
  ~$[SII]$ 6716   ~~~  & ~0.23 ~ &~ 0.39 ~& ~-     ~& ~-   ~\\
  ~$[SII]$ 6731   ~~~  & ~0.38 ~ &~ 0.32 ~& ~-     ~& ~-   ~\\
  ~OI 8446        ~~~  & ~b    ~ &~ ~-   ~& ~-     ~& ~-   ~\\
  ~CaII 8498	  ~~~  & ~4.8  ~ &~ 0.6  ~& ~6.9   ~& ~3.8 ~\\
  ~CaII 8542	  ~~~  & ~3.8  ~ &~ -	 ~& ~5.1   ~& ~-   ~\\
  ~CaII 8662	  ~~~  & ~2.5  ~ &~ -	 ~& ~3.1   ~& ~-   ~\\
  ~HI (P9) 9229   ~~~  & ~0.8  ~ &~ 0.4  ~& ~0.2   ~& ~0.4 ~\\
  \hline
  \end{tabular}
\end{table}

Recapitulating, our low-resolution spectroscopy yields a clear result: Our highly variable 
VLM objects are probably accreting material from a disk and are thus the VLM analogues of CTTS. 
The T Tauri nature of young VLM objects is supported by recent publications, since spectroscopic 
signatures of typical T Tauri behaviour have been detected by other groups (e.g., 
Jayawardhana et al. \cite{jmb02}, White \& Basri \cite{wb03}, Barrado y Navascu\'es et al. 
\cite{bbm03}). The existence of disks around Brown Dwarfs is confirmed by independent observations 
of Natta \& Testi (\cite{nt01}), Natta et al. (\cite{ntc02}), Testi et al. (\cite{tno02}), Apai 
et al. (\cite{aph02}), Liu et al. (\cite{lnt03}), Jayawardhana et al. (\cite{jas03}), and L\'opez 
Mart\'{\i} et al. (\cite{les03}), mostly via mid-infrared excesses. All these findings set strong 
constraints for possible formation scenarios for VLM objects. In competition with star-like formation 
mechanisms, it was recently proposed that Brown Dwarfs form preferably in orbits around more massive 
stars, either in protoplanetary disks (Pickett et al. \cite{pdc00}) or as stellar embryos in multiple 
stellar systems (Reipurth \& Clarke \cite{rc01}). Both scenarios need an ejection mechanism to 
explain the large number of isolated Brown Dwarfs found in stellar clusters and star forming 
regions. It seems unlikely that the objects maintain a significant fraction of their accretion 
reservoir during this ejection process. Thus, the current observational picture favours 
star-like formation for VLM stars and substellar objects.

\section{Conclusions}
\label{conc}

We present the first photometric variability study for VLM members of the young
$\sigma$\,Ori cluster. With multi-filter photometry, 135 cluster member candidates
were identified, including 90\% VLM objects ($M<0.4\,M_{\odot}$). This preliminary
member list still contains about 30\% contaminating field stars and thus needs
spectroscopic verification.

We monitored these targets in two I-band time series in January and December 2001. The
first campaign, using the 2-m Tautenburg Schmidt telescope, covered 110 candidates,
spanning masses from 0.02 to 0.5\,$M_{\odot}$. The second time series, obtained with
the 1.23-m telescope on Calar Alto, registered 52 targets spanning a mass range from
0.02 to 1.4\,$M_{\odot}$. 27 Objects were observed in both runs. After reduction and
relative calibration, we obtained differential lightcurves for all targets. We found
that the difference image analysis package from G{\"o}ssl et al. (\cite{gr02}) improves
the precision of the Calar Alto lightcurves by several mmag.

Time series analysis was focused on the period search, but we included a test to detect
irregular variations as well. We examined the variability of the targets based on the 
scattering of the lightcurves and found that the very young targets contain significantly more 
variable objects than field stars. The fraction of variable objects as well as the 
variability amplitude does not significantly change over the entire mass range. 

We detected periodic variability for 14 CA and 13 TLS targets, with periods spanning from
4 to 240 hours. Four targets show a periodicity in both campaigns. The periodically variable 
candidates clearly fall into two groups: 16 of them show low-level variability which is
in most cases well-approximated by a sine wave. We argue that this type of variability is most
probably caused by magnetically induced photospheric spots co-rotating with the objects. We see
evidence for the spot evolution on VLM objects, since most of these targets show the periodicity 
in only one campaign.

The lightcurves of the remaining objects have amplitudes $>$0.25\,mag and show obvious
deviations from the sine shape. Their photometric behaviour is very similar to those of
classical T Tauri stars. The quasi-periodic oscillation is therefore most probably caused
by hot spots formed by matter flow from an accretion disk. Thus, our results imply the 
existence of accretion disks around VLM stars and Brown Dwarfs. The fact that most of
the highly variable objects show a near-infrared colour excess confirms this finding.
Low-resolution spectroscopy of a subsample of these targets assures their cluster membership and
their youth. We find strong emission features, in particular a dominating H$\alpha$ emission 
line, characteristic for accreting objects. The extension of T Tauri behaviour to VLM stars 
and Brown Dwarfs strongly supports star-like formation scenarios for these objects.

Highly variable and thus accreting objects rotate on average more slowly than low-level variable
targets. This is expected in terms of a disk-locking scenario, where magnetic coupling 
between star and disk removes angular momentum during the T Tauri phase.
In agreement with previous publications, we find that strong accretors are rare in the 
$\sigma$\,Ori cluster, but they certainly exist among VLM stars as well as Brown Dwarfs. 
Comparison with studies of younger clusters and more massive stars give evidence that disk 
dissipation timescales are significantly decreased with decreasing object mass.

In agreement with previous publications, we find a tendency of faster rotation towards 
decreasing object masses, possibly caused by decreasing effectiveness of angular momentum 
removal. This could either be explained with shorter disk lifetimes or with weaker
magnetic fields, as inferred from the low activity level of our targets. 

\begin{acknowledgements}
      It is a pleasure to acknowledge the constructive cooperation with the WeCAPP team. 
      The successful application of OIS would not have been possible without the intensive 
      support in particular from Arno Riffeser. We are very grateful to D. Roberts 
      who provided an implementation of the CLEAN algorithm. Software and support for the 
      pooled variance analysis were kindly provided by M. K{\"u}rster. We thank Arno Riffeser 
      and J{\"u}rgen Fliri for their assistance during the observations on Calar Alto.
      This work was supported by the German \emph{Deut\-sche For\-schungs\-ge\-mein\-schaft, DFG\/} 
      project number Ei~409/11--1. The publication makes use of data products from the Two Micron All 
      Sky Survey, which is a joint project of the University of Massachusetts and the Infrared Processing 
      and Analysis Center/California Institute of Technology, funded by the National Aeronautics and Space 
      Administration and the National Science Foundation.
\end{acknowledgements}

\end{document}